\newif\ifsubmode
\submodefalse

\ifsubmode
\documentclass[12pt,preprint]{aastex}
\else
  \documentclass{emulateapj}

  \usepackage{apjfonts}
\fi

\newcommand{\hdf}{HDF--N}
\newcommand{\hdfn}{HDF--N}

\newcommand{\hst}{\textit{HST}}
\newcommand{\spitzer}{\textit{Spitzer}}

\newcommand{\chandra}{\textit{Chandra}}

\newcommand{\wfu}{\hbox{$U_{300}$}}
\newcommand{\wfb}{\hbox{$B_{450}$}}
\newcommand{\wfv}{\hbox{$V_{606}$}}
\newcommand{\wfi}{\hbox{$I_{814}$}}
\newcommand{\acsb}{\hbox{$B_{435}$}}
\newcommand{\acsv}{\hbox{$V_{606}$}}
\newcommand{\acsi}{\hbox{$i_{775}$}}
\newcommand{\acsz}{\hbox{$z_{850}$}}
\newcommand{\nicj}{\hbox{$J_{110}$}}
\newcommand{\nich}{\hbox{$H_{160}$}}

\newcommand{\ks}{\hbox{$K_s$}}
\newcommand{\AAA}{\hbox{\AA}}
\newcommand{\lya}{Lyman~$\alpha$}

\newcommand{\ha}{\hbox{H$\alpha$}}
\newcommand{\lsim}{\lesssim}
\newcommand{\gsim}{\gtrsim}

\newcommand{\mcal}{\hbox{$\mathcal{M}$}}

\newcommand{\etal}{et al.}
\newcommand{\eg}{e.g.}
\newcommand{\ie}{i.e.}

\newcommand{\msol}{\hbox{$\mathcal{M}_\odot$}}

\newcommand{\lsol}{\hbox{$L_\odot$}}
\newcommand{\kms}{\hbox{km~s$^{-1}$}}

\newcommand{\jmk}{\hbox{$J - K_s$}}
\newcommand{\uJy}{\hbox{$\mu$Jy}}
\newcommand{\ujy}{\hbox{$\mu$Jy}}
\newcommand{\lir}{\hbox{$L_{\mathrm{IR}}$}}
\newcommand{\mone}{\hbox{$[3.6\mu\mathrm{m}]$}}
\newcommand{\mtwo}{\hbox{$[4.5\mu\mathrm{m}]$}}
\newcommand{\mthree}{\hbox{$[5.8\mu\mathrm{m}]$}}
\newcommand{\mfour}{\hbox{$[8.0\mu\mathrm{m}]$}}

\newcommand{\figcapzhist}{Distribution of redshifts for the DRG and
\hdf\ galaxy samples.  The top panel shows the redshift distribution
of DRGs with no 24~\micron\ detections.  The middle panel shows the
redshift distribution of DRGs with $f_\nu(24\micron) \geq 50$~\uJy.
The bottom panel shows the redshift distribution of the 24 galaxies
from the \hdfn\ with $1.5 \leq z \leq 3.5$ and $\ks \leq 23.2$~mag.
The dashed lines indicate the redshift limits of the sample.  The
shaded histogram shows the distribution for those \hdfn\ galaxies with
$f_\nu(24\micron) \geq 10$~\ujy.   \label{fig:zhist}}

\newcommand{\figcapsed}{Ensemble photometry of the DRGs in the
GOODS--S sample.  The data points are the measured flux densities from
the ACS (\acsb, \acsv, \acsi, \acsz), ISAAC ($J$, $H$, \ks), IRAC
(\mone, \mtwo, \mthree, \mfour), and MIPS 24~\micron\  images,
normalized to a common flux density at 1.6~\micron\ rest--frame.  Open
symbols show the photometry for galaxies detected at 24~\micron\ with
MIPS, and solid symbols show galaxies with no 24~\micron\ detection.
Stars correspond to galaxies detected in the X--ray data.   The shaded
region is defined by the SEDs for ordinary galaxies from Coleman,
Weedman, \& Wu (1980; CWW80), and span Hubble types from elliptical
galaxies through Sab, Sbc, and Scd spirals, to the bluest Magellanic
irregular, Im, templates.  The solid cyan curves show the SEDs from
0.1 to 3~\micron\ for stellar populations from the Bruzual
\& Charlot (2003; BC03) models formed with constant star formation for
100~Myr with dust extinction of 1.5, 4.5, and 7.5~mag at 1600~\AA\
using the \citet{cal00} extinction law.  The solid blue curve  shows
the SED of the local ULIRG Arp 220 from an empirical model  for
$1-5$~\micron\ \citep{sil98} and a composite spectrum covering
$5-3000$~\micron\ \citep{spo04}.  The latter shows the prominent PAH
emission feature at $7.7$~\micron, and silicate absorption at
$9.7$~\micron. The short--dashed, red curve shows the SED of the local
ULIRG and AGN Mrk~231, and the long--dashed magenta curve shows that
for the local Seyfert 2 galaxy NGC~1068 \citep{lef01}, both of which
rise smoothly through the near-- and mid--IR.  \label{fig:sed}}

\newcommand{\figcapjmkcc}{The $\ks-\mtwo$ versus $\acsi - \ks$,
color--color diagram for the GOODS--S DRGs.  Symbols show the derived
colors from the ACS, ISAAC, and IRAC photometry.  Black, open circles
show DRGs not detected at 24~\micron\ with MIPS.  Red, open squares
denote DRGs with 24~\micron\ detections.  Filled stars correspond to
sources detected in the soft or hard X--ray bands with
\textit{Chandra}.   Arrows denote upper limits on the ACS $\acsi - \ks$
colors.  The lines show the expected colors of stellar populations at
$z=2.2$ (the median redshift in the DRG sample) using the
\citet{bru03} models.   The dashed line shows the colors of a
passively evolving stellar population formed in a single burst with an
age of 50~Myr to 2.8~Gyr; hash marks indicate the colors at ages of
0.5, 1, and 2~Gyr, as labeled.   The solid lines show the colors of a
stellar population forming with constant star formation for the same
ages as above, but with a dust extinction of 2 and 6~mag at 1600~\AA,
as labeled.
The arrow shows the expected differential color for 2 magnitudes of
extinction at 1600~\AA. The histograms on the top and left panels show
the color distributions of those DRGs detected at 24~\micron\ (red
solid lines) and undetected at 24~\micron\ (shaded
regions).\label{fig:jmkcc}}

\newcommand{\figcaplirz}{Total IR luminosities, $\lir \equiv
L(8-1000\micron)$, of galaxies inferred from their observed MIPS
24~\micron\ emission.  Red circles show the IR luminosities for the
GOOD--S DRGs detected at 24~\micron, black stars show DRGs with both
X--ray and 24~\micron\ detections, and blue squares show objects
detected by MIPS 24~\micron\ in the \hdf.    The solid line denotes
the 50\% completeness limit of the GTO 24~\micron\ data in the CDF--S,
$f_\nu(24\micron) = 60$~\uJy.   Error bars include only uncertainties
on the 24~\micron\ flux density.   We estimate that the systematic
uncertainties on the conversion between $f_\nu(24\micron)$ and
$L(8-1000\micron)$ are typically $\approx 0.5$~dex (inset, and see
text).\label{fig:lirz}}

\newcommand{\figcapuvir}{Relation between the IR/UV luminosity ratio
and the UV spectral slope, $\beta$, where $f_\lambda \sim
\lambda^\beta$.  The solid line shows the relationship for local
UV--luminous starburst galaxies from \citet{meu99}.  Red pentagons
denote the GOODS--S DRGs, black stars indicate DRGs with X-ray
detections, and the symbol size scales with IR luminosity (see plot
inset).  Blue squares denote galaxies with 24~\micron\ detections from
the \hdfn.  The top abscissa shows the extinction at 1600~\AA\ that
corresponds to $\beta$ \citep[see][]{meu99}. Error bars on the IR/UV
luminosity ratios only correspond to the uncertainties on flux density
measurements. The systematic uncertainty in the conversion between
$f_\nu(24\micron)$ and \lir\ is $\approx 0.5$~dex, and is indicated by
the inset error bar. \label{fig:uvir}}

\newcommand{\figcapsfrsfr}{Comparison between the SFRs derived from the
extinction--corrected UV luminosity and the SFRs derived from the sum
of the UV and IR luminosities as described in the text.   Red circles
denote DRGs detected at 24~\micron, and gray triangles show upper
limits for DRGs with undetected at 24~\micron.  Black stars indicate
DRGs  detected in X--rays.   Blue squares denote galaxies from the
\hdfn\ at $1.5 \leq z \leq 3.5$.   Error bars on the UV + IR--derived
SFR are $\approx 0.5$~dex, as indicated by the inset error bar.  The
diagonal lines indicate constant ratios of 1, 10, and 100, as
labeled. \label{fig:sfrsfr}}

\newcommand{\figcapsedfitone}{Illustration of
spectral synthesis model fitting results for one of the GOODS--S DRGs
which has an indication of both young and old stellar populations.
The star formation history used here is parameterized as a monotonic,
decaying exponential with $e$--folding time $\tau$, as described in
the text.  The top panels show the 68 and 95\% confidence intervals on
various quantities plotted against the stellar population age: dust
extinction (parameterized as the color excess, $E[B-V]$), and the
stellar mass.   The cross hairs show the most--likely parameter values
in each two--dimensional projection of the full probability
distribution function. The bottom panel shows the best--fit model
spectrum in the observed frame, and the best--fit set of parameters
for this galaxy. The data points show the ACS \acsb\acsv\acsi\acsz,
ISAAC $JH$\ks, and IRAC 3.6--8.0~\micron\ photometry and
errors.\label{fig:sedfitone}} 

\newcommand{\figcapsedfittwo}{Same as
Figure~\ref{fig:sedfitone}, but for one of the DRGs with a
near--power-law-like rest--frame UV--to--near--IR SED.  The
best--fitting models generally prefer a SED that is dominated by young
stars and heavily extincted by dust.
\label{fig:sedfittwo}} 

\newcommand\figcapsedfittwocomp{Best--fit stellar--population models
with two components to the two example DRGs shown in
Figure~\ref{fig:sedfitone} and \ref{fig:sedfittwo}.  The solid grey
line shows the best--fitting two--component model, and the data points
correspond to the observed photometry.   The two galaxy components
correspond to a monotonically evolving stellar population (blue,
short--dashed line) and a stellar population formed in a single burst
at $z = \infty$ (red, long--dashed line).   The inset best--fitting
parameters refer to the monotonically evolving stellar population. The
mass fraction is the percent contribution of the younger,
monotonically evolving stellar population to the total stellar
mass. The top panels show the 68 and 95\% confidence regions on the
age and dust extinction parameters and correspond to the galaxy in the
spectrum that lies below each plot.   The cross hairs show the most
likely value in the two--dimensional projection of the full
probability distribution function. \label{fig:sedfittwocomp}}

\newcommand{\figcapchisq}{Comparison of the minimum reduced $\chi^2$
from fitting the DRGs with models of different star--formation
histories.  The abscissa shows the minimum reduced $\chi^2$ derived
for models with single--component, exponentially decaying
star--formation histories.  The ordinate shows the minimum reduced
$\chi^2$ for the two--component models where one component corresponds
to the exponentially decaying star--formation histories and the other
corresponds to an instantaneous burst of star--formation at
$z_\mathrm{form} = \infty$, that evolves passively
thereafter. \label{fig:chisq}}

\newcommand{\figcapmassz}{Stellar masses derived from modeling the
galaxy SEDs, plotted as a function of redshift.  The left panel shows
the masses derived using the single--component, monotonically evolving
star--formation histories.  The right panel shows the masses derived
using two--component models.  In both panels filled red circles denote
the GOODS--S DRGs undetected at 24~\micron, and open red circles show
those DRGs with 24~\micron\ detections.  Blue squares denote the
galaxies in the \hdf. Black stars show those DRGs with X--ray
detections.  Error bars are not shown for clarity, but mean error bars
are shown as a function of galaxy stellar mass.  The short--dashed
line shows the value of the characteristic stellar mass of the
present--day mass function \citep{col01}.  In the right panel, the
long--dashed line shows the stellar--mass for a passively evolving
stellar population formed instantaneously at $z=\infty$ with $\ks =
23.2$, which is the stellar--mass limit for the GOODS--S \ks--band
data for such systems. \label{fig:massz}}

\newcommand{\figcapsfrmass}{The SFRs derived from the sum of the UV
and IR luminosities plotted as a function of galaxy stellar mass.  Red
circles denote the GOODS--S DRGs with 24~\micron\ detections; gray
triangles show upper limits for DRGs undetected at $24$~\micron,
assuming they have $f_\nu(24\micron) \leq 60$~\ujy.
Black stars indicate those DRGs with X-ray detections.   Blue squares
correspond to the 24~\micron--selected galaxies in the \hdf.  All
galaxy samples in this plot are restricted to the redshift range $1.5
\leq z \leq 3.0$.   Systematics dominate the uncertainties on the SFRs
and are $\approx 0.5$~dex, as indicated by the inset error bars. The
error bars also show the mean uncertainty on the stellar masses in
three mass bins.\label{fig:sfrmass}}

\newcommand{\figcapspecsfrmass}{The specific SFR as a function of
galaxy stellar mass.  The top panel shows the results for the DRG
sample and the \hdf\ galaxies, all restricted to $1.5\leq z \leq 3.0$,
with symbols and errors as in Figure~\ref{fig:sfrmass}. The bottom
panels  show the specific SFRs as a function of stellar mass for
lower--redshift galaxy samples from the COMBO--17 survey (as
labeled). Open circles show COMBO--17 galaxies detected with \spitzer\
at 24~\micron, and small filled circles show upper limits for
COMBO--17 galaxies undetected at 24~\micron. \label{fig:specsfrmass}}

\newcommand{\figcapspecsfrevol}{Evolution of the integrated specific
SFR, \ie, the ratio of the total SFR to the total stellar mass.  The
curves show the expected evolution of the ratio of the total SFR to
the total galaxy stellar mass densities from an empirical fit to the
evolution of the SFR density \citep[solid lines, thick line includes
correction for dust extinction;][]{col01}, and the model of
\citet[dashed line;][]{her03}.   The data points show results for
galaxies with $\mathcal{M} \geq 10^{11}$~$\mathcal{M}_\odot$. The
filled diamonds show the mean values derived for COMBO--17 galaxies,
and the filled circle shows the mean value for the DRGs.  The data
point for the DRGs sets the SFR of those galaxies without MIPS
detections to zero.  The lower bound of the box around the DRG point
shows the value if we exclude those DRGs with direct X--ray detections,
and those DRGs with $\lir \geq 10^{13}$~\lsol\ or IR colors indicative
of AGN (see text).  The upper bound shows the value for the DRGs with
no MIPS detection have $f_\nu(24\micron) = 60$~\ujy.  The error bars
on the data points themselves are derived by a bootstrap resampling of
the dataset.   These do not include systematic uncertainties in the
SFRs, which are indicated by the inset error
bar. \label{fig:specsfrevol}}

\shorttitle{SPITZER OBSERVATIONS OF MASSIVE, RED GALAXIES AT HIGH REDSHIFT}
\shortauthors{PAPOVICH ET AL.}

\begin{document}

\slugcomment{Accepted for Publication in the Astrophysical Journal}
\title{\textit{SPITZER} OBSERVATIONS OF MASSIVE, RED GALAXIES AT HIGH REDSHIFT\altaffilmark{1}} 


\author{\sc C.~Papovich\altaffilmark{2,3}, 
L.~A.~Moustakas\altaffilmark{4}, 
M.~Dickinson\altaffilmark{5}, 
E.~Le~Floc'h\altaffilmark{2,6},
G.~H.~Rieke\altaffilmark{2}, 
E.~Daddi\altaffilmark{3,5},
D.~M.~Alexander\altaffilmark{7},
F.~Bauer\altaffilmark{8},
W.~N.~Brandt\altaffilmark{9},
T.~Dahlen\altaffilmark{10,11}, 
E.~Egami\altaffilmark{2},
P.~Eisenhardt\altaffilmark{4},
D.~Elbaz\altaffilmark{12},
H.~C.~Ferguson\altaffilmark{10}, 
M.~Giavalisco\altaffilmark{10}, 
R.~A.~Lucas\altaffilmark{10},
B.~Mobasher\altaffilmark{10}, 
P.~G.~P\'erez--Gonz\'alez\altaffilmark{2},
A.~Stutz\altaffilmark{2},
M.~J.~Rieke\altaffilmark{2}, 
and H.~Yan\altaffilmark{13}} 

\altaffiltext{1}{This work is based in part on observations
made with the \textit{Spitzer Space Telescope}, which is operated by
the Jet Propulsion laboratory, California Institute of Technology,
under NASA contract 1407; on observations taken with the NASA/ESA
Hubble Space Telescope, which is operated by the Association of
Universities for Research in Astronomy, Inc.\ (AURA) under NASA
contract NAS5--26555; and observations collected at the Kitt Peak
National Observatory (KPNO), National Optical Astronomical
Observatories (NOAO), which is operated by AURA, Inc., under
cooperative agreement with the National Science Foundation.
Observations have also been carried out using the Very Large Telescope
at the ESO Paranal Observatory under Program ID: LP168.A-0485}
\altaffiltext{2}{Steward Observatory, University of Arizona, 933 North Cherry
  Avenue, Tucson, AZ 85721; papovich@as.arizona.edu}
\altaffiltext{3}{Spitzer Fellow}
\altaffiltext{4}{Jet Propulsion Laboratory, California Institute of
Technology,  Mail Stop 169-327, 4800 Oak Grove Dr, Pasadena, CA  91109}
\altaffiltext{5}{National Optical Astronomy Observatory, 950 North
Cherry Avenue, Tucson, AZ 85721}
\altaffiltext{6}{Associated to Observatoire de Paris, GEPI, 92195
Meudon, France} 
\altaffiltext{7}{Institute of Astronomy, Madlingley Road, Cambridge
CB30HA, UK}
\altaffiltext{8}{Chandra Fellow, Columbia Astrophysics Laboratory, Columbia
University, 500 W.\ 120th St., New York, NY 10027}
\altaffiltext{9}{Department of Astronomy and Astrophysics, 525 Davey
Laboratory, Pennsylvania State University, University Park, PA 16802}
\altaffiltext{10}{Space Telescope Science Institute, 3700 San Martin
Drive, Baltimore, MD 21218}
\altaffiltext{11}{Department of Physics, Stockholm University, SE-106
91, Stockholm, Sweden} 
\altaffiltext{12}{CEA Saclay/Service d'Astrophysique, Orme des
Merisiers, F-91191 Gif-sur-Yvette Cedex, France} 
\altaffiltext{13}{Spitzer Science Center, California Institute of
Technology, MS 220-6, Pasadena, CA 91125}


\begin{abstract}
We investigate the properties of massive galaxies at $z\sim 1-3.5$ using
\textit{Hubble Space Telescope} observations at optical wavelengths,
ground--based near--infrared (IR) imaging, and \textit{Spitzer Space
Telescope} observations at 3--24~\micron.   From \ks--selected
galaxies over a $\simeq 130$~arcmin$^2$ field in the southern Great
Observatories Origins Deep Surveys (GOODS--S), we identify  153
distant red galaxies (DRGs) with $(\jmk)_\mathrm{Vega} \ge 2.3$.  This
sample is approximately complete in stellar mass for passively
evolving galaxies above $10^{11}$~\msol\ and $z\leq 3$.    Of the
galaxies identified by this selection, roughly half are objects whose
optical and near--IR rest--frame light is dominated by evolved stars
combined with ongoing star formation (at $z_\mathrm{med}\sim 2.5$),
and the others are galaxies whose light is dominated by heavily
reddened ($A_{1600} \gsim 4-6$~mag) starbursts (at $z_\mathrm{med}
\sim 1.7$).  Very few of the galaxies ($\lsim 10$\%) have no
indication of current star formation.   The total star--formation
rates (SFRs) including the reradiated IR emission for the DRGs are
up to two orders of magnitude higher than those derived from the UV
luminosity corrected for dust reddening.  We use population synthesis
models to estimate stellar masses and to study the stars that dominate
the rest--frame UV through near--IR light in these galaxies. DRGs at
$z\sim 1.5-3$ with stellar masses $\geq 10^{11}$~\msol\ have specific
SFRs (SFRs per unit stellar mass) ranging from 0.2 to 10~Gyr$^{-1}$,
with a mean value of $\sim 2.4$~Gyr$^{-1}$.  Based on the X--ray
luminosities and rest--frame near--IR colors, as many as one--quarter
of the DRGs may contain AGN, implying that the growth of supermassive
black holes coincides with the formation of massive galaxies at
$z\gsim 1.5$.   The DRGs with $\mcal \geq 10^{11}$~\msol\ at $1.5 \leq
z \leq 3$ have integrated specific SFRs greater the global value over
all galaxies at this epoch.   In contrast, we find that galaxies at
$z\sim 0.3-0.75$ with $\mcal \geq 10^{11}$~\msol\ have integrated
specific SFRs less than the global value, and more than an order of
magnitude lower than that for massive DRGs at $z\sim 1.5-3$.  At
$z\lsim 1$, lower--mass galaxies dominate the overall cosmic mass
assembly.   This suggests that the bulk of star formation in massive
galaxies occurs at early cosmic epochs and is largely complete by
$z\sim 1.5$.  Further mass assembly in these galaxies takes place with
low specific SFRs.
\end{abstract}
 
 

\section{Introduction}\label{section:intro}

Most of the stellar mass in galaxies today apparently  formed during
the relatively short period between $z\sim 3$ and 1
\citep[\eg][]{dic03,rud03,fon04,gla04}.   Some early--type galaxies
appear as soon as $z\sim 1.5-2$ \citep{dun96,spi97,mccar04b,dad05}.
By $z\sim 1$, there is a significant population of galaxies with red
colors and morphologies consistent with passively--evolving
early--type galaxies, implying they formed their stellar populations
at $z_\mathrm{form} \gsim 2-3$ \citep[\eg,][for a
review]{cim02,mou02,mou04,pap05,treu05,mccar04}.   The cosmic
star--formation rate (SFR) density has declined by roughly a factor 10
from $z\sim 1$ to the present--day \citep[\eg][and references
therein]{hop04}.   During the time since $z\sim 1$ the stellar mass in
passively--evolving, early--type galaxies has increased by less than a
factor 2 \citep[\eg][]{bri00,bel04}, and at the present epoch, roughly
one--third of all stars exist in such galaxies \citep{bal04}.

Although massive galaxies appear to have formed most of their stars at
epochs prior to $z\sim 1-2$, we have few constraints on how they
assembled their stellar mass.  One hypothesis is that galaxies
``downsize'' and star formation shifts from galaxies at the high to
the low end of the mass function with decreasing redshifts
\citep[\eg,][]{cow99,fon03,hea04,kau04,bauer05,cap05,jun05,per05}.
Another possibility is that massive galaxies assemble their stellar mass
early--on, either \textit{in situ}, ``closed--box'' formation events  with
subsequent passive evolution \citep[\eg,][]{egg62}, or in low--mass
systems, which then coalesce to form large galaxies with little
subsequent star formation \citep[\eg,][]{bau98,kau98,cim02b}.   

Testing these proposals has been frustrated by difficulties in
conducting a complete census of star--forming galaxies and massive
galaxies at $z\gsim 2$, when such systems are expected to experience a
peak in their stellar assembly rates \citep[\eg][]{delucia05,nag05}.
Ultraviolet (UV)--luminous star--forming galaxies at these redshifts
are identified by the characteristic ``break'' in their colors due to
neutral hydrogen absorption that attenuates the flux shortward of
\lya\ (1216~\AA) and the 912~\AA\ Lyman limit \citep[\eg][for a
review]{ste96,giav02}.   These UV--dropout, Lyman--break galaxies
(LBGs) dominate the UV luminosity density at $z\sim 2-6$, and possibly
the global SFR density at these redshifts
\citep[\eg,][]{ste99,bou04,gia04b}.  However, the UV--dropout
technique is primarily sensitive to galaxies with ongoing, relatively
unreddened star formation.  Surveys with SCUBA have identified a
population of sub--mm galaxies at $z\gsim 2$ that emit the bulk of
their bolometric luminosity at infrared (IR) wavelengths
\citep[see][for a review]{bla02}, and may contribute substantially to
the cosmic SFR \citep{bar00}.   Their inferred space densities and
SFRs suggest that they could be the progenitors of the most massive
present--day galaxies \citep{bla04,cha05}.  Neither the LBG nor the
sub-mm--galaxy populations necessarily provide a full sample selected
by stellar mass, so how they participate in the formation of
present--day massive galaxies is poorly understood.  Hence, it is
unclear what fraction of present--day galaxies pass through such
stages during their assembly.

Theoretically, the star--formation histories of massive galaxies are
also poorly known.   Galaxies within massive halos have short cooling
times and tend to convert all their gas into stars rapidly, unless
feedback is invoked from stellar winds and supernovae to reheat the
gas or prevent it from cooling \citep[\eg][]{her03,spr05a,spr05b}.
Models predict that massive galaxy halos continue to accrete smaller
aggregates up to the current epoch, which rejuvenates star--formation
and predicts galaxy colors that are too blue compared to
observations \citep[\eg][]{som01} unless very large dust extinctions
or non--standard stellar initial mass functions (IMFs) are invoked
\citep{bau05,nag05}.  Some recent theoretical models suppress
star--formation at late times in massive galaxies using feedback from
active--galactic nuclei \citep[AGN;][]{gran01,dimat05,hop05,spr05a}.   This
process may provide the impetus for the present--day
black-hole--bulge-mass correlation \citep{mag98,geb00,kau04}.
We require observations of star--formation and AGN activity
in high--redshift, massive galaxies to improve our understanding of
how such systems form.

Surveys using deep, near--IR observations have identified
high--redshift, massive--galaxy candidates with red near--IR colors
\citep[\eg,][]{dic00,tot01,im02,dad04,yan04a}.  \citet{fra03} used
deep  $JHK_s$ observations from VLT/ISAAC to identify a population of
galaxies with $(\jmk)_\mathrm{Vega} > 2.3$~mag in the Faint IR
Extragalactic Survey (FIRES).    In principle, this color selection
identifies galaxies that have a strong Balmer/4000~\AA\ break between
the $J$ and \ks\ bands at $z\sim 2-3.5$, down to an approximately
complete limit in stellar mass. Subsequent analysis has concluded that
these distant red galaxies (DRGs) are mostly massive, old, and
actively forming stars at $z\sim 1.5-3.5$
\citep{vandok03,for04,rub04,knu05,red05} although some appear to be
completely devoid of star--formation and passively evolving
\citep{lab05}, while others appear to host powerful AGN
\citep{vandok04}.   The inferred stellar masses of DRGs at $z\sim 2-3$
are similar to those of local early--type galaxies \citep{for04}.
They are generally higher than those inferred from LBG samples at
similar redshifts \citep{saw98,pap01,sha01}, although some overlap
between the two clearly exists \citep{sha05,red05}.  The estimated
stellar population ages of the DRGs suggests that they have been
forming stars since $z\sim 5-6$ \citep{for04}.  Thus DRGs may
represent the older stellar populations formed in higher--redshift
LBGs \citep[\eg][]{pap04a}, and these galaxies possibly link the
UV--luminous LBGs to the sub--mm galaxies \citep{vandok04}.

In addition, extinction may contribute to the red colors of some of
the galaxies. \citet{sma02} find that dust--extincted starbursts
at $z\sim 1-2$ selected as extremely red objects (EROs) typically have
red $J-K$ colors similar to DRGs.   Using current  hierarchical model
predictions, \citet{nag05} suggest that massive galaxies in formation
at $z\gsim 1$ may be heavily reddened by dust.   If so, then most of
their emission should appear in the thermal IR. In fact, up to half of
EROs (selected with red $R-K$ or $R-[3.6\micron]$ colors) at $z\gsim
1$ are detected in the thermal IR by \spitzer/MIPS at 24~\micron\
\citep{wil04,yan04b}.   In addition, the inferred evolution of the
galaxy population responsible for the IR number counts implies a
substantial population of galaxies at $z\sim 1-3$
\citep{pap04b,cap05,per05}, and the evolution in the IR luminosity
function suggest that  luminous IR galaxies (LIRGs, $\lir =
10^{11-12}$~\lsol), and ultraluminous IR galaxies ($\lir \geq
10^{12}$~\lsol) dominate the IR luminosity density at $z\gsim 1$
\citep{lef05,per05}.

For this study, we selected galaxies at $z\sim 1 - 3.5$ with
$(\jmk)_\mathrm{Vega} > 2.3$~mag ([$\jmk]_\mathrm{AB} > 1.37$~mag) from
a $\ks$--band selected catalog in the southern Great Observatories
Origins Deep Survey (GOODS-S) field.  At the magnitude limit of $\ks
\leq 23.2$~[AB], this sample is approximately complete in
stellar mass for passively evolving galaxies with  $\mcal >
10^{11}$~\msol\ for $z\leq 3$.  We also use IR and X--ray observations
to constrain the star--formation and AGN processes in these galaxies.
The AGN connection in these galaxies is explored in more detail in a
forthcoming paper (L.~A.~Moustakas et al., in preparation).  In
\S~\ref{section:data}, we summarize the data and we define the galaxy
samples.  In \S~\ref{section:jmk}, we describe broad properties of the
DRGs using their rest--frame UV to near--IR colors. In
\S~\ref{section:lirsfr}, we compare the IR luminosities for the
galaxies with those derived from their UV luminosity and measured
extinction, and we compare the DRGs to other galaxies at $1.5 \leq z
\leq 3.5$.  In \S~\ref{section:sedfit}, we use population synthesis
models  to estimate the properties of the galaxies' stellar
populations.   In \S~\ref{section:discussion}, we discuss the
relationship between the SFRs and stellar population properties for
the ensemble of galaxies, and we comment on the presence of AGN.   We
also compare the distribution of SFR as a function of galaxy stellar
mass at high redshift, and we compare with results at lower redshifts
($z\sim 0.3-0.75$).  In \S~\ref{section:conclusions} we present our
conclusions.   

Throughout this paper we use a cosmology with $\Omega_\mathrm{Total} =
1$, $\Omega_\mathrm{M} = 0.3$, $\Lambda = 0.7$, and $H_0 = 70$~\kms\
$h_{70}$ Mpc$^{-1}$ where $h_{70} \equiv 1$.   Unless otherwise
specified, we present all magnitudes in the AB system, $m_\mathrm{AB}
= 23.9 -  2.5\log( f_\nu/$1~$\mu$Jy$)$.  We denote galaxy magnitudes
from the \hst\ ACS bandpasses F435W, F606W, F775W, and F850LP as
\acsb, \acsv, \acsi, and \acsz, respectively.   Similarly, where
applicable we  denote magnitudes from the \hst\ WFPC2 and NICMOS
bandpasses F300W, F450W, F606W, F814W, F110W, and F160W as \wfu, \wfb,
\wfv, \wfi, \nicj, and \nich, respectively.   We also denote
magnitudes from the four \spitzer\ IRAC channels as \mone, \mtwo,
\mthree, and \mfour, respectively.


\section{The Data and Sample Definitions}\label{section:data}

The GOODS--S field center is located in the southern \textit{Chandra}
Deep X--ray field (CDF--S) at
$3^\mathrm{h}32^\mathrm{m}30^\mathrm{s}$, 
$-27^\circ48^\prime20^{\prime\prime}$, which provides 1~Ms of imaging
in the soft (0.5--2~keV) and hard (2--8~keV) X--ray bands in this
field.   The available observations include imaging with \hst/ACS in
four passbands, \acsb, \acsv, \acsi, and \acsz\ over 160~arcmin$^2$,
ground--based near--IR imaging from VLT/ISAAC in the $J\ks$ bands over
130~arcmin$^2$ (with $H$--band imaging over $\simeq 50$~arcmin$^2$),
and IR imaging from \spitzer/IRAC in four bands, \mone, \mtwo,
\mthree, \mfour.      

The ACS observations and data reduction are described in
\citet{gia04a}.  The images have a PSF FWHM $\simeq 0\farcs125$, and
the limiting $10\sigma$ sensitivities are $\acsb = 27.8$, $\acsv =
27.8$, $\acsi = 27.1$, $\acsz = 26.6$, measured in
$0\farcs2$--diameter circular apertures.   The \spitzer/IRAC images
have a PSF FWHM ranging from $\simeq 1\farcs5$ at 3.6~\micron\ to
$\simeq 2$\arcsec at 8~\micron, and for isolated point sources achieve
$5\sigma$ limiting sensitivities ranging from 0.11~\ujy\ at
3.6~\micron\ to 1.66~\ujy\ at 8~\micron\ (M.\ Dickinson, et al., in
preparation).

The near--IR ISAAC imaging is from the version 1.0 release (B.~Vandame
et al., in preparation),  available on the ESO/GOODS
webpages.\footnote{http://www.eso.org/science/goods/releases/20040430/}
The ISAAC data have excellent image quality (full--width at half
maximum, FWHM $\approx0\farcs45$) with mean exposures times of 14000
and 24000~s in $J$ and \ks, respectively, reaching limiting magnitudes
of $J = 24.7$, $H=24.1$ and $\ks = 24.1$ (10~$\sigma$) in
1\arcsec--diameter apertures, although the depth varies over the
GOODS--S field.

\spitzer\ imaged the CDF--S field with MIPS at 24, 70, and
160~\micron\ under \spitzer/Guaranteed Time Observer (GTO) time.
Here, we focus exclusively on the 24~\micron\ imaging, which was
reduced using the instrument team Data Analysis Tool \citep{gor05}.
The GTO MIPS imaging covers $1^\circ \times 0.5^\circ$  with a FWHM
$\approx 6$~arcsec, and covers all of the GOODS--S field and most of
the ESO imaging and COMBO--17 surveys (see \S~\ref{section:combo17}).

\subsection{GOODS Source Cataloging and DRG Sample
Selection}\label{section:select_drgs} 

We use a source catalog selected from the ISAAC \ks--band  data.  We
rebinned the ACS data to the pixel scale of ISAAC, and convolved the
ACS images to match the image quality of the ISAAC images.  Source
catalogs were then constructed using the SExtractor software
\citep{ber96} by first locating sources on the \ks--band image, then
measuring photometry in matched apertures on the \hst/ACS and ISAAC
images.   We measured photometry in each band in isophotal apertures
defined from the \ks--band image.   We then scaled these to total
magnitudes using the difference between the \ks--band isophotal
magnitude (SExtractor MAG\_ISO) and the \ks--band magnitude measured
in a ``total'', elliptical aperture defined by the Kron radius
(SExtractor MAG\_AUTO).  Photometric uncertainties are derived by
SExtractor after adjusting the image rms maps to account for the
correlated noise properties introduced by drizzling.\footnote{see
ftp://archive.stsci.edu/pub/hlsp/goods/v1/h\_goods\_v1.0\_rdm.html}
The SExtractor--derived uncertainties still likely underestimate the
true errors, because they do not account for systematic errors in the
measurements themselves.  Therefore, we have included an additional
error of $\sigma_\mathrm{sys}/f_\nu \approx 3$\%, added in quadrature
to the uncertainties on the ACS and ISAAC photometry.

We detected objects in the \spitzer/IRAC images using a weighted--sum
image of IRAC channels 1 and 2.   Magnitudes were then measured in
each IRAC band in 4\arcsec--diameter apertures, and we applied
aperture corrections of 0.30, 0.34, 0.53, and 0.67~mag to the bands
\mone, \mtwo, \mthree, and \mfour, respectively, to correct to total
magnitudes.  We did not attempt to measure photometry on
versions of the ISAAC data PSF--matched to the IRAC data, because of
the significantly poorer image quality of the IRAC data. The aperture
corrections are based on Monte Carlo simulations using artificial,
compact sources added to the real images, and are appropriate for
sources with half--light radii $< 0\farcs5$, such as those of interest
here. These simulations also allow us to estimate the
error on the IRAC photometry as a function of flux density.  The
uncertainties are similar for IRAC channels 1 and 2, and range from
$\Delta m \approx 0.03$~mag at $m\approx 21$~mag to $\Delta m \approx
0.3$~mag at $m\approx 25$~mag.   For IRAC channels 3 and 4, they range
from  $\Delta m \approx 0.05$~mag at $m\approx 21$~mag to $\Delta m
\approx 0.4$~mag at $m\approx 25$~mag.

We used $(\jmk)_\mathrm{Vega} > 2.3$~mag to identify DRGs.   In the
GOODS--S field, we found 153 of them to a signal--to--noise ratio
$\mathrm{S/N}(\ks) \geq 10$ limit, which is the median S/N of objects
at $\ks \leq 23.2$~mag within the ``total'', MAG\_AUTO apertures, and
with \jmk\ colors measured in the seeing--matched MAG\_AUTO apertures.
Restricting sources to the S/N requirement is appropriate as the ISAAC
depth varies over the GOODS--S field.  Using a S/N limit also ensures
that we can derive robust colors from the ACS and ISAAC data, greatly
improving the accuracy of our SED modeling and photometric redshifts.
We inspected the ACS image for each DRG at their
native resolution to identify objects resulting from chance
galaxy--galaxy alignments along the line of sight.  In one case the
DRG does appear to involve multiple ACS sources, blended at the K--band
resolution.  We exclude this object although its inclusion
does not effect the results in this paper.  A unique IRAC source
exists for 132 of the 153 DRGs with a matching radius of $r\leq
0\farcs5$.   We visually inspected each object to verify that the
matched IRAC source corresponds to the DRGs in the $\ks$--band image.
The unmatched objects suffer from crowding in the IRAC images from
other sources within $\simeq$~1--2\arcsec. The IRAC flux from these
non--detected DRGs is either completely blended within the isophote of
the neighbor, or confusion with the neighbor offsets the centroid of
the IRAC flux past the matching criterion.

\ifsubmode
\else
\begin{figure}
\plotone{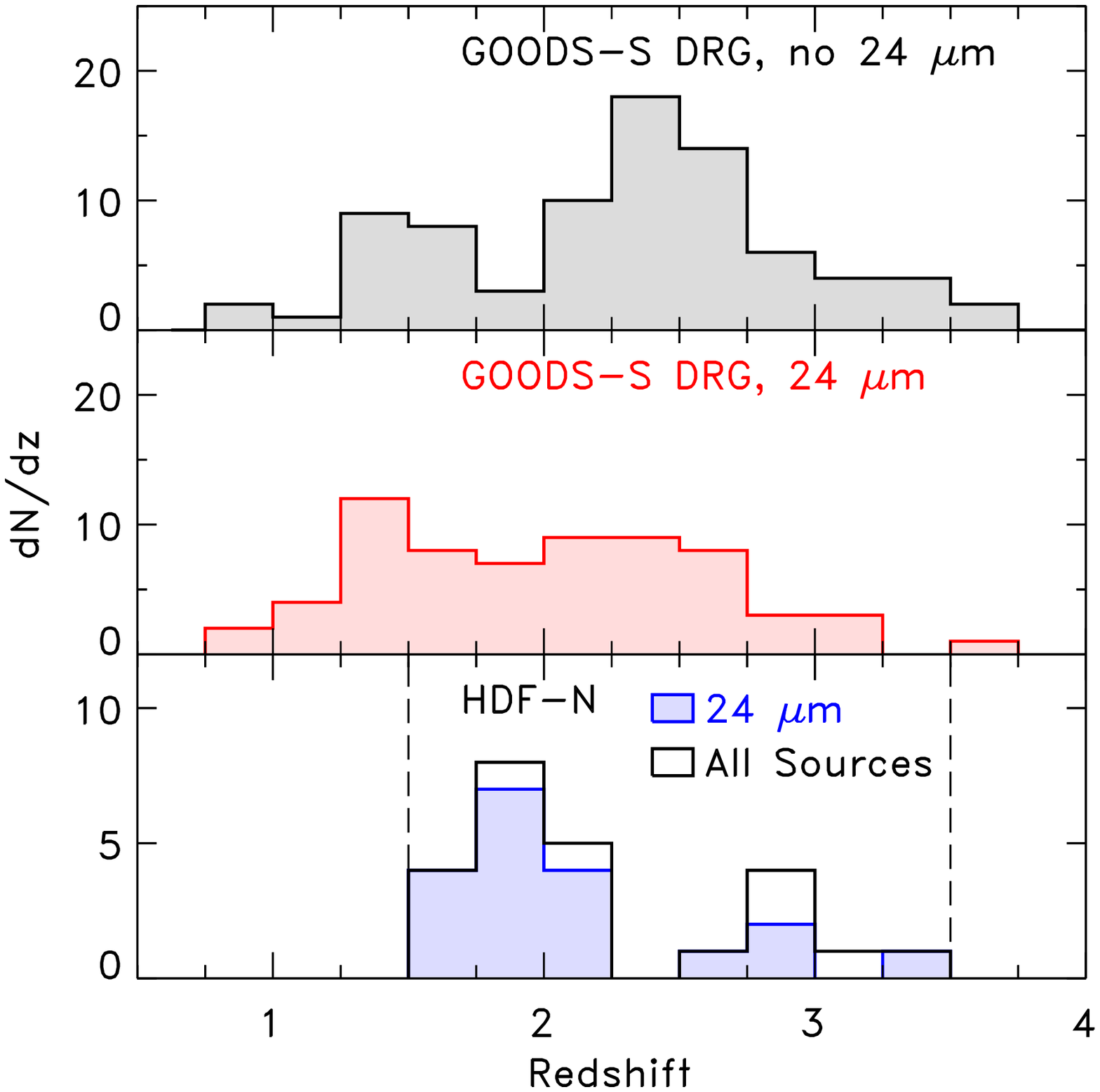}
\caption\figcapzhist
\end{figure}
\fi
\ifsubmode
\begin{figure}
\plotone{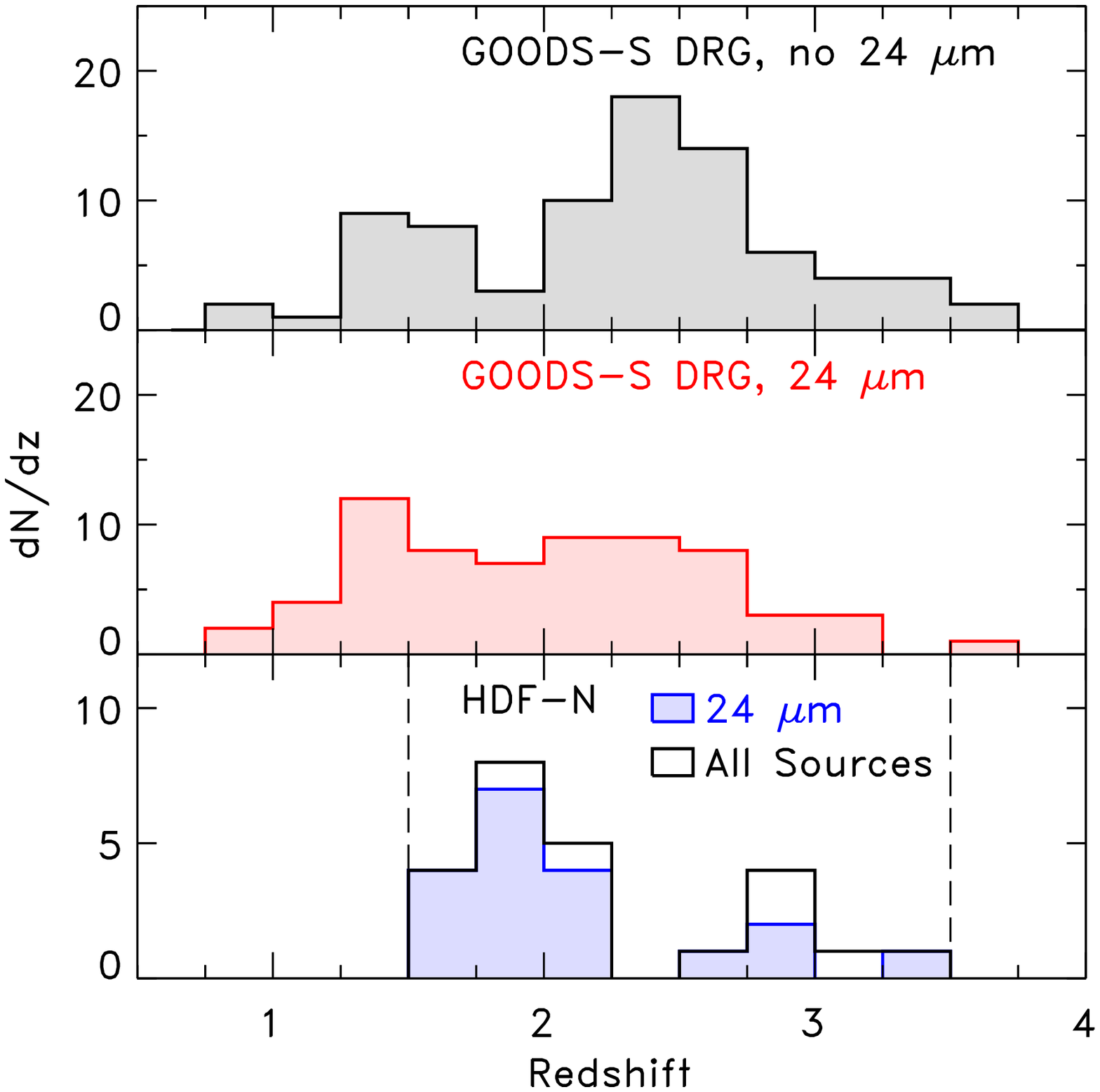}
\caption\figcapzhist
\end{figure}
\fi

Spectroscopic redshifts are available for 12 of the DRGs (Szokoly
\etal\ 2004; Mignoli \etal\ 2005; Vanzella \etal\ 2005; D.~Stern 2005,
private communication). We therefore supplement the redshift
information using photometric redshifts from Mobasher et al. (2004 and
in preparation).   The high--quality dataset allows for accurate
photometric redshifts.  The median photometric--redshift uncertainty
for all \ks--band sources with spectroscopic redshifts is $\delta
z/(1+z) \simeq 0.1$, similar to the accuracy found by \citet{cap05}
using similar data in the same field.  For the 12 DRGs with
spectroscopic redshifts, the photometric--redshift accuracy is even
better, $\langle \delta z/(1+z) \rangle \simeq 0.04$.   A photometric
redshift is unavailable for one of the DRGs, whose ACS and ISAAC
$J$--band photometry have S/N$< 1$.   We ignore this DRG for any
analysis requiring a redshift (e.g., luminosities, stellar masses,
etc.).  Figure~\ref{fig:zhist} shows the redshift distribution.   The
redshifts range from $z\simeq 0.8-3.7$, with a median redshift
$\langle z \rangle = 2.2$.   This is in broad agreement with the
redshift distribution for DRGs reported in the FIRES surveys
\citep{fra03,for04}, although there is a larger fraction of DRGs with $z
\lsim 2$ in the GOODS--S field compared to the FIRES samples (see
\S~\ref{section:jmk}).    We believe the lower median redshift of the
GOODS--S DRGs primarily results from the larger areal coverage
combined with the somewhat brighter flux limit of the GOODS--S field.
In addition, cosmic variance between the GOODS--S and FIRES fields may
contribute to this difference.

However, roughly 20\% (30/152) of the DRGs have redshifts $z\leq 1.5$
(including two with spectroscopic redshifts), significantly lower than
the typical redshifts reported for the FIRES
samples.\footnote{\citet{vandok03} find that one of the six DRGs in
their spectroscopic sample has $z_\mathrm{spec} = 1.19$.   They
estimate that the red $\jmk$ color selection has a $\sim 20$\%
contamination of (dusty) galaxies at lower redshift, probably
consistent with the fraction of DRGs at $z\leq 1.5$ in our sample.}
We have inspected the SEDs of this low--redshift DRG subsample and
found that the ACS through IRAC colors supports the derived redshifts
with high fidelity for the majority (27/30) of galaxies. For these
DRGs, the IRAC photometry shows a fairly robust turnover at
rest--frame 1.6~\micron\ near the expected peak in the stellar
emission, and the SEDs at the inferred redshift are otherwise
consistent with the ACS and ISAAC colors.   In the remaining three
cases, their ACS photometry has low S/N, leading to dubious results.
Ignoring these three galaxies has no effect on the analysis and
conclusions derived here, but we include them in the sample for
completeness.   In \S~\ref{section:discussion}, we conclude that the
lower redshift ($z\lsim 2$) DRGs are mostly heavily extincted
starbursts, and are probably part of the class of dusty EROs at
$z\gsim 1$, which typically have red $J-\ks$ colors that satisfy the
DRG selection criterion \citep{sma02,fra03}.

\subsection{\spitzer\ 24~\micron\ Source Detection and Cataloging}  

\citet{pap04b} describe the data reduction and point--source
photometry methods applied to the \spitzer/MIPS 24~\micron\ image.
They show that the GTO 24~\micron\ data reach a 50\% completeness
limit at 60~$\mu$Jy. We reanalyzed the simulations
discussed in Papovich \etal\ to estimate the flux uncertainties as a
function of MIPS 24~\micron\ flux density.  The median 24~\micron\ S/N
is $\approx$4 for objects with $f_\nu(24\micron) = 80-90$~\ujy\ (the
80\% completeness limit).  We cross--correlated 24~\micron\ sources
with $f_\nu(24\micron) \ge 50$~\ujy\ (the S/N $\approx 3$ limit) to the
\ks--band catalog and identified matches with a radius of 2\arcsec.
Roughly one--half (74/153) of the DRGS are detected by \spitzer/MIPS
at 24~\micron.  The majority of these 24~\micron--detected DRGs
(71/74) have IRAC counterparts associated with each DRG in the
\ks--band image.  The remaining three appear associated with IRAC
sources at distances $0\farcs5 \leq r \leq 2\farcs0$ from the
\ks--band source.  For the source density at the flux limit of the
24~\micron\ data \citep{pap04b}, we expect a random--association
probability of 0.03 within a  2\arcsec\ radius, which is consistent
with the three sources with no IRAC counterpart.   Therefore, these
sources are likely change alignments.

Figure~\ref{fig:zhist} shows the redshift distribution of the DRGs
detected at 24~\micron.  The distribution of this sub--population is
similar to those DRGs not detected at 24~\micron, except for an
apparent spike in the redshift distribution at $z\simeq 2.2$ for the
24~\micron--undetected DRGs.   The 24~\micron--detected DRGs have a
mean redshift $\langle z \rangle$ = 2.0, slightly lower than the mean
redshift for the DRGs with no 24~\micron\ detection, $\langle z
\rangle$ = 2.3.  Using a Kolmogoroff--Smirnov statistic, there is a
moderate likelihood (90\% confidence) that the two sub--populations of
DRGs are drawn from different parent samples.   Different redshift
distributions in part may arise because the 24~\micron\ subsample
includes a large number of lower--redshift galaxies.  However, at
$z\sim 2$ the 7.7~\micron\ emission feature from polycyclic aromatic
hydrocarbons (PAHs) lies in the 24~\micron\ bandpass, and thus one
might expect \textit{more} IR--detected galaxies to appear.   The
sharp increase in the number of 24~\micron--undetected DRGs at
$z\simeq 2.2$ may result from the fact that at this redshift the DRG
selection begins to pick up galaxies with strong Balmer/4000~\AA\
breaks that shift between the $J$ and
\ks--bands. Such galaxies presumably have low specific SFRs (SFR per
unit stellar mass), and perhaps have lower IR emission, so fewer would
be detected at 24~\micron.

\subsection{X--ray Source Detection and Cataloging}

The CDF-S has aim--point flux limits (S/N$=3$) in the 0.5--2.0~keV and
2--8~keV bands of $\approx 2.5\times10^{-17}$~erg~cm$^{-2}$~s$^{-1}$
and $\approx 1.4\times10^{-16}$~erg~cm$^{-2}$~s$^{-1}$,
respectively. The completeness limit over $\approx$~90\% of GOODS-S
field is $\approx 1.3\times10^{-16}$~erg~cm$^{-2}$~s$^{-1}$
(0.5--2.0~keV) and $\approx 8.9\times 10^{-16}$~erg~cm$^{-2}$~s$^{-1}$
(2--8~keV).   Assuming an X--ray spectral slope of $\Gamma=$~2.0, a
source detected with a flux of $\approx 10^{-16}$~erg cm$^{-2}$
s$^{-1}$ would have both observed and rest frame luminosities of
$\approx 5.2\times 10^{41}$~erg~s$^{-1}$ and  $\approx 7.6\times
10^{42}$~erg~s$^{-1}$ at $z=1$ and $z=3$, respectively, assuming no
Galactic absorption.

In this study we use the main and supplementary {\it Chandra} catalogs
of \citet{ale03}.   The median  positional accuracy for the sources in
the GOODS-S field in the main {\it Chandra} catalog is $0\farcs6$.
We matched all \ks--band sources to the X--ray catalog within a radius
of 1\arcsec\ and identified all matches in both the soft-- and
hard--band catalogs.   Nearly one--seventh (22/153) of the DRGs have
X-ray detections in either the soft or hard bands (or both). Of these,
12 are detected at 24~\micron, while the remaining 10 are not.  All of
the X--ray detected DRGs have IRAC counterparts.

\subsection{High--Redshift Sample of IR--luminous Galaxies from the
\hdf}\label{section:hdfn} 

We construct a comparison sample of
galaxies within the northern Hubble Deep Field
\citep[HDF--N,][]{wil96}, which spans all types of galaxies (not just
the reddest galaxies identified by the DRG--selection), and extends
our analysis to fainter 24~\micron\ fluxes. This
allows us to study how the DRGs are drawn from the general galaxy
population at similar redshifts.  Galaxies were selected from the
NICMOS HDF catalog of M.~Dickinson et al.\ \citep{dic00,pap01,dic03},
and matched to the deep IRAC and MIPS 24~\micron\ observations of the
northern GOODS (GOODS--N) field (M.~Dickinson et al., in preparation;
R.~Chary et al., in preparation).   From the NICMOS--selected catalog,
we identify 24 galaxies in the \hdf\ with $\ks \leq 23.2$~mag and
redshift, $1.5 \leq z \leq 3.5$, roughly in the same range as in the
GOODS--S DRGs.   Spectroscopic redshifts are available for 13 of the
galaxies in this sample \citep[see][and references therein]{dic03}.
For the remaining 11 galaxies, we use the photometric redshift catalog
from \citet{bud00}.   Of this sample, 19 are detected with the MIPS
24~\micron\ imaging to $f_\nu(24\micron) \geq 10$~\ujy.

Figure~\ref{fig:zhist} shows the redshift
distribution of the \hdfn\ galaxies.   We note that four of these
galaxies with $\ks \leq 23.2$~mag have $(\nicj-\ks)_\mathrm{AB} >
1.6$, and possibly satisfy the DRG color criterion (allowing for
differences in the \hdfn\ and GOODS--S filter sets).   Of these, three
have $f_\nu(24\micron) \geq 50$~\ujy, and so could have been detected
in the 24~\micron\ of the CDF--S.
%

\subsection{Low--Redshift Galaxy Samples from
COMBO--17}\label{section:combo17}  

In \S~\ref{section:discussion}, we compare the SFRs and stellar masses
of the high--redshift DRGs to those of lower--redshift samples.   The
GTO \spitzer/MIPS 24~\micron\ imaging intersects roughly
700~arcmin$^2$ of the COMBO--17 survey \citep{wol03} in a
substantially larger region encompassing the 130~arcmin$^2$ GOODS--S
field.   COMBO--17 provides photometric redshifts for galaxies with
$R\lsim 23.5$~mag to $z\lsim 1.3$ \citep{wol04}.   Where possible, we
replaced many of these photometric redshifts with spectroscopic ones
\citep{leferve04,vanz04}.  We then constructed samples of galaxies
with $R\leq 23.5$~mag in two redshift slices, $0.3 \leq z < 0.5$ and
$0.65\leq z <  0.75$.  We cross--correlate the COMBO--17 catalogs with
the GTO MIPS 24~\micron\ catalogs to identify matches within a
2\arcsec\ radius.   We find 1495 galaxies in the full $z\sim 0.7$
sample, of which 464 are detected in the MIPS 24~\micron\ catalog to
24~\micron\ 50\% completeness flux limit.   Similarly, there are 1269
sources in the full $z\sim 0.4$ sample, with 276 sources detected in
the MIPS 24~\micron\ catalog (see also the discussion in Bell et al.\
2005; Le~Floc'h et al.\ 2005). Each sample contains approximately the
same co-moving volume in these redshift intervals ($\simeq
10^6\,h_{70}^3$~Mpc$^3$), although we find relatively fewer objects at
$z\sim 0.4$ than $z\sim 0.7$ to the same magnitude limit.  This
results from large--scale clustering in this field, which is known to
be underdense at $z\sim 0.4$ \citep{wol03}.  While IRAC imaging also
exists for most of the COMBO--17 field, we have not included it in our
analysis of the galaxy stellar masses as it provides observations
longward of rest--frame 2~\micron, past the peak of the stellar
emission (see discussion in \S~\ref{section:spec_sfr}).  



\section{The Rest--Frame Optical and Near--IR Colors of
DRGs}\label{section:jmk}  

Galaxies at $z\sim 2-3.5$ with a strong 4000~\AA/Balmer break should
have colors that satisfy the $(\jmk)_\mathrm{Vega} > 2.3$~mag criterion
\citep{fra03}.  This color selection is also sensitive to starburst
galaxies at $z\gsim 1$ whose light is heavily obscured by dust.
\citet{for04} note evidence for both galaxy types in the $\wfi-J_s$,
$J_s-H$, and $H-\ks$ color distributions of DRGs, which is similar to
the reputed nature of $BzK$--selected objects \citet{dad04}.  We find
that the full rest--frame UV to near--IR colors of DRGs provide
further support for this dual population.

\ifsubmode
\else
\begin{figure*}
\epsscale{.9}
\plotone{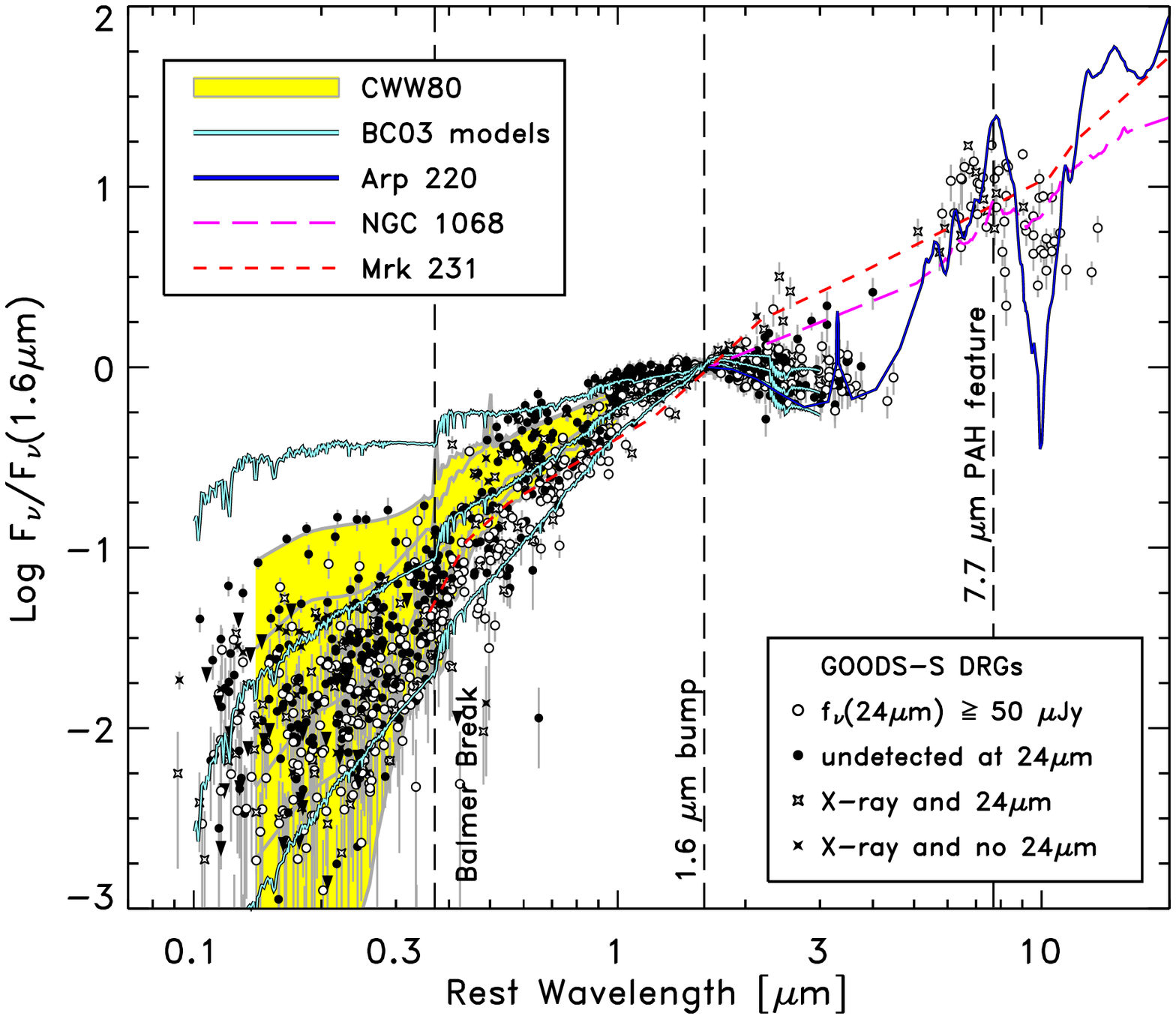}
\caption\figcapsed
\epsscale{1.0}
\end{figure*}
\fi

In Figure~\ref{fig:sed} we compare the SEDs of the GOODS--S DRGs with
empirical and theoretical UV--IR SEDs for local galaxies.  The DRGs
have colors that fit within the envelope defined by various galaxy
types.   Many of the DRGs have colors close to the locus of that for
early--type galaxies.  There is also a class of DRG with optical
colors redder than even the elliptical template of
\citet{col80}.  Most of them are 24~\micron\ sources, and they have
very red UV--optical rest--frame colors, consistent with
dust--obscured starbursts.  The flux ratio between the 24~\micron\ and
near--IR emission is consistent with that of galaxies with massive
starbursts, such as the local ULIRG Arp~220.   Most of the DRGs show a
prominent inflection in their SEDs at rest--frame 1.6~\micron\ as
expected in the SEDs of galaxies dominated by the light of composite
stellar populations \citep[\eg,][]{sim99,saw02}.  The ensemble DRG
photometry in the mid--IR shows slight evidence for a peak at $\sim
8$~\micron\ rest--frame coincident with the PAH 7.7~\micron\ emission
feature, suggesting that the emission from the majority of DRGs stems
from stellar processes rather than nuclear activity.  There is a
slight decline in the flux density at $9-10$~\micron, possibly
consistent with silicate absorption, which is observed in both
starburst galaxies and AGN.  A small subset of the DRGs have red
colors around $\lambda \sim 1-5$~\micron\ rest--frame, a signature of
AGN emission \citep[\eg,][and discussion in
\S~\ref{section:agn}]{rie78,neu79}. 

\ifsubmode
\begin{figure}
\epsscale{0.8}
\plotone{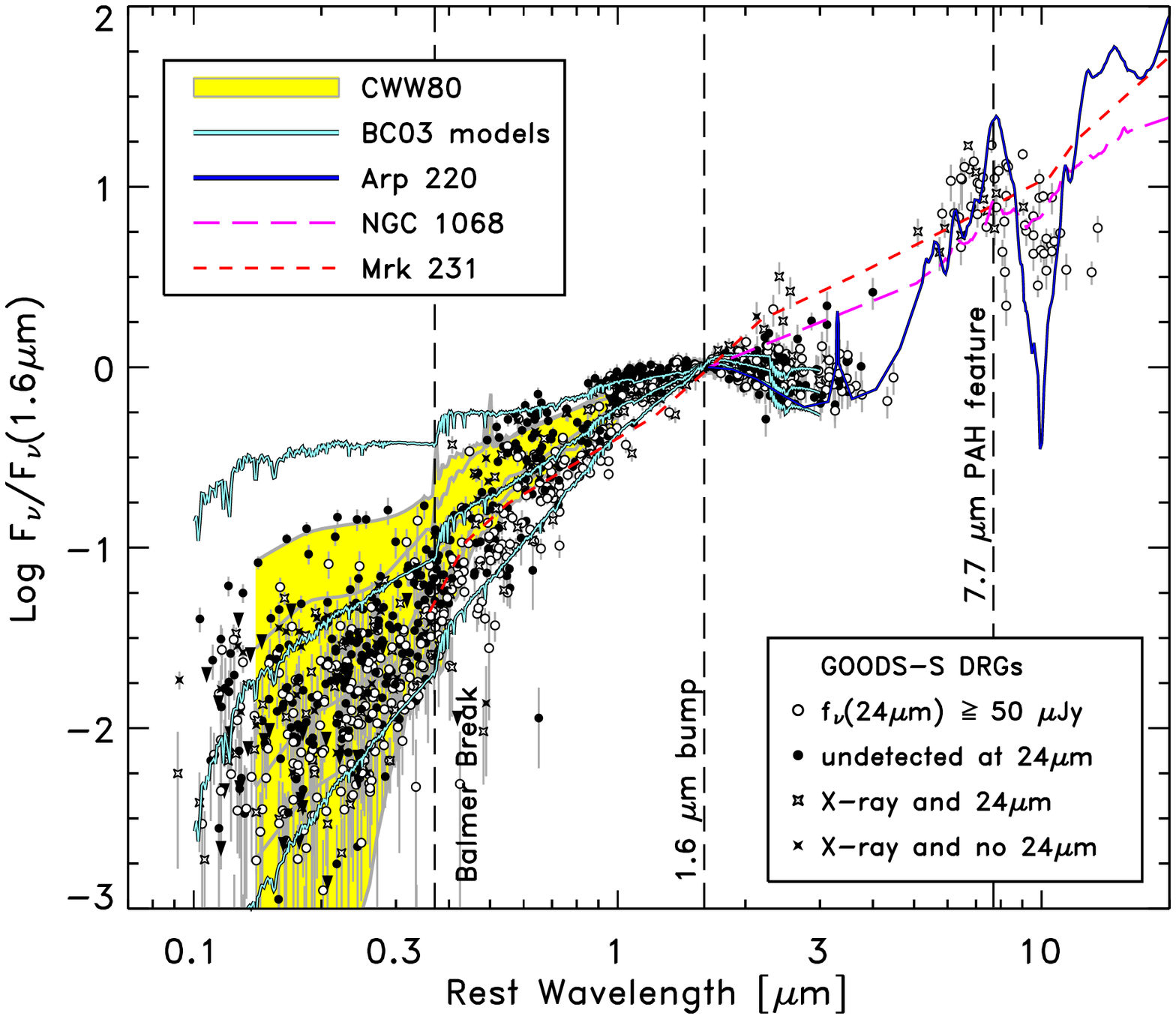}
\epsscale{1.0}
\caption\figcapsed
\end{figure}
\fi

\ifsubmode
\else
\begin{figure*}
\plotone{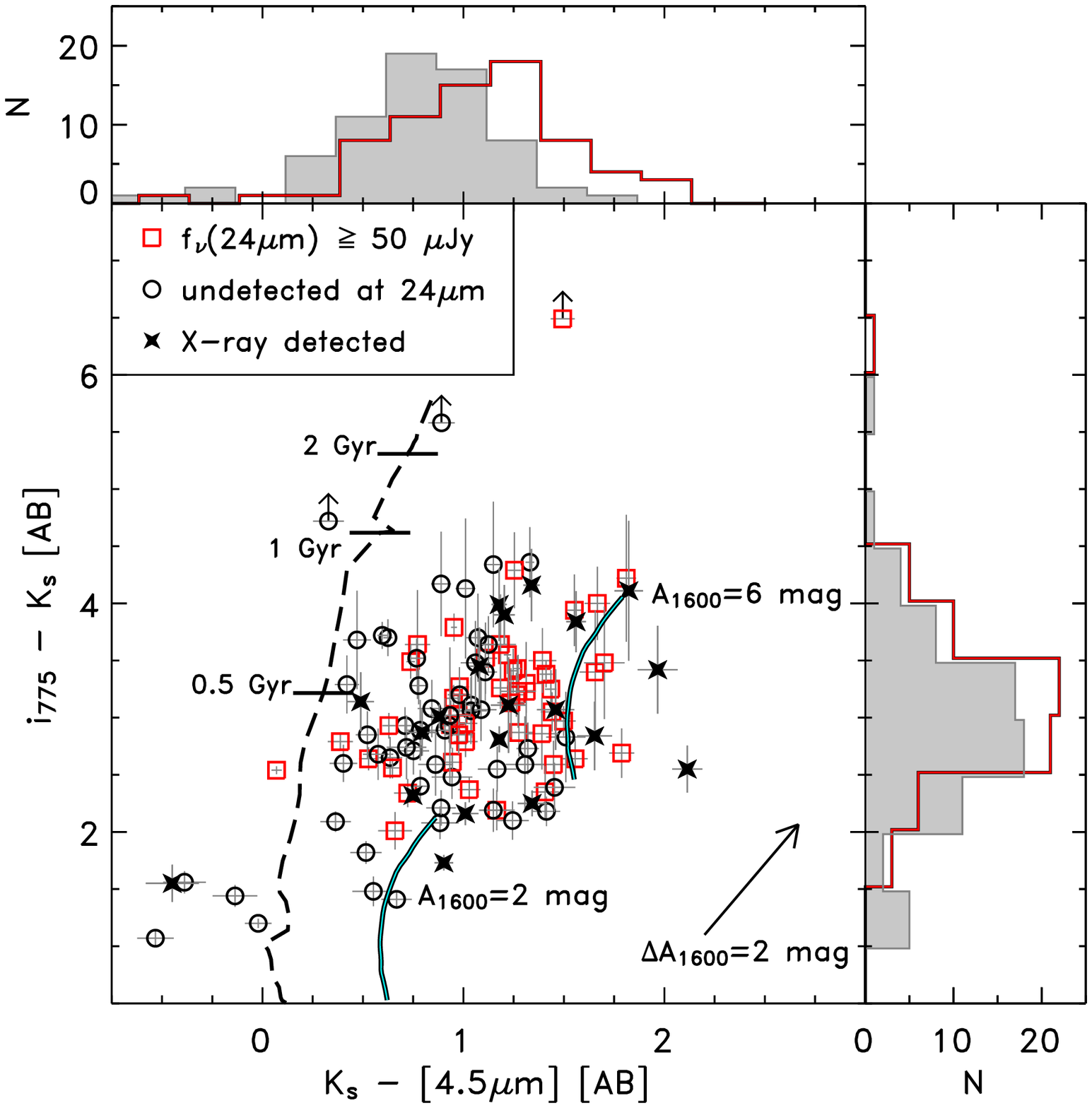}
\caption\figcapjmkcc
\end{figure*}
\fi

\ifsubmode
\begin{figure}
\epsscale{0.75}
\plotone{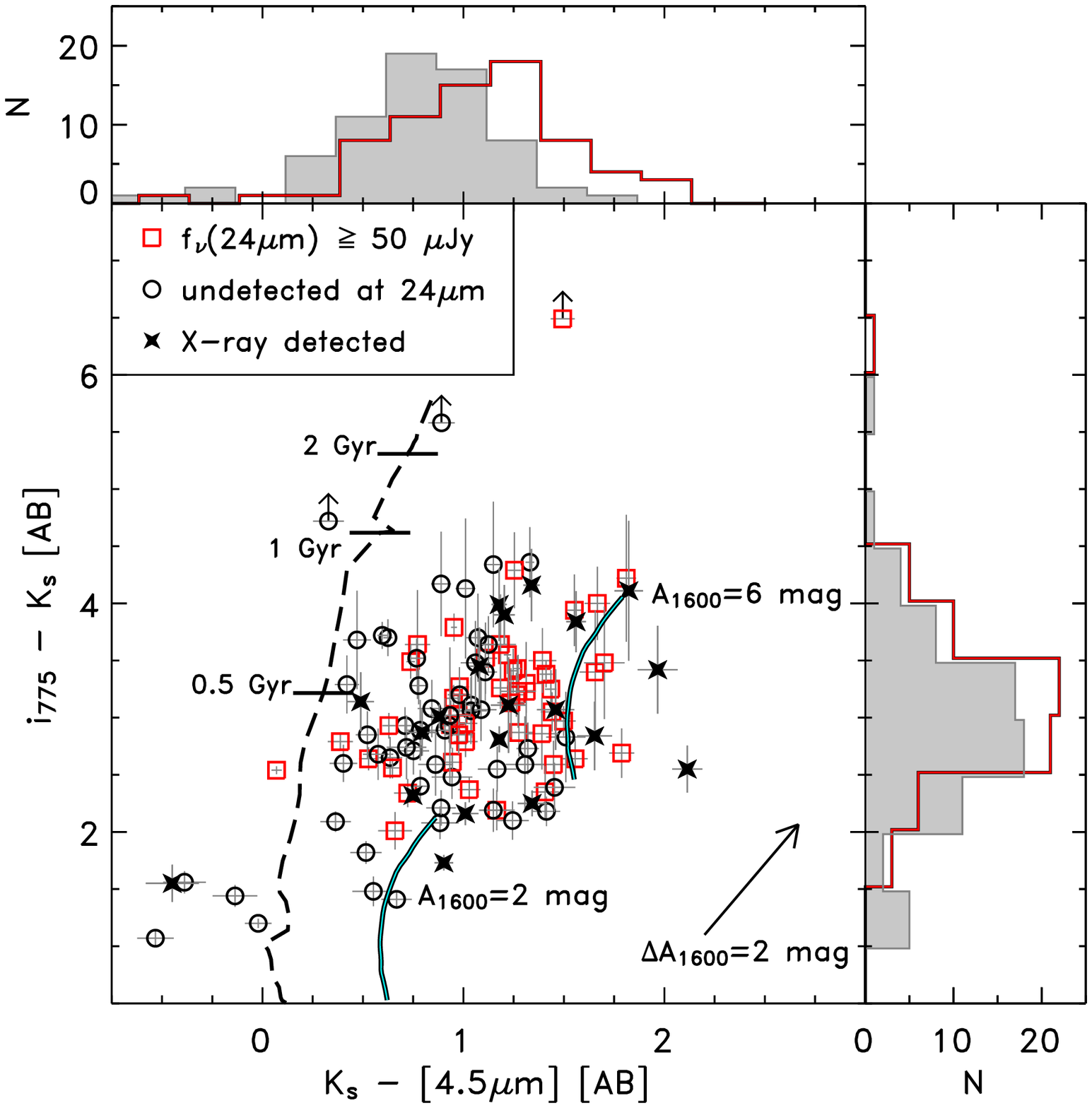}
\epsscale{1.0}
\caption\figcapjmkcc
\end{figure}
\else
\fi

In Figure~\ref{fig:jmkcc}, we show an optical and near--IR
color--color diagram to study the stellar populations and dust
extinction in the DRGs.   The simple models plotted in the figure
bound the range of colors observed in the GOODS--S DRGs, comparable to
the findings of \citet{lab05} for a smaller FIRES sample.    DRGs
with the blue $\ks-\mtwo$ colors require a substantial population of
mature stellar populations with a strong 4000~\AA/Balmer break, which
produces the red $\jmk$ color. DRGs with redder $\ks-\mtwo$ colors
require ongoing star formation with substantial dust extinction.
Changes in the redshift of the model stellar populations from $z=2.2$
to 3.5 or 1.5 shift the expected colors by $\pm 0.3$~mag in
$\ks-\mtwo$~mag, respectively, but have little effect on the
$\acsi-\ks$ colors.  The models bound the majority of the colors of
the DRGs (with the exception of several X--ray sources, whose optical
to near--IR colors are possibly influenced by AGN).  Thus, the colors
of the DRGs imply a mix of old and young stellar populations, in some
cases with substantial extinction.   The MIPS 24~\micron\ data support
this interpretation.   The top panel of Figure~\ref{fig:jmkcc} shows
the $\ks-[4.5\micron]$ colors of the DRGs detected at 24~\micron\ and
those with no 24~\micron\ detection.  The DRGs with 24~\micron\
detections  have redder $\ks-[4.5\micron]$ colors, suggesting that the
emission from many DRGs is attenuated by a large dust opacity.

Of the sample of 132 DRGs with IRAC detections, there are only 6--9
DRGs (within the photometric uncertainties) that have $\acsi-\ks \geq
4$~mag and $\ks - \mtwo < 1.5$~mag (including two sources with
24~\micron\ detections and one X--ray source).  These color
thresholds should identify passively evolving galaxies with ages
greater than $\approx 0.75$~Gyr over the redshift range $z\sim 2-3$,
and these colors bound the ``red and dead'' DRGs in the FIRES sample
from the southern Hubble Deep Field (HDF--S) \citep{lab05}.
Assuming the GOODS--S DRGs with these colors are uniformly distributed
over the GOODS--S area and redshift range $2 \leq z \leq 3$ implies
they have a number density of $0.7-2\times 10^{-6}\,h_{70}^{-3}$
Mpc$^{-3}$.   This contrasts with \citet{lab05} who found a higher
number density, $1.9 \times 10^{-4}\,h_{70}^{-3}$ Mpc$^{-3}$, for DRGs
in the HDF--S assuming the same redshift interval.  The difference
between the DRGs in these fields is not removed by relaxing the
color--selection criteria further.  Based on the full photometry of
our DRG sample (see \S~\ref{section:sedfit}), we find that 15 DRGs
($\sim 10$\%) have best fit models consistent with old ages ($>
1$~Gyr), little dust ($E[B-V] < 0.1$), and passive evolution.   They
have $\acsi-\ks > 2.6$~mag and $\ks -\mtwo < 1.1$~mag, both somewhat
bluer than the original limits.   These galaxies suggest that the
number density of massive, passively evolving galaxies at $2 \leq z
\leq 3$ is $3.2\times 10^{-5}\,h_{70}^{-3}$ Mpc$^{-3}$, still nearly
an order of magnitude lower than that in the HDF--S.  However, some of
the discrepancy likely arises from the fact that the FIRES data
achieve fainter near--IR flux densities, and parenthetically we note
that two of the three candidates for passively evolving DRGs in the
Labb\'e et al.\ sample have $\ks > 23.2$~mag.   Nevertheless, the
difference supports the notion that these sources are highly clustered
\citep{dad03}, and that the HDF--S itself has an unusual overdensity
of them.  The low number density of massive, passively evolving
galaxies in our sample supports the assertion that the density of
passive galaxies is rising strongly at $z<2$ \citep[see,
\eg,][]{dad05,lab05}.

There are few DRGs with $\acsi - \ks \lsim 2$ \citep[see
also,][]{lab05}.   Four of the GOODS--S DRGs have $\acsi-\ks < 2$ and
$\ks-\mtwo < 0$.  One of these galaxies is an X--ray source at $z\sim
2.3$ whose colors may be affected by an AGN.   The remaining three
have redshifts $z\gsim 3$ and relatively flat SEDs from UV to near--IR
rest--frame wavelengths. The red \jmk\ color arises from a weaker
4000~\AA/Balmer break apparently augmented by photometric errors or
emission lines in the passbands,\footnote{van Dokkum \etal\ (2004) find
small emission--line corrections of $\simeq 0.1-0.2$~mag to the
\ks--magnitudes in a sample of seven, bright ($\ks[\mathrm{Vega}]\leq
20$~mag) DRGs.} which push the $J$--band fainter and \ks--band
brighter by small amounts, and conspire to produce
$(\jmk)_\mathrm{Vega} > 2.3$~mag.  In general we retain these sources
in the sample for completeness (although they have a negligible effect
on our results).  In our analysis below, we also consider a restricted
sample with $1.5 \leq z\leq 3.0$, thus excluding these objects.
 
\section{Total Luminosities and Star--formation Rates of
High--Redshift Galaxies}\label{section:lirsfr} 

\subsection{Estimating the Total Infrared Luminosities of Star--forming
Galaxies}\label{section:lir}

At the redshifts of the DRGs, $z\sim 1-3.5$, the \spitzer\ 24~\micron\
probes the rest--frame mid--IR, which broadly correlates with the
total thermal IR luminosity, $\lir \equiv L(8-1000\micron)$
\citep[\eg,][]{spi95,rou01,cha01,dal01,elb02,pap02}.   We convert the
observed 24~\micron\ flux density to a rest--frame luminosity density
at $24/(1+z)$~\micron.  We then correct these values
to a total IR luminosity using the \citet{dal02} IR template SEDs
assuming that a given rest--frame IR luminosity density translates
uniquely to a single SED template.    If we instead used the IR
templates of \citet{cha01}, then we would derive IR luminosities a
factor of 2--3 higher relative to those of Dale \& Helou for galaxies
at $\lir \sim 10^{12.5-13}$~\lsol\ (with smaller differences for
lower--luminosity galaxies).   A recent study of IR--luminous galaxies
at $z<1.2$ indicates that IR--luminosities estimated from the Chary \&
Elbaz models have a scatter of a factor of two compared to IR
luminosities derived from the radio--far-IR correlation \citep{mar05}.
In contrast, the IR luminosities estimated using the Dale \& Helou
models provide a tighter correlation with IR--luminosities derived
from the radio--far-IR correlation, with a scatter of 40\%, suggesting
these templates possibly better reflect reality.   Some scatter is
inherent in this estimation of the total IR luminosity: \citet{cha03}
find that the temperature--luminosity distribution in IR--luminous
galaxies has a scatter of roughly a factor of 2--3 in IR luminosity
for galaxies with fixed dust temperature.  However, \citet{dad05b}
find that the \citep{cha01} IR model template with $\lir =
10^{12.2}$~\lsol\ fits the average SED of 24~\micron--detected $BzK$
objects at $\langle z\rangle =1.9$, suggesting that the uncertainty in
the templates is not severe.   Nevertheless, we add 0.3~dex as a
systematic error on the inferred $\lir$ to account for the systematic
scatter in this conversion.

The uncertainty of the photometric redshifts leads to another source
of uncertainty in the conversion from $L_\nu(24\micron/[1+z])$ to the
total IR luminosity.   Owing to the large bolometric
corrections from the mid--IR to the total IR luminosities, small
changes in the redshift have a significant effect
\citep[see][]{pap02}.   We find that taking the 68\% confidence range
on the photometric redshifts of the DRGs leads to variations in the
inferred \lir\ of 0.4 dex.  We add this source of error in quadrature 
with the uncertainty from the IR templates, bringing the total
error budget on the derived IR luminosities to 0.5~dex.

\ifsubmode
\begin{figure}
\plotone{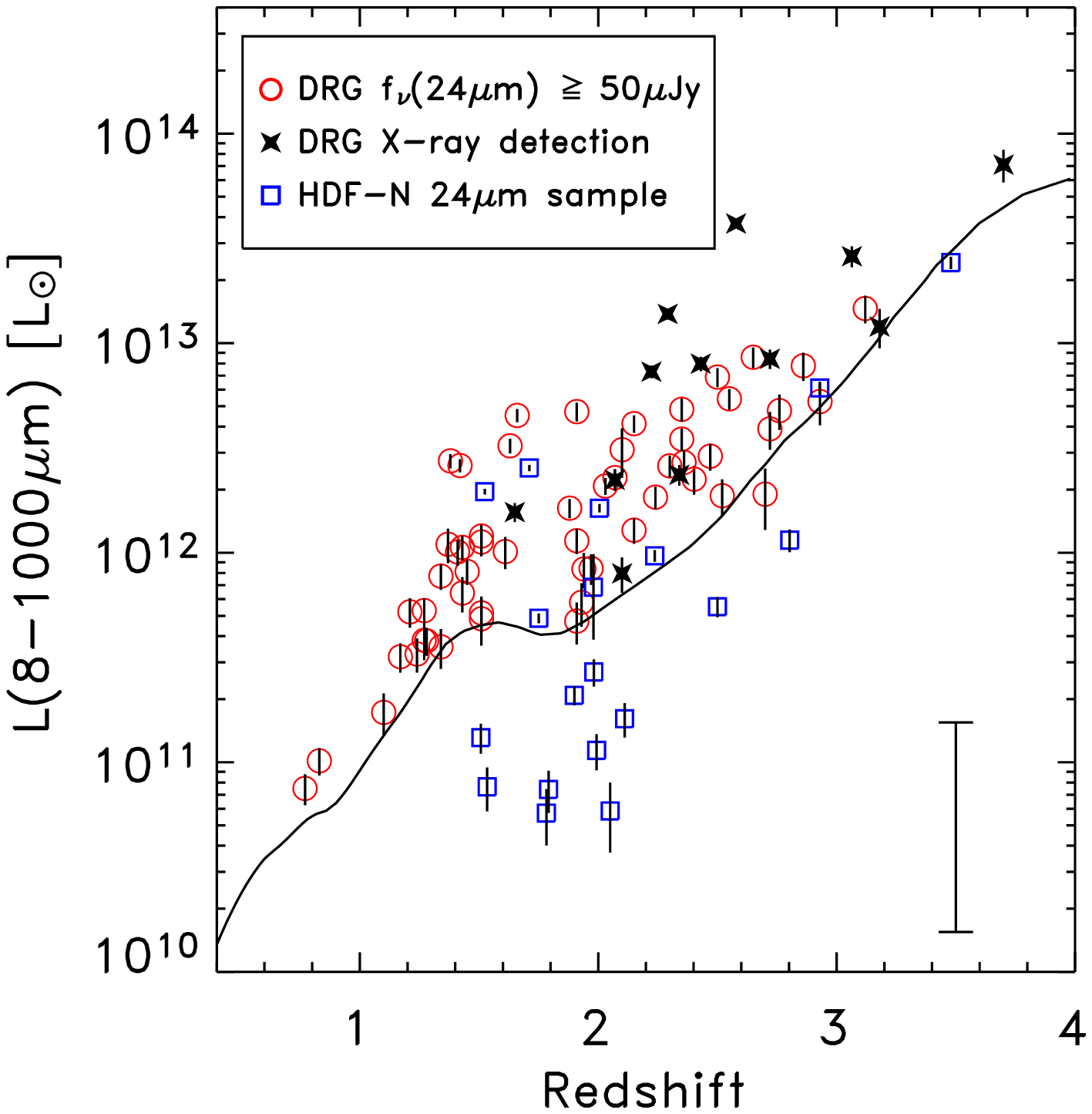}
\caption\figcaplirz
\end{figure}
\fi

Figure~\ref{fig:lirz} shows the total IR luminosities of the DRGs.
The completeness limit (converted to a total IR luminosity in the same
way as for the 24~\micron\ detected galaxies) is indicated in
Figure~\ref{fig:lirz} as the solid line.   The figure also shows the
total IR luminosity for the \hdf\ galaxies in this redshift range
derived from deeper 24~\micron\ observations (see
\S~\ref{section:hdfn}).  The total IR luminosities of the
24~\micron--detected DRGs are $10^{11-14}$~$L_\odot$, or
$10^{11-13.5}$~\lsol\ if we exclude X--ray detected sources. The
majority of these objects have IR luminosities comparable to local
ULIRGs, $L_\mathrm{IR} \gsim 10^{12}$~\lsol, which if attributed to
star--formation implies SFRs greater than 100~\msol\ yr$^{-1}$.

\ifsubmode
\else
\begin{figure}
\plotone{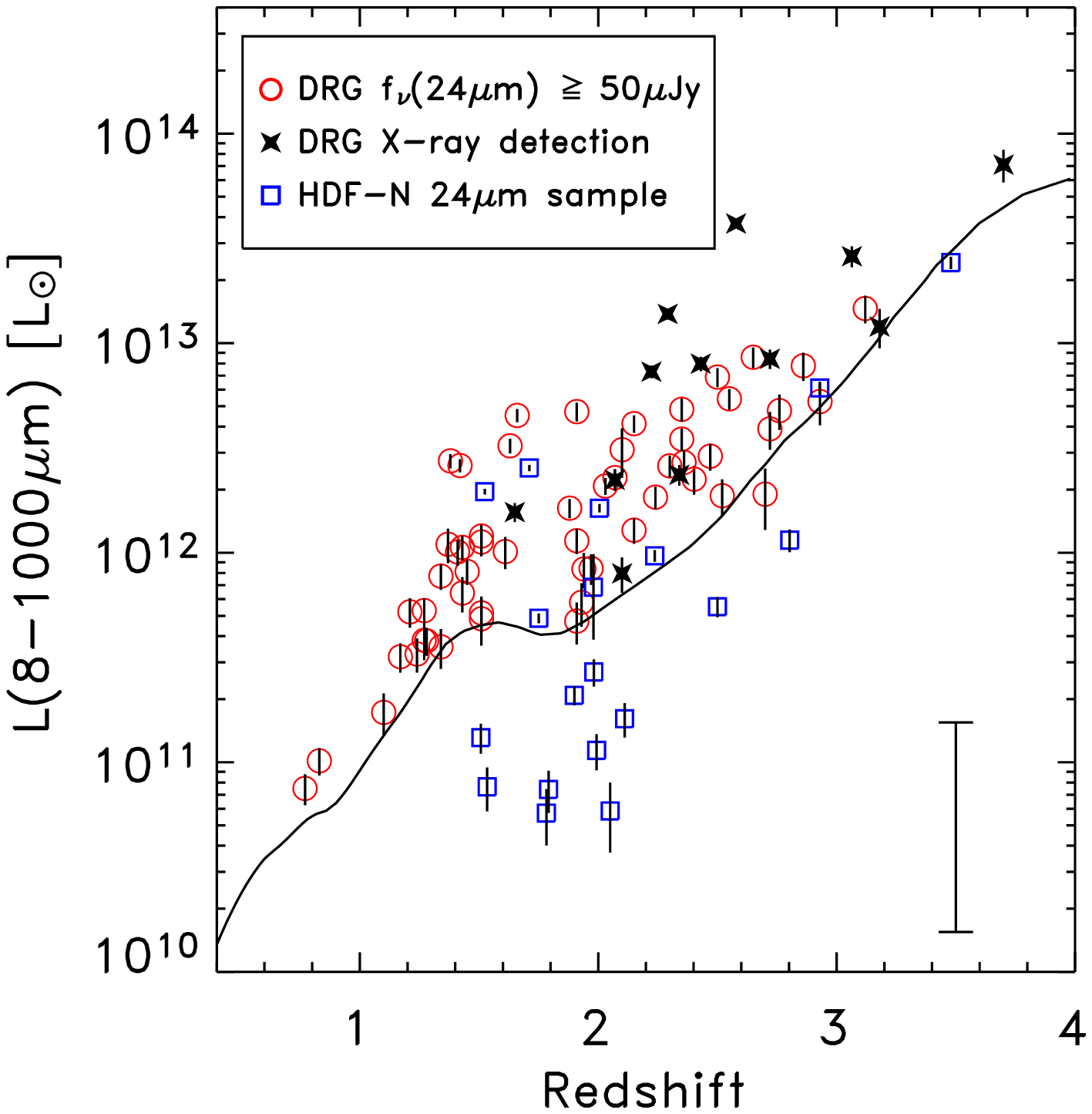}
\caption\figcaplirz
\end{figure}
\fi

Several of the DRGs (6/152) have IR luminosities, $L_\mathrm{IR} \geq
10^{13}$~\lsol\ (so--called Hyper luminous IR galaxies, HyLIRGs).
These IR luminosities are comparable to those of PG quasars at $z\gsim
1$, which have warm thermal dust temperatures \citep{haa03}.  Locally,
HyLIRGs --- and many ULIRGs with $\lir \geq 10^{12.3}$~\lsol\ have
rest--frame optical emission spectra characteristic of Seyfert
galaxies \citep[\eg][]{soi95,vei95,san96,vei99}, and the IR emission
possibly originates from AGN processes.  We suspect that the most
IR--luminous DRGs may have a contribution to their bolometric emission
from AGN.  The majority (5/6) of the DRGs with inferred $L_\mathrm{IR}
\geq 10^{13}$~\lsol\ are detected by \textit{Chandra}, compared to the
10\% X-ray--detection fraction over the whole sample.  This X--ray
detection fraction is consistent with the limit on the AGN fraction for the
coeval sub--mm galaxies \citep{alex05}, a source population that also
has inferred $\lir \sim 10^{13}$~\lsol.   HyLIRGs are also present in
high--redshift IRAC--selected AGN samples, and $\sim 50$\% are
undetected in deep X--ray data \citep{alo05}. Therefore, the X--ray
emission from AGN in many of these IR--luminous objects may be
attenuated by dust below the detection limit of the surveys.  However,
if AGN contribute to the emission in DRGs with $\lir \gsim
10^{13}$~\lsol, then we may be overestimating the galaxies' IR
luminosity.   Although the \citet{dal02} IR templates include galaxies
with $\lir \gsim 10^{13}$~\lsol, using a template for Mrk~231 with a
known AGN and warmer dust temperature would reduce the inferred IR
luminosity by factors of $\sim$2--3.    To limit the effects of any bias
caused by IR template uncertainties for the highest luminosity DRGs,
we consider below how restricting our sample to galaxies without
X--ray detections, $\lir \leq 10^{13}$~\lsol, and IR colors indicative
of AGN (see \S~\ref{section:agn}) affects our analysis.
 
\subsection{The Relation Between the UV Spectral Slope and the
Infrared Excess}\label{section:uvir}

Locally, UV--selected starburst galaxies have a relation
between their UV spectral slope, $\beta$, where $f_\lambda \sim
\lambda^\beta$, and the ratio of their IR to UV luminosity
\citep[also termed  the ``infrared excess''; \eg,][]{meu99}. This
relation links an observed increase in $\beta$ with greater dust
extinction of the intrinsic spectrum of a young stellar population.
The dust--absorbed  UV light is radiated in the thermal infrared,
yielding an anticorrelation between the UV and IR.   While this
relation holds for a range of UV--luminous starburst galaxies
\citep{meu99}, it does not apply to other types of star--forming
galaxies, including normal galaxies \citep{kon04,bua05}, and ULIRGs
\citep{gol02}, apparently due to geometry effects between the UV and
IR emitting regions.  Using the IR luminosities from
\S~\ref{section:lir}, we test whether the local relation between the
UV spectral slope and the IR excess applies to the DRGs.  We also test
if it applies to other high--redshift galaxies detected by MIPS at
24~\micron.

Following \citet{meu99}, we calibrated measurements of $\beta$ by
comparing the UV spectral slope derived from the full UV SEDs of local
starburst galaxies, $\beta_\mathrm{spec}$,  with that measured using
only a broad--band photometric color, $\beta_\mathrm{phot}$.   The
correction for the difference between $\beta_\mathrm{spec}$ and
$\beta_\mathrm{phot}$ is necessary as starburst--galaxy spectra
contain numerous absorption and emission features.   A spectroscopic
UV slope is fit to spectral windows to avoid these features
\citep{cal94}, but they are unavoidable in measuring $\beta$ from
broad--band colors.   We fit the spectral slope,
$\beta_\mathrm{spec}$, using the UV SEDs of starburst galaxy templates
\citep{kin96}, which have dust extinction varying from $E(B-V) < 0.10$
to $0.61 < E(B-V) < 0.70$. We then shifted these spectra to the
observed frame at $z=0.5-4$, and measured the observed ACS
$\acsb-\acsv$, $\acsb-\acsi$, $\acsv - \acsi$, and WFPC2 $\wfv - \wfi$
colors.  We measured the approximate UV spectral slope from the ACS
and WFPC2 broadband photometry, $\beta_\mathrm{phot}$, using the
effective wavelengths of the filters,  $\beta_\mathrm{phot} =
3.20(\wfv-\wfi) - 2.0$ for the WFPC2 bands, and $\beta_\mathrm{phot} =
2.91(\acsb-\acsv) - 2.0$, $\beta_\mathrm{phot} = 1.59(\acsb-\acsi) -
2.0$, and $\beta_\mathrm{phot}=2.14(\acsv-\acsz) - 2.0$ for the ACS
bands.  We then fit a quadratic polynomial to the difference,
$\beta_\mathrm{spec} - \beta_\mathrm{phot}$,  as a function of
redshift.

\ifsubmode
\else
\begin{figure*}
\plotone{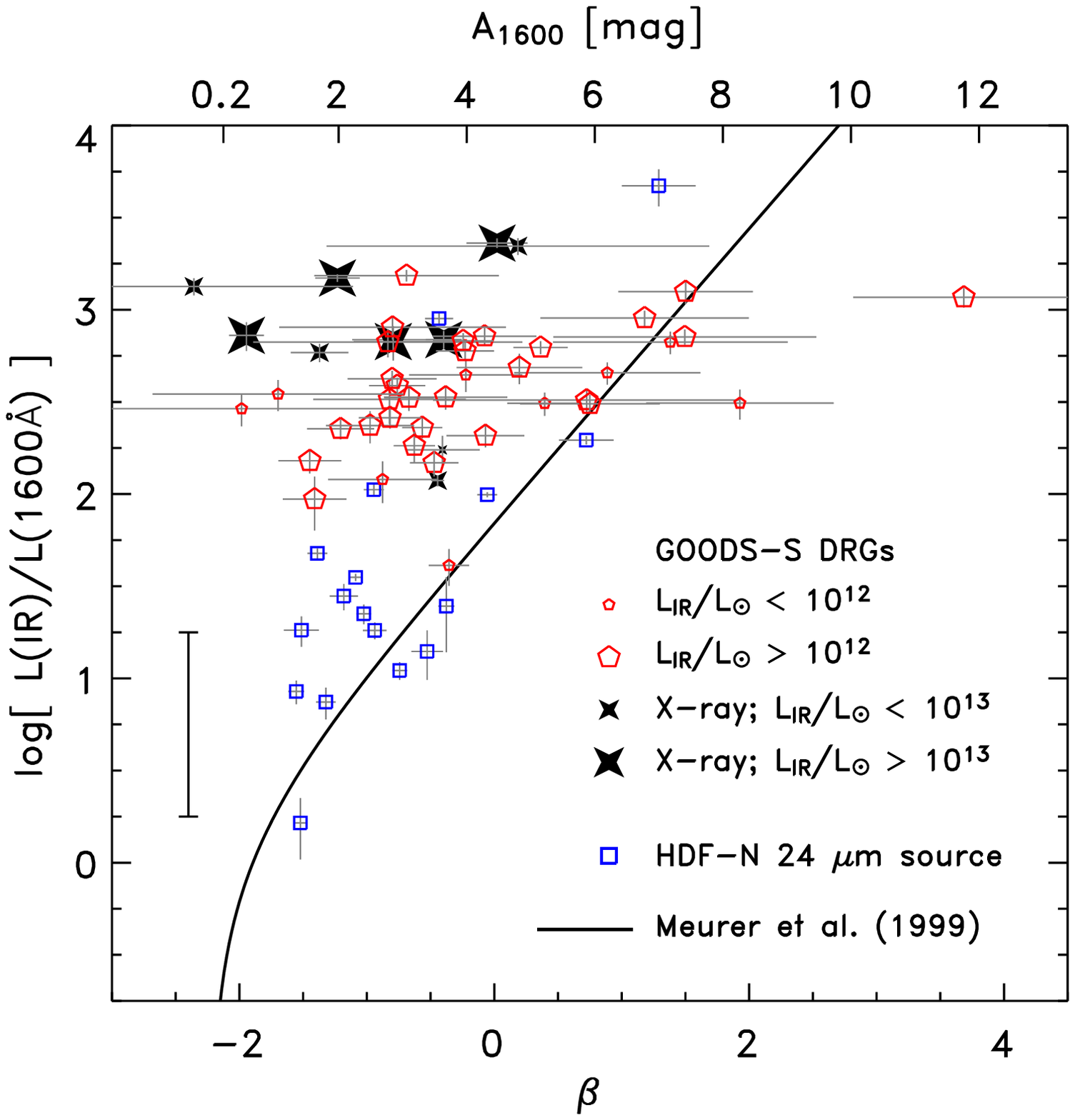}
\caption\figcapuvir
\end{figure*}
\fi

To derive the UV spectral slopes of the DRGs, we want to span the
longest wavelength baseline in the rest--frame wavelength range
1250--2800~\AA.   At wavelengths longer than 2800~\AA\ there may be a
significant contribution to the SED from A-- and later--type stars
from previous star--formation episodes
\citep[\eg,][]{cal94}. Therefore, for galaxies at $z > 2.2$ in the
GOODS--S field, we will use the $\acsv-\acsz$ color to measure
$\beta$.  For galaxies with $1.7\leq z\leq 2.2$, we  will instead use
the $\acsb-\acsi$, where the lower redshift bound results from the
fact that below this redshift the $\acsi$--band probes $\lambda >
2800$~\AA.  For galaxies with $1.2 \leq z < 1.7$, we use $\acsb-\acsv$
to derive the UV spectral slope.  For $z < 1.2$, there are no ACS
colors that measure the UV spectral slope at $\lambda < 2800$~\AA.

For galaxies from the \hdf\ with WFPC2 photometry and redshift, $z$,
we derive the UV spectral slope, $\beta$, using the quadratic fit to the
empirical relationship \citep[cf.,][]{meu99},  
\begin{equation}
\beta = 3.20(\wfv - \wfi) - 4.45 + 2.03 z - 0.423 z^2.
\end{equation}
For the DRGs with ACS photometry we use
\begin{equation}
\beta = \left\{ \begin{array}{ll}
2.91(\acsb - \acsv) - 3.35 + 1.49 z - 0.438 z^2, &
\mbox{$1.2 \le z < 1.7$} \\
1.59(\acsb - \acsi) - 3.92 + 1.97 z - 0.512 z^2, &
\mbox{$1.7 \le z \leq 2.2$} \\
2.14(\acsv - \acsz) - 4.76 + 2.14 z - 0.418 z^2, &
\mbox{$z>2.2$.} \\
\end{array}
\right.
\end{equation}
We find that the different formulae give consistent results within the
photometric errors for galaxies in redshift intervals where several
ACS colors can be used to measure $\beta$.   Using these formulae, we
then calculate the UV spectral slopes of the DRGs and \hdfn\ galaxies
using the appropriate ACS and WFPC2 bands, respectively.  We also use
uncertainties of the observed colors and the covariant uncertainties
of the polynomial fits to estimate the error in $\beta$.

\ifsubmode
\begin{figure}
\epsscale{0.95}
\plotone{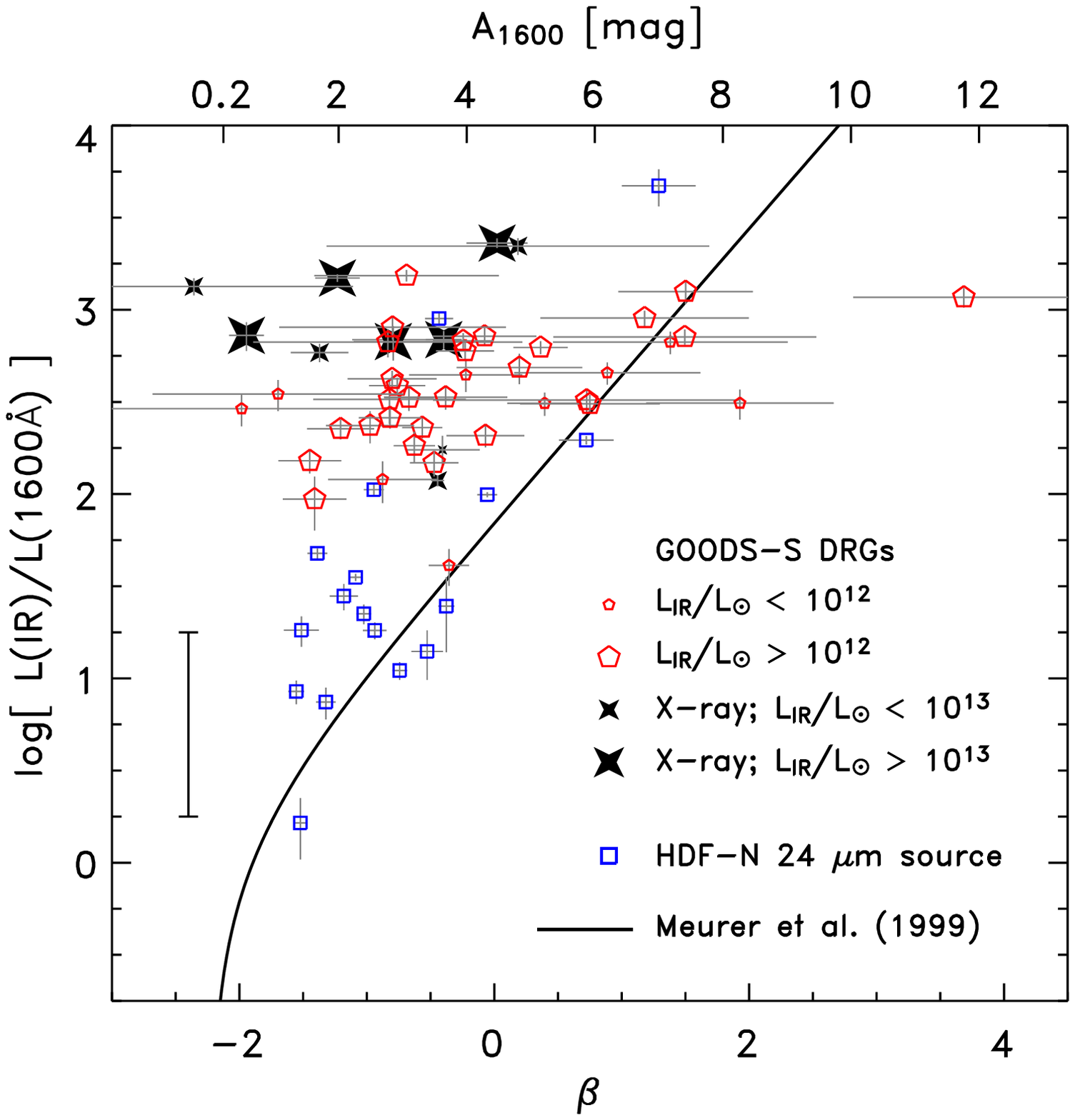}
\epsscale{1.0}
\caption\figcapuvir
\end{figure}
\else
\fi

Figure~\ref{fig:uvir} shows the relation between $\beta$ and the IR/UV
luminosity ratio for the high--redshift galaxies, and compares them to
the local relation from \citet{meu99}.   Many of the \hdf\
24~\micron--detected galaxies at $1.5 < z < 3.5$ have UV spectral
slopes and IR/UV luminosity ratios that lie near the local relation.
These systems have IR luminosities in the range $L_\mathrm{IR} \sim
10^{10-11}$~\lsol,  comparable to the IR luminosities in the Meurer et
al.\ sample.   There is a trend for galaxies with higher IR/UV
luminosity ratios to move away from the local relation in the sense
that they have more IR luminosity than otherwise predicted from their
UV luminosity and spectral slopes.

This trend in galaxies with higher IR luminosities having higher UV/IR
ratios is continued for the 24~\micron--detected DRGs.   The DRGs span
$\beta \sim -2$ to 2 (with the exception of one object with $\beta
\simeq 3.7$ and large photometric uncertainty). Many of DRGs with 
$\lir \leq 10^{12}$~\lir\ and some with $\lir > 10^{12}$~\lsol\ have
IR/UV ratios near the Meurer et al.\ relation.  However, most of the
DRGs have IR/UV luminosity ratios of more than one order of magnitude
in excess of what would be predicted from their spectral slopes.
That is, the total amount of star formation in these galaxies will be
underestimated from their UV rest--frame luminosities and spectral
slopes alone.  This result is qualitatively unchanged if we restrict
the DRG sample to redshifts $1.5 < z < 2.5$, where the total
correction factor from observed $f_\nu(24\micron)$ to \lir\ spans a
relatively narrow range compared to that for the full sample.

The DRGs occupy a very similar range of $\beta$ and IR/UV flux ratios
as has been observed for local ULIRGs \citep{gol02}.   In local ULIRGs
the geometry of the forming star clusters and dust is highly complex.
The UV and IR emitting regions are typically displaced from one
another, or ``patchy'', such that the regions that dominate the UV
emission are unassociated with the regions producing the large IR
luminosity \citep[as is the case for local IR--luminous galaxies with
\textit{ISO} IR and \hst\ UV imaging; see][]{cha04a}.  A similar
situation probably holds for the 24~\micron--detected DRG population.
It is also plausible that the galaxies with the largest IR
luminosities have more complicated geometries to account for the fact
that these galaxies have the largest offsets from the local relation.
Charmandaris \etal\ noted that for many of the local LIRGs, the
\textit{ISO} source is offset from the regions that dominate the UV
emission by as much as several kpc, and such objects can have large IR
excesses.  This effect should be even stronger for ULIRGs.  At
redshifts typical of the DRGs, this corresponds to less than 1\arcsec,
and will be unresolvable with MIPS at 24~\micron.

Most of the X-ray--detected DRGs have IR/UV luminosity ratios and UV
spectral slopes comparable to the rest of the DRG sample, although
several have the most extreme IR/UV luminosity ratios or UV spectral
slopes ($\log \lir/L[1600\AAA] \sim 4$, and/or $\beta \sim -2$), lying
away from the other DRGs.   We suspect that an AGN contributes
substantially to the IR emission, the UV emission, or both.  In this
case, the \citet{dal02} IR templates may overestimate the IR
luminosity in these objects (see \S~\ref{section:lir}).

\subsection{Inferring the Star Formation Rates of High--Redshift
Galaxies from Bolometric Luminosities}\label{section:sfr}

Nearly all of the bolometric luminosity from star--forming regions is
emitted in the UV and IR \citep[\eg,][]{bel03}.  Therefore, we
estimate the instantaneous star--formation rates (SFRs) for galaxies
in our samples using the combination of their UV and IR luminosities.

We use the SFR conversion from \citet{bel05}, based on the UV and IR
calibration presented by \citet{ken98},  
\begin{equation}\label{eqn:sfr}
\Psi / \msol \,\,\mathrm{yr}^{-1} = 1.8 \times 10^{-10} \times
(L_\mathrm{IR} + 3.3\,\, L_{2800}) / \lsol,
\end{equation}
where $L_\mathrm{IR}$ is the total IR luminosity, and $L_{2800} \equiv
\nu\, L_{\nu}(2800\AAA)$ is the monochromatic luminosity at
rest--frame 2800~\AA\ interpolated from the ACS photometry. We have
adjusted the SFR to correspond to a single power--law, Salpeter IMF
with mass cutoffs of 0.1 and 100~\msol. \citet{bel03} tested these
UV+IR--derived SFRs against extinction--corrected \ha\ and
radio--derived measures, finding excellent agreement with $\lsim
0.3$~dex scatter and no offset.   Nevertheless, it should be noted
that the UV and IR--calibrations are based on local galaxy
correlations.  Although the indications are these hold at higher
redshifts \citep[see, \eg,][]{elb02,app04}, they should be used with
some caution.

This calibration explicitly assumes that star--formation processes
account for the bolometric UV and IR emission.  However, the presence
of AGN may also contribute some (or all) of this bolometric emission
from accretion processes onto supermassive black holes (SMBHs).
Therefore, the SFRs derived using equation~\ref{eqn:sfr} may be upper
limits if AGN are present (modulo uncertainties in the conversion from
the 24~\micron\ to IR luminosity).

\ifsubmode
\else
\begin{figure}
\plotone{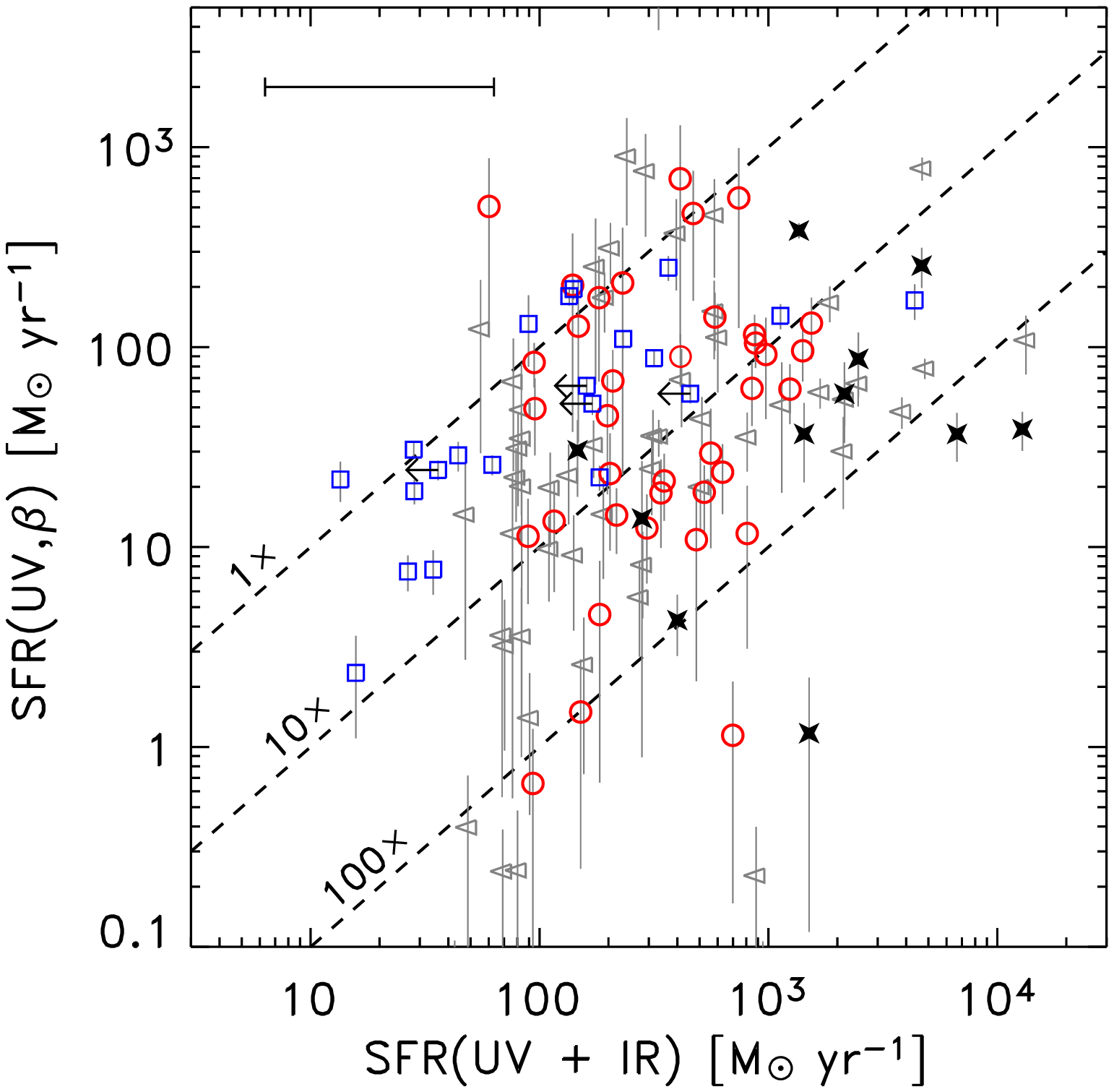}
\caption\figcapsfrsfr
\end{figure}
\fi

Figure~\ref{fig:sfrsfr} compares the SFRs derived from the sum of the
UV and IR emission using equation~\ref{eqn:sfr} versus those derived
using solely the UV rest--frame luminosity corrected for extinction
using the derived spectral slopes \citep[see \eg,][]{meu99,ade00}.  To
calibrate the UV--derived SFRs, we compared the 1600~\AA\ luminosity
(averaged over the range 1400--1800~\AA) to the SFR from \citet{bru03}
spectral templates over a wide variety of star--formation histories.
For a single power--law Salpeter IMF with mass limits 0.1 and
100~\msol, we find that a SFR of 1~\msol\ yr$^{-1}$ corresponds to a
luminosity density of $l_\nu(1600\AAA) = 8.7\times 10^{27}$~erg
s$^{-1}$ Hz$^{-1}$ for galaxies with ongoing star formation for $\gsim
10$~Myr \citep[see, e.g.,][]{mad98}.   Figure~\ref{fig:sfrsfr} shows
this relation for the GOODS--S DRGs, and the \hdf\ galaxies.  The
HDF--N galaxies extend to smaller SFRs, which are in closer agreement
with the rest--frame UV data.   The DRGs and many of the \hdfn\
galaxies with larger IR luminosities have significantly larger SFRs
derived from the UV+IR than those estimated from the UV only.  The
difference can be up to two orders of magnitude.

\ifsubmode
\begin{figure}
\plotone{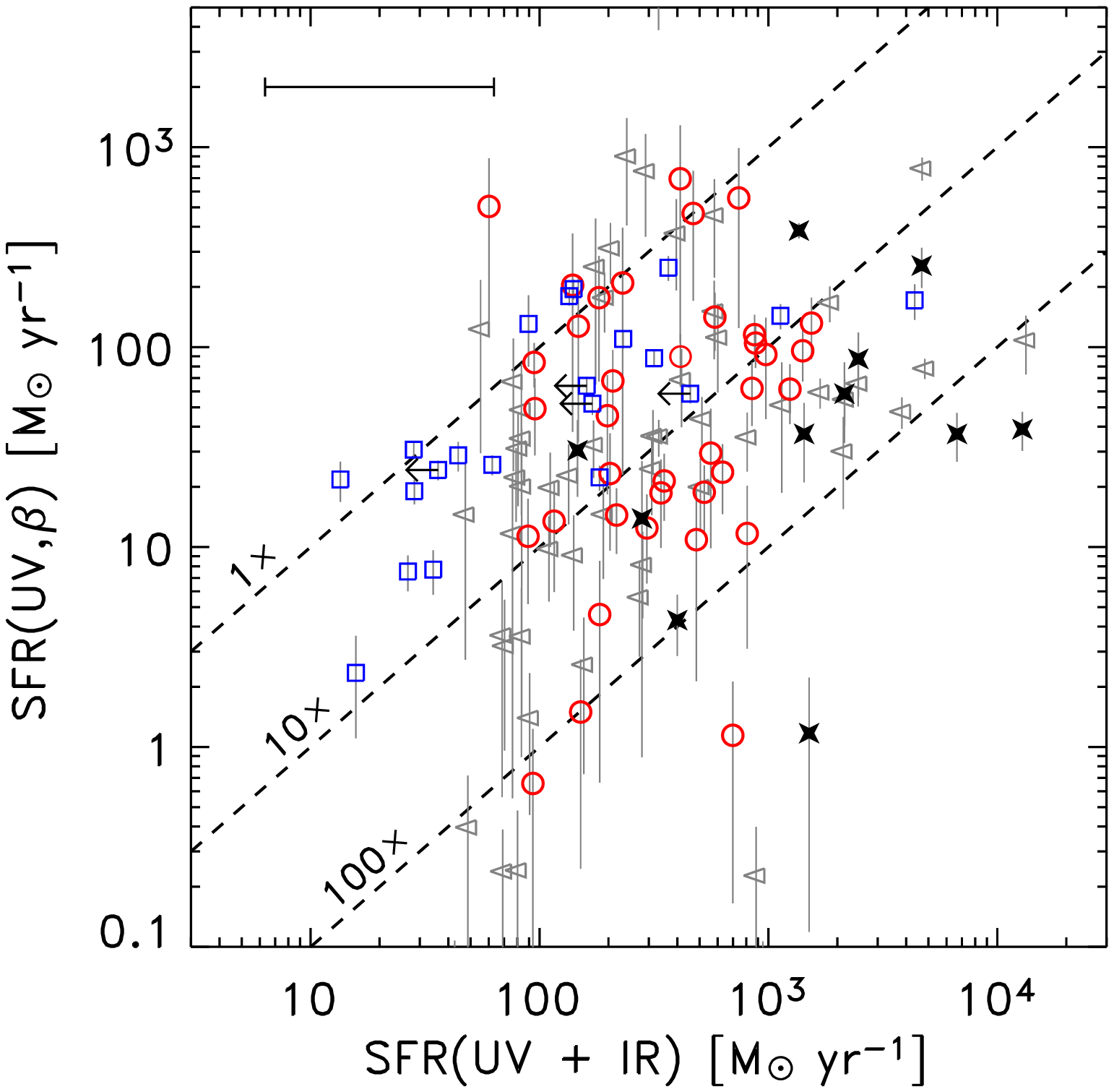}
\caption\figcapsfrsfr
\end{figure}
\else
\fi

\section{Stellar Populations and Star Formation in High Redshift
Galaxies}\label{section:sedfit} 

In \S~\ref{section:jmk}, we showed that the UV, optical, and near--IR
colors of \jmk--selected galaxies are consistent with a multi--variate
population of heavily dust--enshrouded starbursts and galaxies whose
rest--frame optical and near--IR light are dominated by later--type
stars.   Here, we extend this analysis by comparing the full
photometry of the DRG and \hdfn\ samples to stellar population
synthesis models.  As in \citet{pap01}, we first consider a model of a
single, monotonically evolving stellar population with a SFR that
decays exponentially with a characteristic $e$--folding timescale,
$\tau$.   In reality, the star--formation histories of high--redshift
galaxies are presumably more complex, involving stochastic events from
mergers, interactions, feedback from star formation and AGN, as well
as quiescent star formation \citep[\eg,][]{som01,nag05,delucia05}.
The models here should be considered as fiducial averages of past
star--formation histories.   Our definition of galaxy ``age'' is the
time since the onset of star formation.   Our monotonically evolving
models continuously produce new stars with young ages (albeit at a
lower rate than in the past).    For example, under our definition a
stellar population formed with a constant SFR has age, $t$, while the
mean age would be $\int\! \Psi(t)\,t\,dt / \int\! \Psi(t)\, dt = t/2$,
and the luminosity--weighted mean age (weighted heavily toward the
short--lived early--type stars) would be younger still.  The
definition of galaxy age also neglects all previous discrete, episodes
of star--formation.   Older stellar populations from past
star--forming events may very well exist, but be lost in the ``glare''
of the nascent stars.  We consider these effects by using a second
model that adds a maximally old stellar population formed in a single
burst at $z = \infty$ to the stellar populations formed with the
simple exponentially decaying models described above
\citep[see also,][]{pap01,dic03}.  The latter model has a maximal
stellar-mass--to--light ratio.  We will use the single--component and
two--component models to constrain the range of stellar masses and
star--formation histories. 

\subsection{Fitting the Models to the Photometry}

We fit the galaxy photometry with the \citet{bru03} stellar--population
synthesis models.  While the metallicities of the DRGs are poorly known
\citep{for04}, \citet{vandok04} provide evidence for solar and
super--solar metallicities for luminous DRGs. \citet{sha04,sha05} have
estimated the metallicities of massive ($\mcal \gsim 10^{11}$~\msol)
$U_n$--dropout--selected LBGs at similar redshifts to be approximately
solar. Because the expected colors for the majority of massive
$U_n$--dropouts satisfy the DRG selection criteria \citep{red05,sha05}, the
solar--metallicity assumption for DRGs is also reasonable.   Using
different metallicities will affect our fitting results.   However,
the derived stellar masses vary by factors of less than 2--3
\citep{pap01}.

We use models with a single power--law Salpeter IMF with mass limits
0.1 and 100~\msol.   Changing the shape of the IMF  affects the
derived stellar masses (and other stellar population parameters as
well; see Papovich et al.\ 2001).   For example, the Chabrier,
Kennicutt, or Kroupa IMFs have a turnover in the mass function below
1~solar mass and produce stellar populations with roughly the same
colors but with a stellar mass of 0.25~dex lower than that for the
adopted Salpeter IMF.   Although these other IMFs possibly better
reflect nature, we choose to use a Salpeter IMF to facilitate
comparisons of our results with those in the literature.   As of yet
there is no reason to expect the IMF to differ strongly from that
observed in local galaxies \citep[\eg][]{bal03,lar05}, although some
empirical studies and theoretical predictions suggest a steeper IMF may
be required in high--redshift massive starbursts \citep{fer02,bau05}.

\ifsubmode
\else
\begin{figure*}
\plottwo{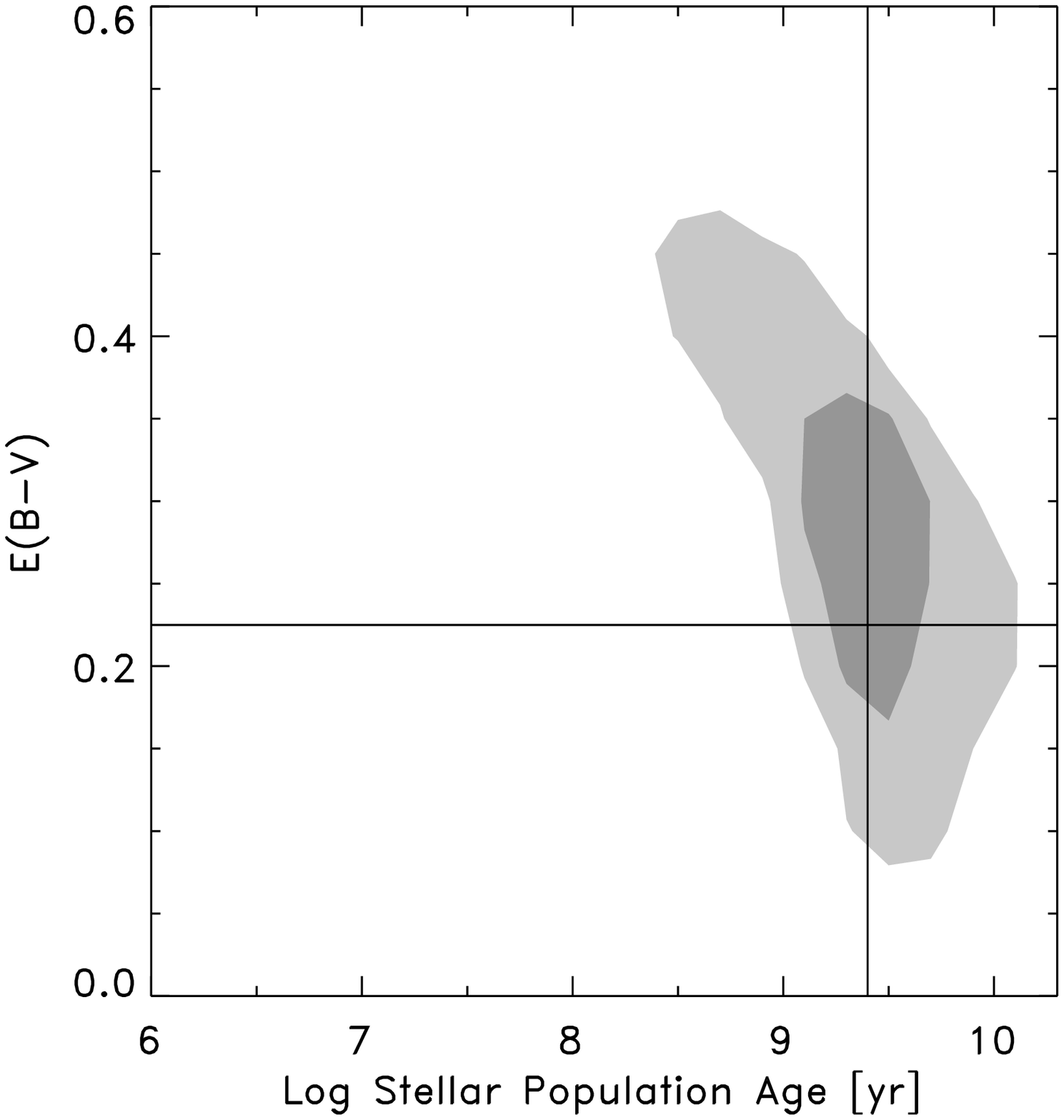}{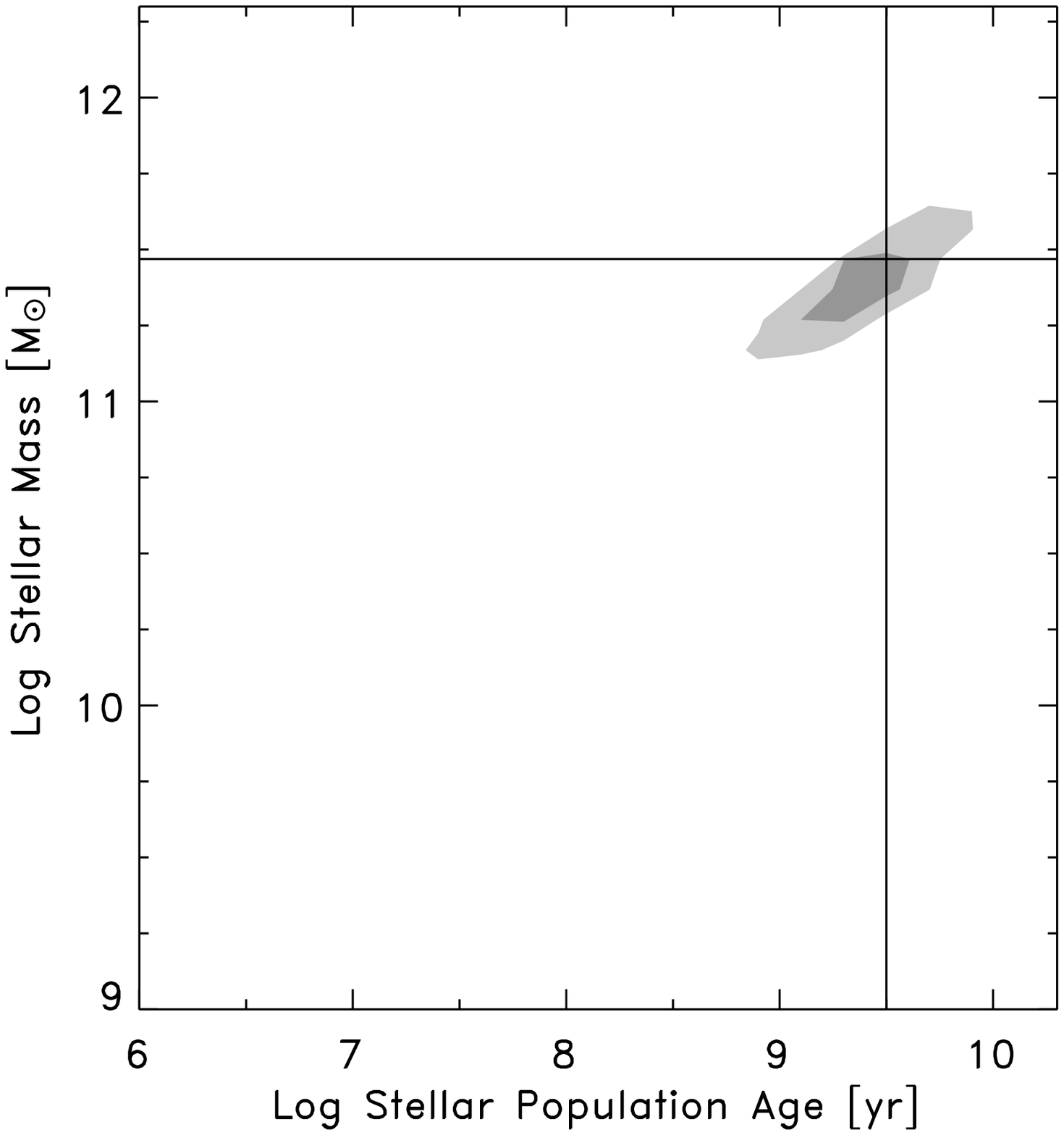}
\epsscale{0.75}
\plotone{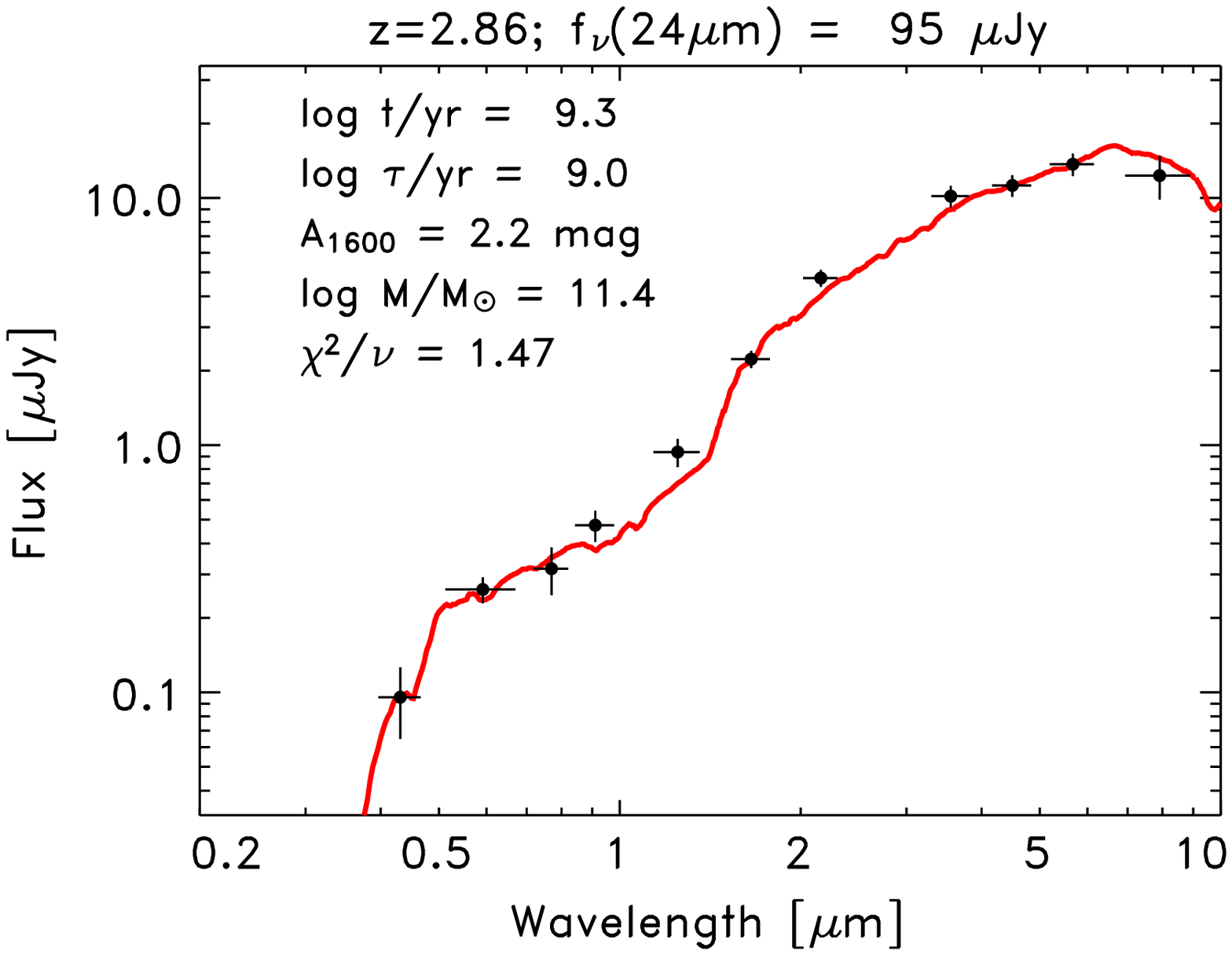}
\epsscale{1.0}
\caption\figcapsedfitone
\end{figure*}
\fi

\ifsubmode
\else
\begin{figure*}
\plottwo{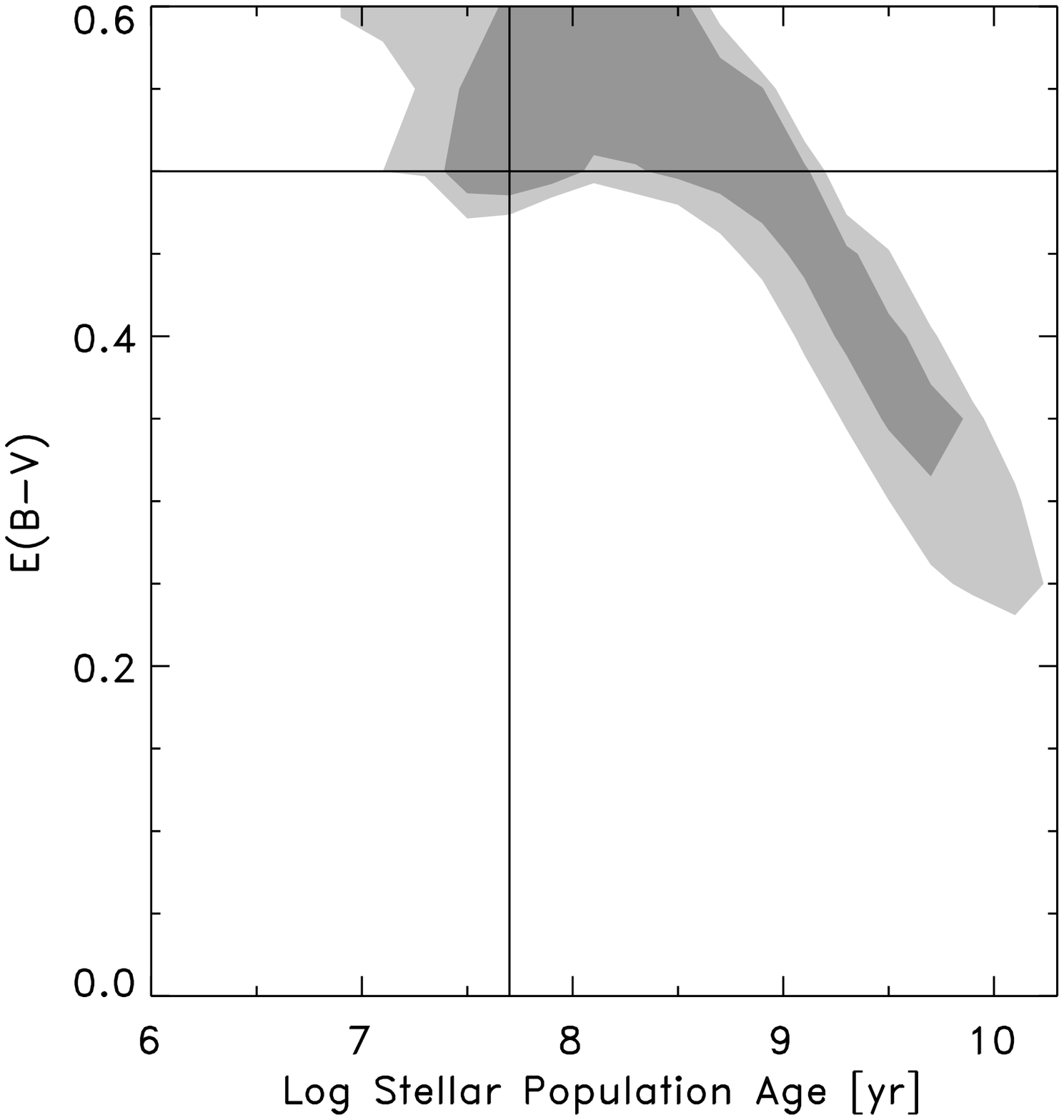}{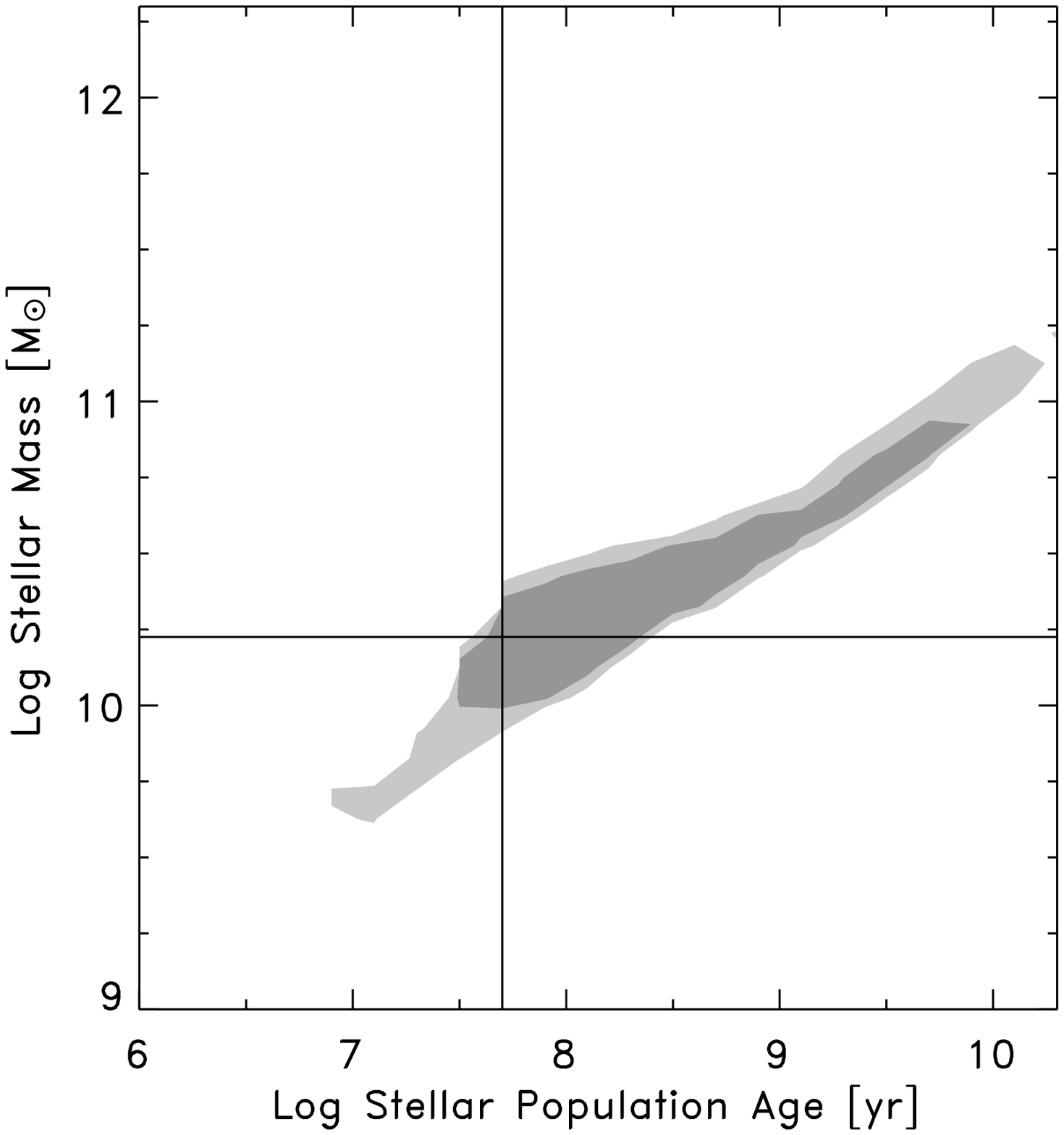}
\epsscale{0.75}
\plotone{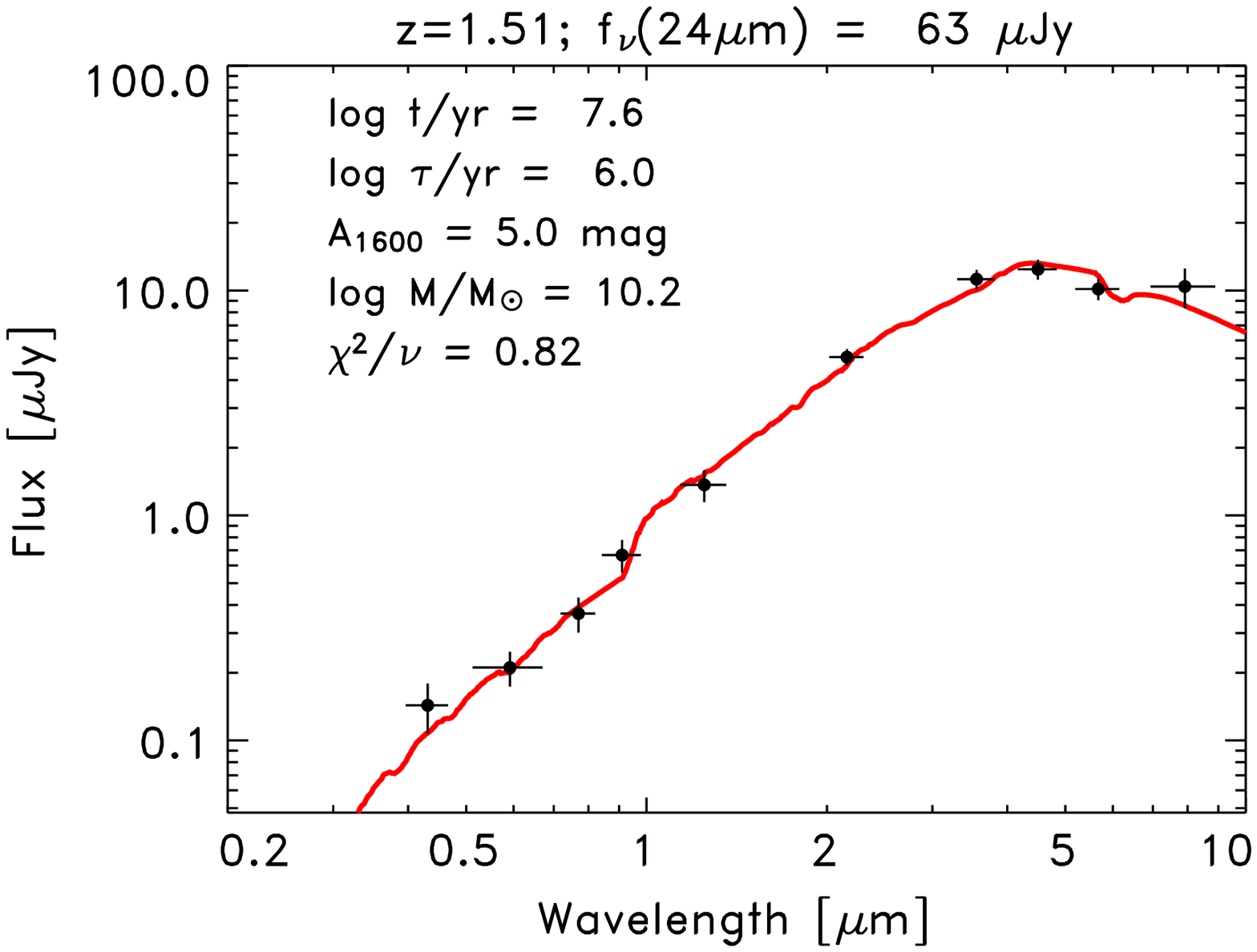}
\epsscale{1.0}
\caption\figcapsedfittwo
\end{figure*}
\fi

We generate a suite of photometry from the \citet{bru03} models for
galaxies spanning the range of redshifts in our sample with a redshift
step--size of $\delta z = 0.05$.  The models range in stellar
population age from $10^6$ to $2\times 10^{10}$~yr in the
quasi--logarithmic steps provided in the models.  We include dust
extinction using the \citet{cal00} law with color--excess values
$E(B-V) = 0.0-0.6$ in steps of $\delta E(B-V) = 0.025$.  For the
Calzetti \etal\ extinction parameterization, these color excesses correspond
to extinctions in the UV (rest--frame 1600~\AA) of $A_{1600} =
0-6$~mag in increments of $\delta A_{1600} = 0.25$~mag.   
We make the assumption that in these high--redshift galaxies the
Calzetti \etal\ extinction law applies to the stellar populations
dominating the light observed in the ACS, ISAAC, and IRAC passbands
(although it may not apply to the ionizing source responsible for the
far--IR emission, see \S~\ref{section:uvir}).   Using different
extinction laws would affect the derived stellar--population ages and
extinction, but would not strongly change the inferred stellar masses
\citep{pap01}.  We also make the assumption that the stellar
populations dominating the rest--frame UV to near--IR also dominate
the stellar mass.  If a substantial fraction of the galaxies' stellar
mass is obscured from view, then the stellar--mass derived from the
SED modeling will be underestimated.  We first allow for a range of
star--formation histories with a SFR parameterized as a decaying
exponential with a $e$--folding time, $\tau$, where the SFR at any
age, $t$, is given by $\Psi(t) \sim \exp( -t/\tau)$.  In our models,
$\tau$ ranges from 1~Myr to 100~Gyr in quasi--logarithmic
steps. Short--duration $\tau$ values correspond to instantaneous
bursts of star formation while long--duration $\tau$ values correspond
approximately to constant star--formation histories.


For each DRG, we fit all available bands among the ACS
\acsb\acsv\acsi\acsz, ISAAC $JH\ks$, and IRAC \mone\ and \mtwo\ data.
In the model fitting, we add a  $\sigma/f_\nu = 4$\% error in
quadrature to the photometric uncertainties on each band to account
for the fact that the population--synthesis models do not continuously
sample the model parameter space \citep[see][]{pap01}.   The fits
provide a normalization between the photometry and the model, and a
minimum $\chi^2$ for each particular model with a distinct set of
parameters.   The IRAC \mthree\ or \mfour\ photometry because the
longer--wavelength IRAC data generally have lower signal--to--noise
ratios, and it is possible that at rest--frame $\gsim 2$~\micron\ PAH
features or emission from obscured AGN shift into the IRAC bandpasses.
Some galaxy SEDs show evidence for such features in these data (see
Figure~\ref{fig:sed}), and so we exclude these points to avoid any
potential bias.  However, we find that a majority of the best--fit
models broadly reproduce the \mthree\ and
\mfour\ photometry, which lends credence to the fits.

The model with the minimum chi--squared value, $\chi^2_0$, is the
model with the best--fit for a given set of parameters ($t$, $\tau$,
$E[B-V]$, and $\mcal$; the latter is derived using the
stellar-mass--to--light ratio of the model, see Papovich et al.\
2001).   Using the $\Delta \chi^2$ difference between the chi--squared
value derived for other models with a different set of parameters and
$\chi^2_0$, we can construct confidence regions on each parameter in
the model.   For each galaxy we generated up to 1000 Monte Carlo
realizations for the photometry by perturbing the measured flux
densities by a random value taken from a Normal distribution with a
standard deviation equal to the flux--density errors. We then refit
the new photometry and re-obtain best fits on all model parameters.
We do not take into account errors in the photometric redshifts, which
can affect the derived star--formation histories but generally have
little effect on the inferred stellar masses (see Dickinson \etal\
2003).   We identify the $\Delta \chi^2$ values from the fit to the
measured data that encompasses 68\% and 95\% of the best--fit values
from the Monte Carlo realizations, which provides the equivalent
confidence range on the model parameters for each source.

\subsection{Stellar population models with single exponentially
decaying star--formation histories}

\ifsubmode
\else
\begin{figure*}
\plottwo{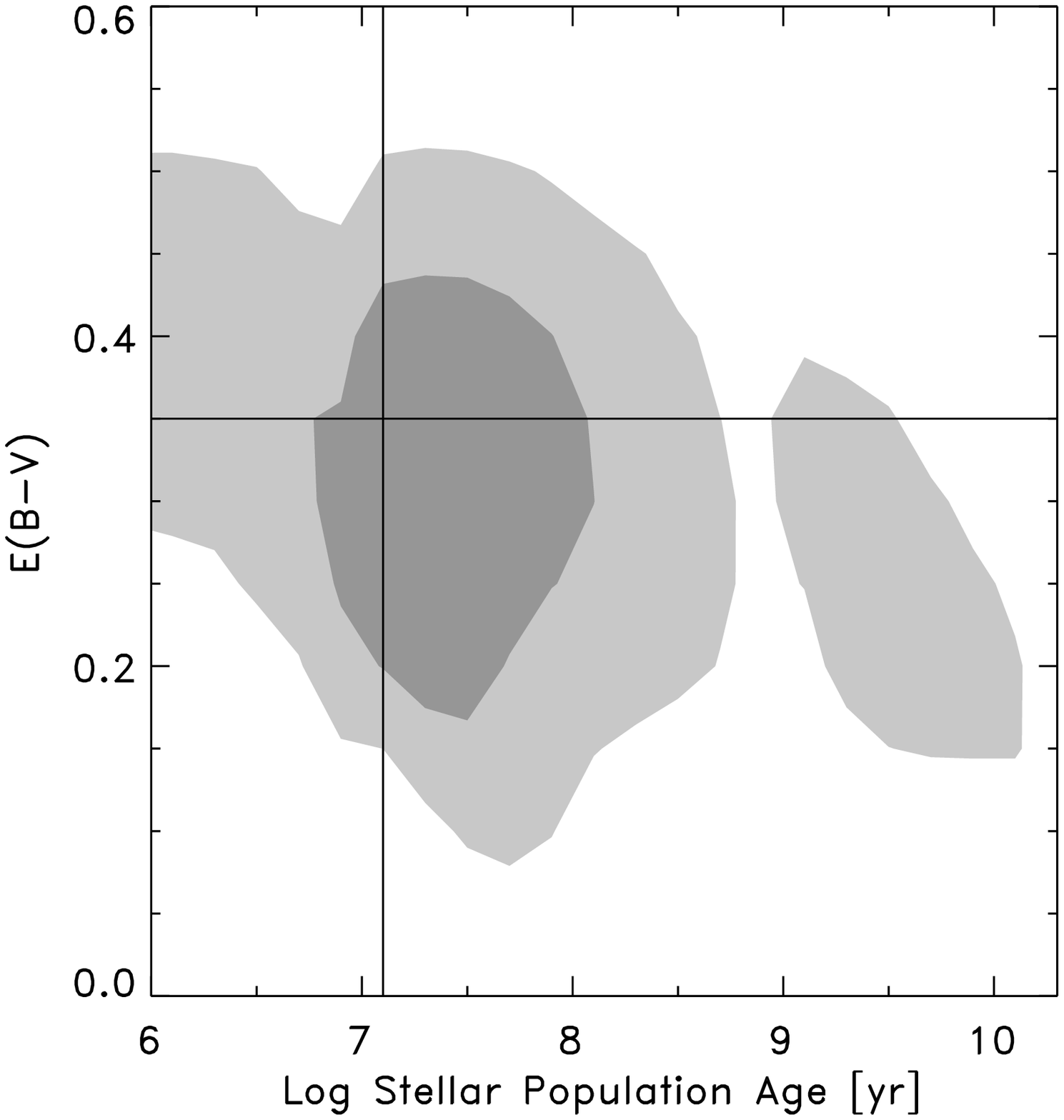}{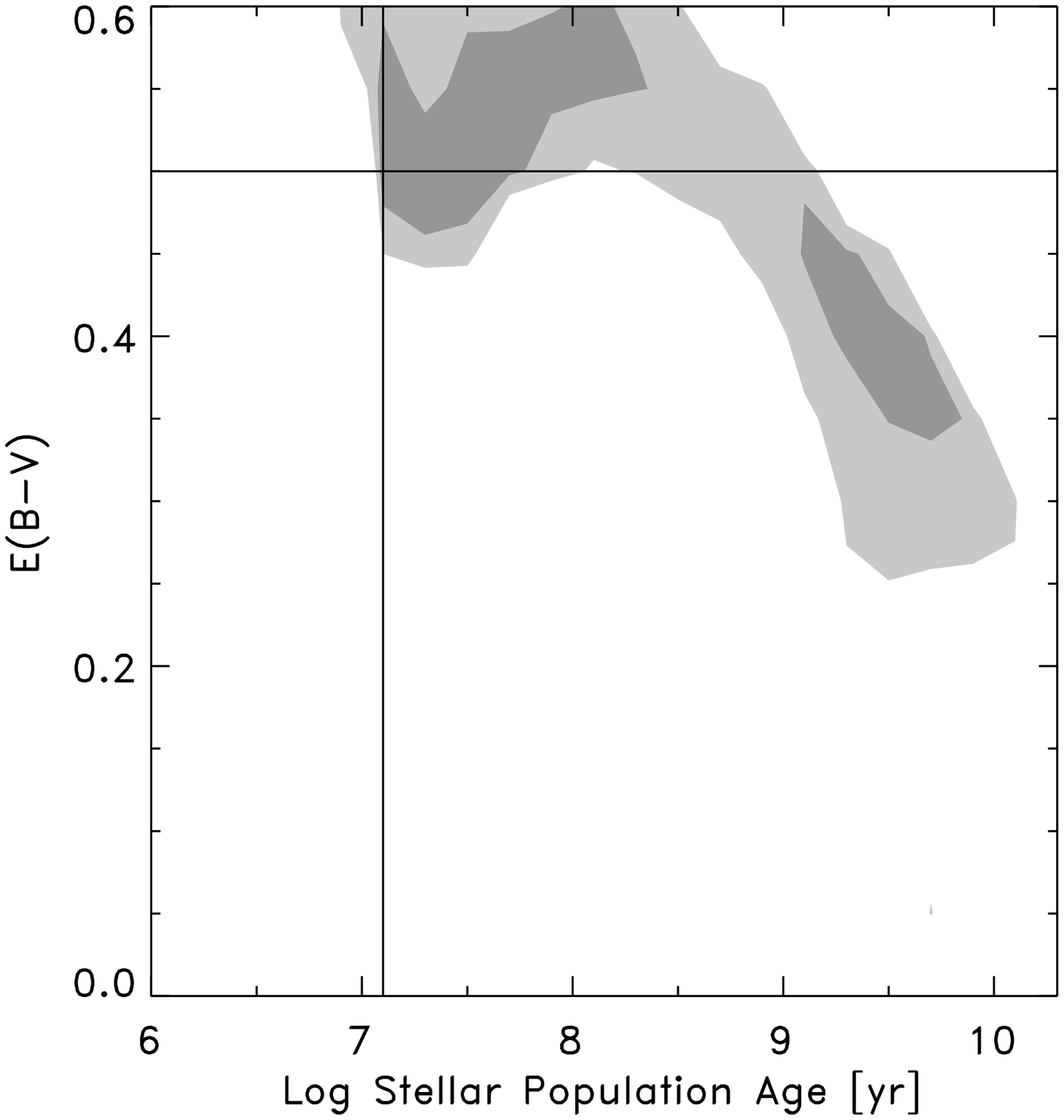}
\plottwo{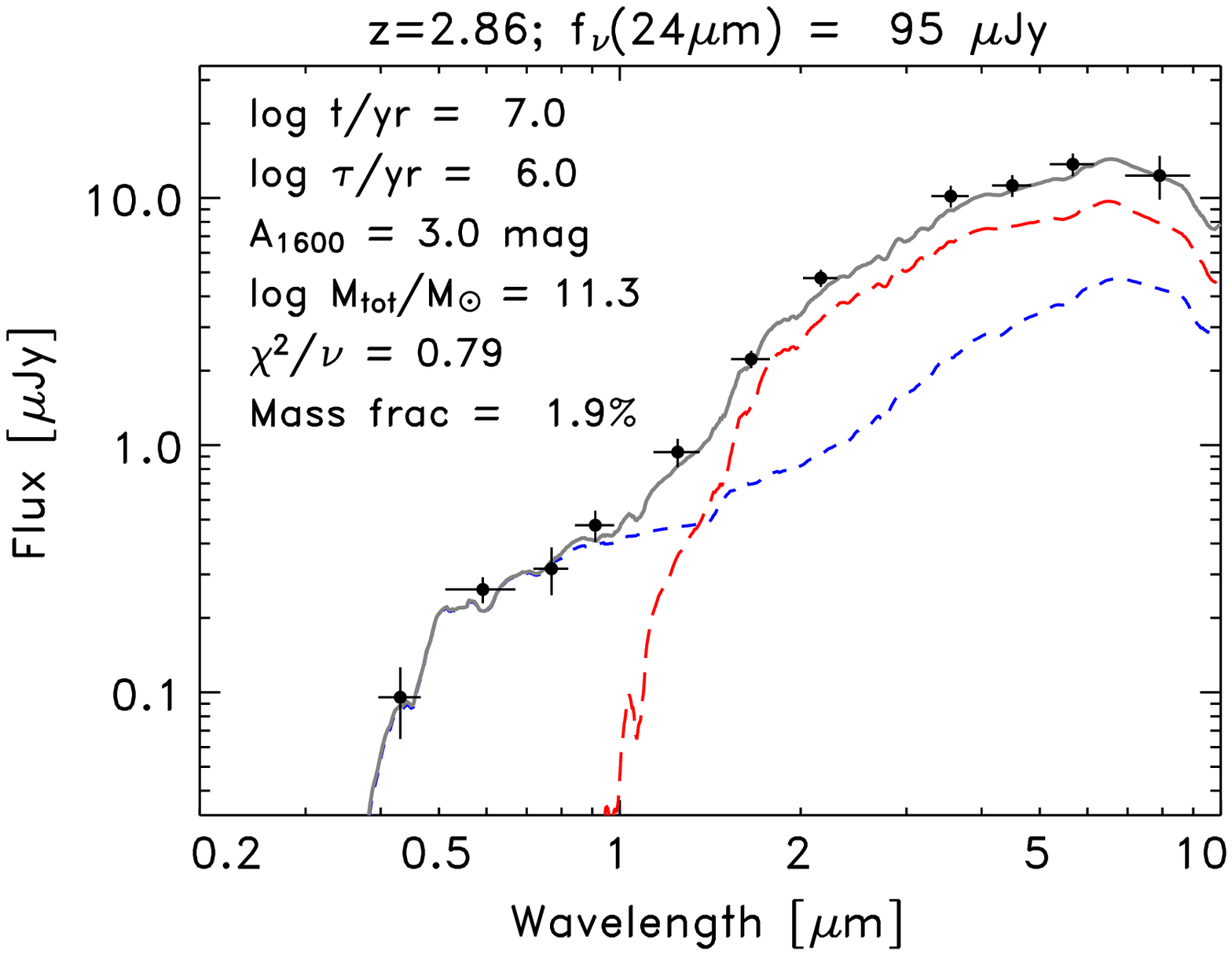}{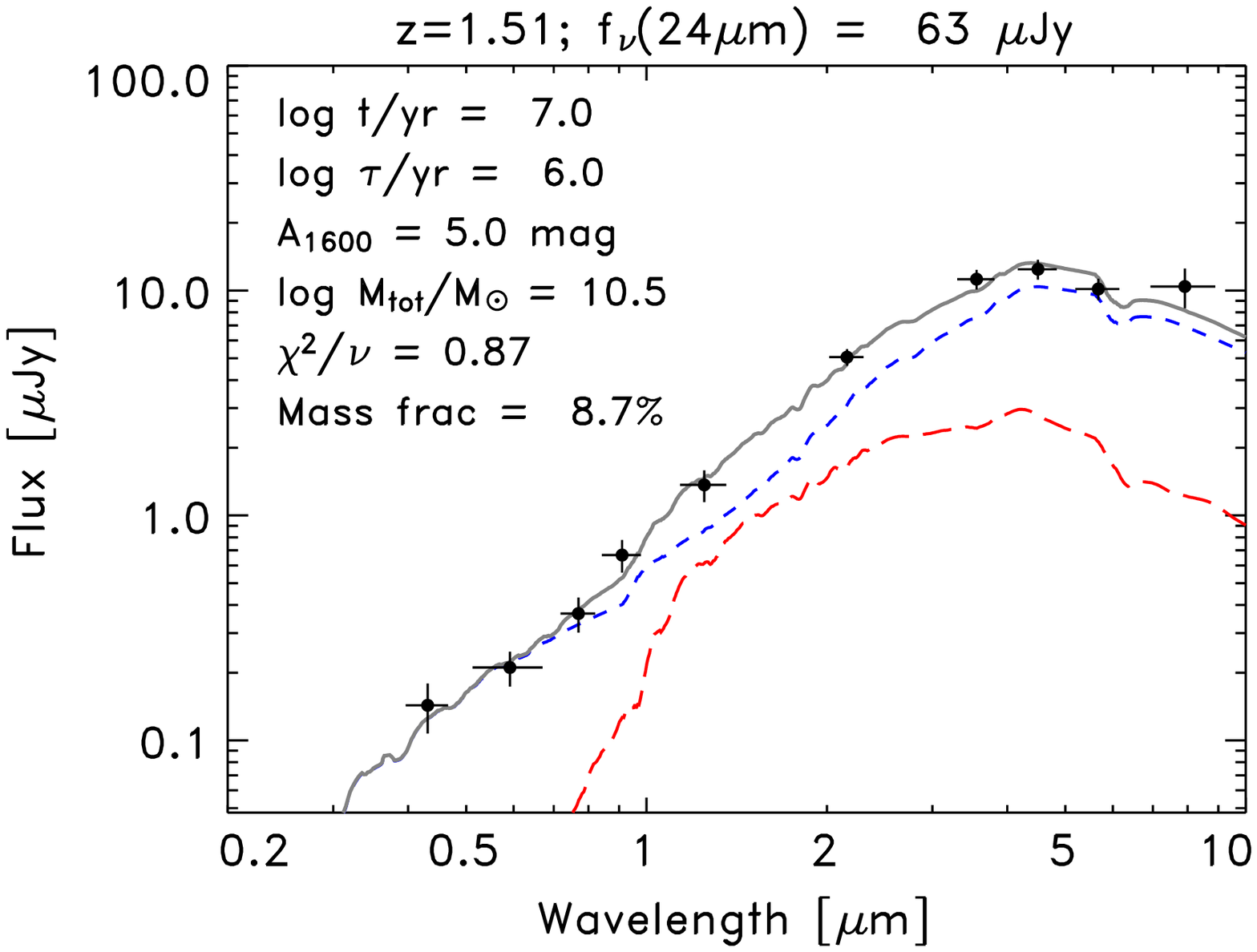}
\caption\figcapsedfittwocomp
\end{figure*}
\fi

\ifsubmode
\begin{figure}
\epsscale{0.9}
\plottwo{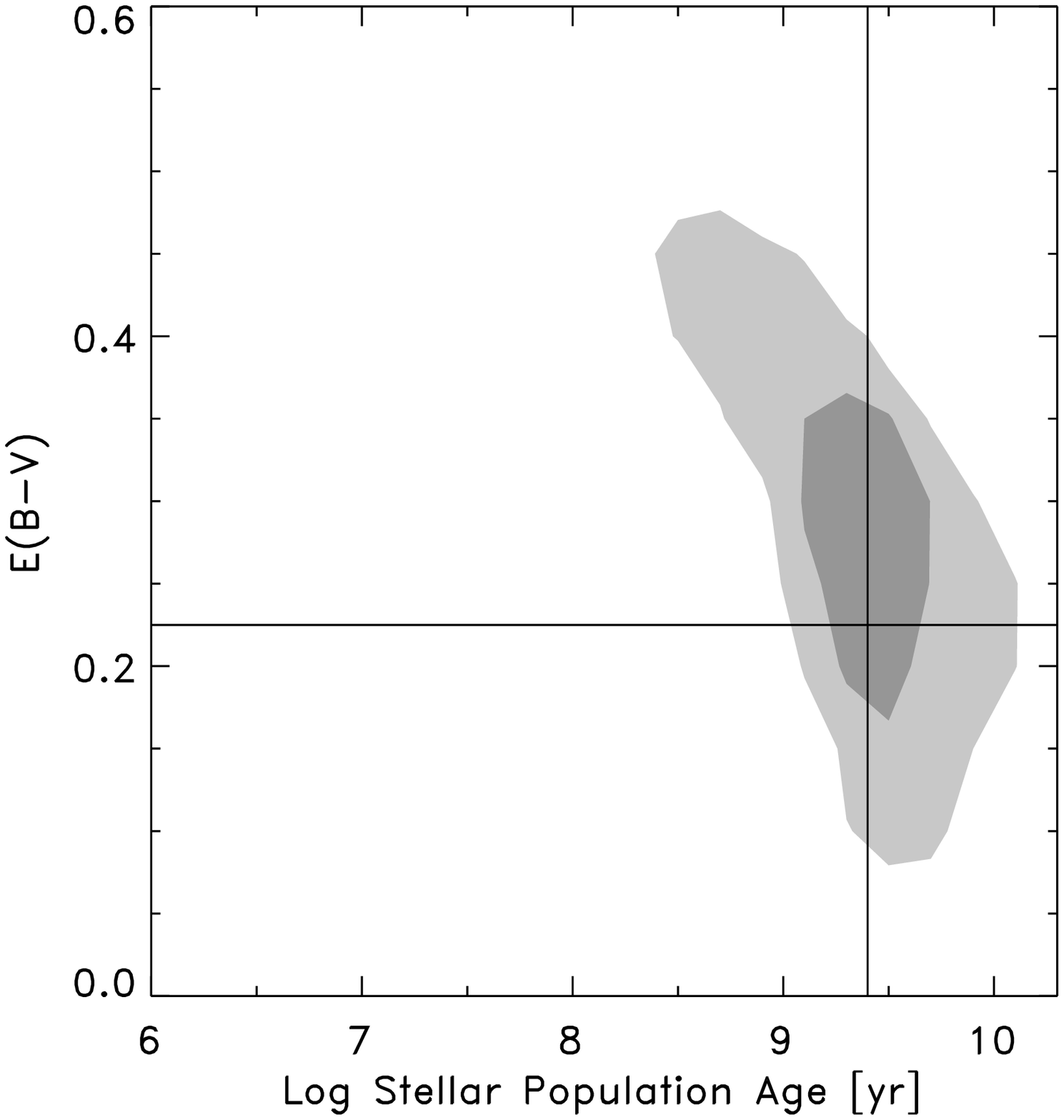}{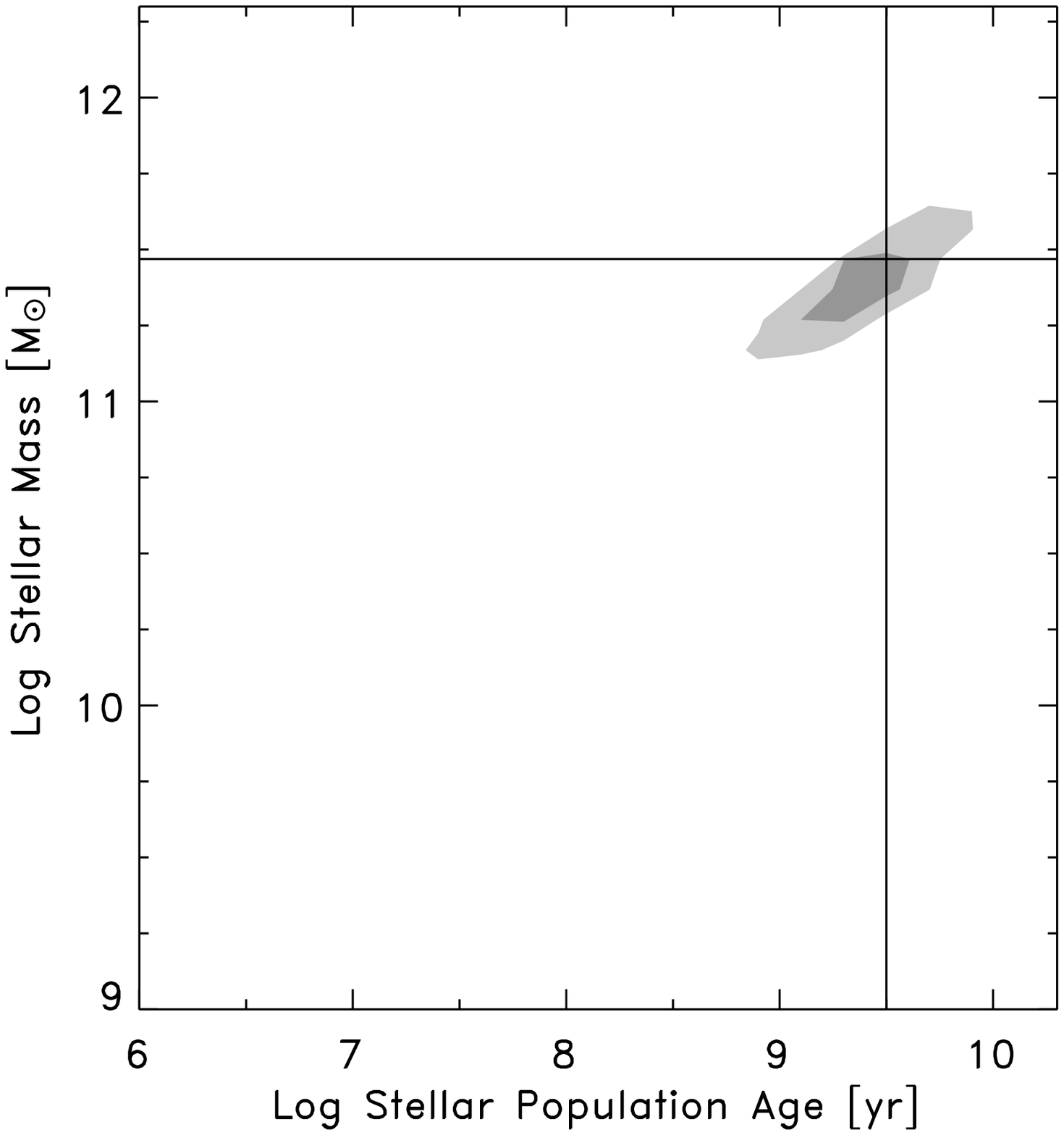}
\epsscale{0.75}
\plotone{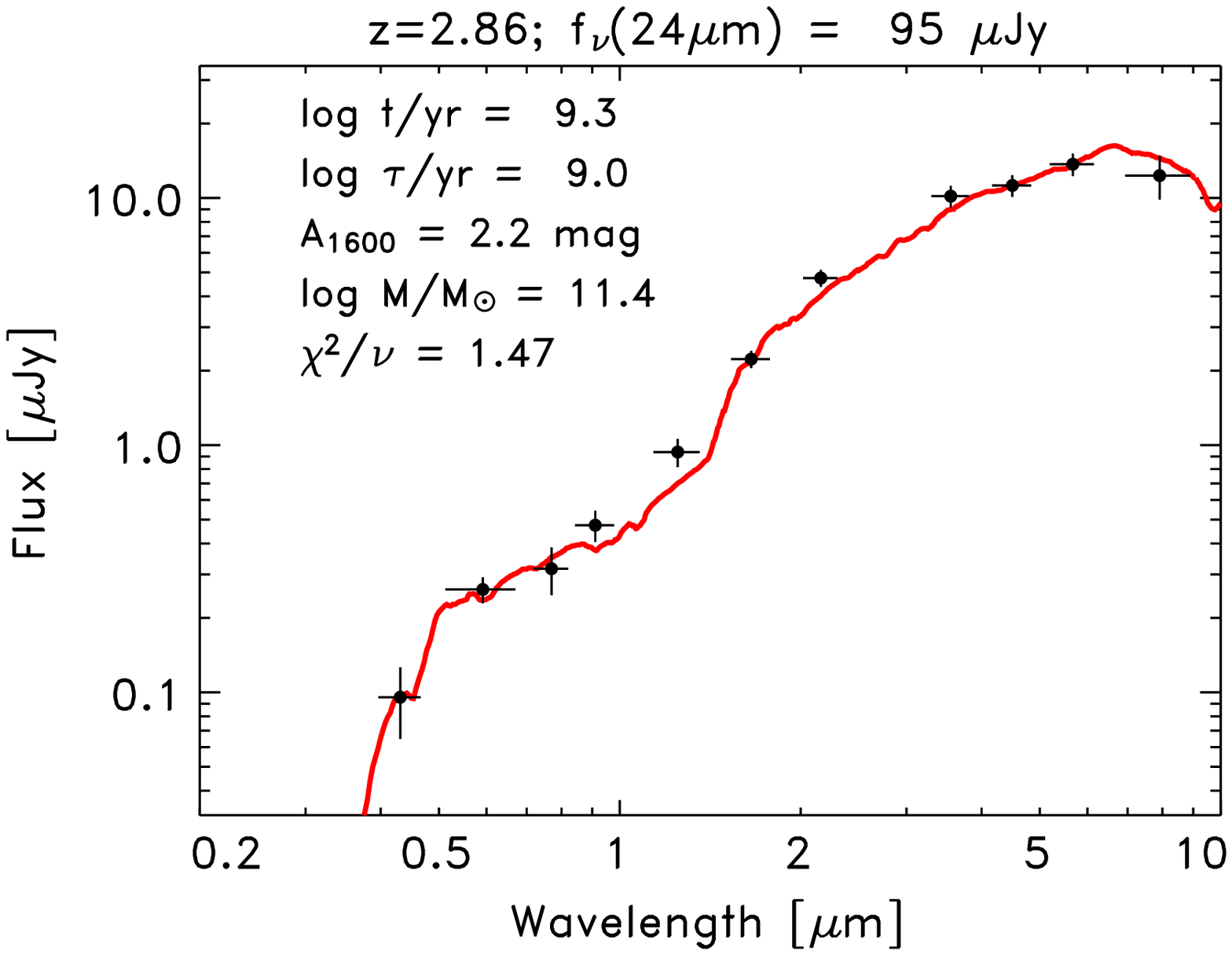}
\epsscale{1.0}
\caption\figcapsedfitone
\end{figure}
\fi

\ifsubmode
\begin{figure}
\epsscale{0.9}
\plottwo{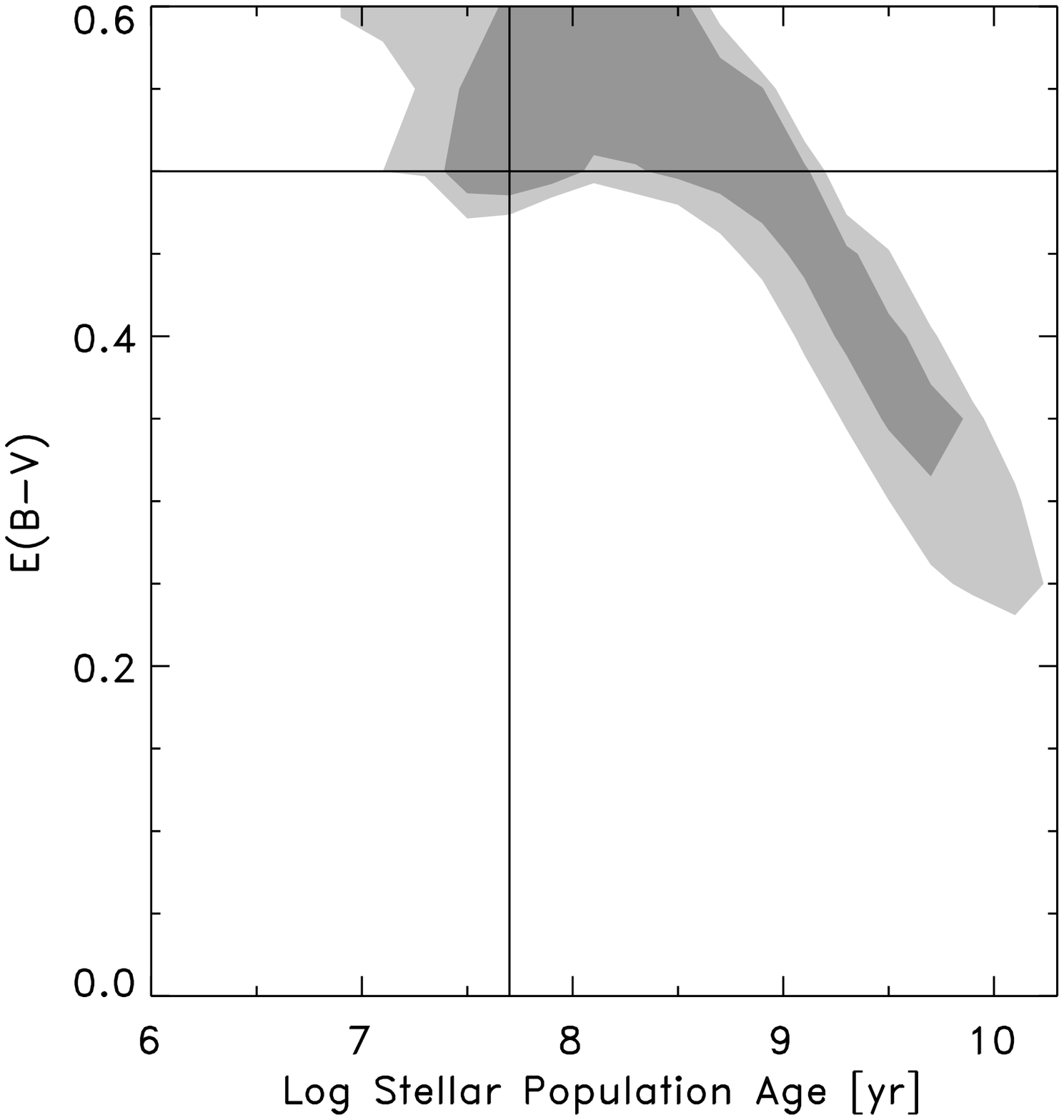}{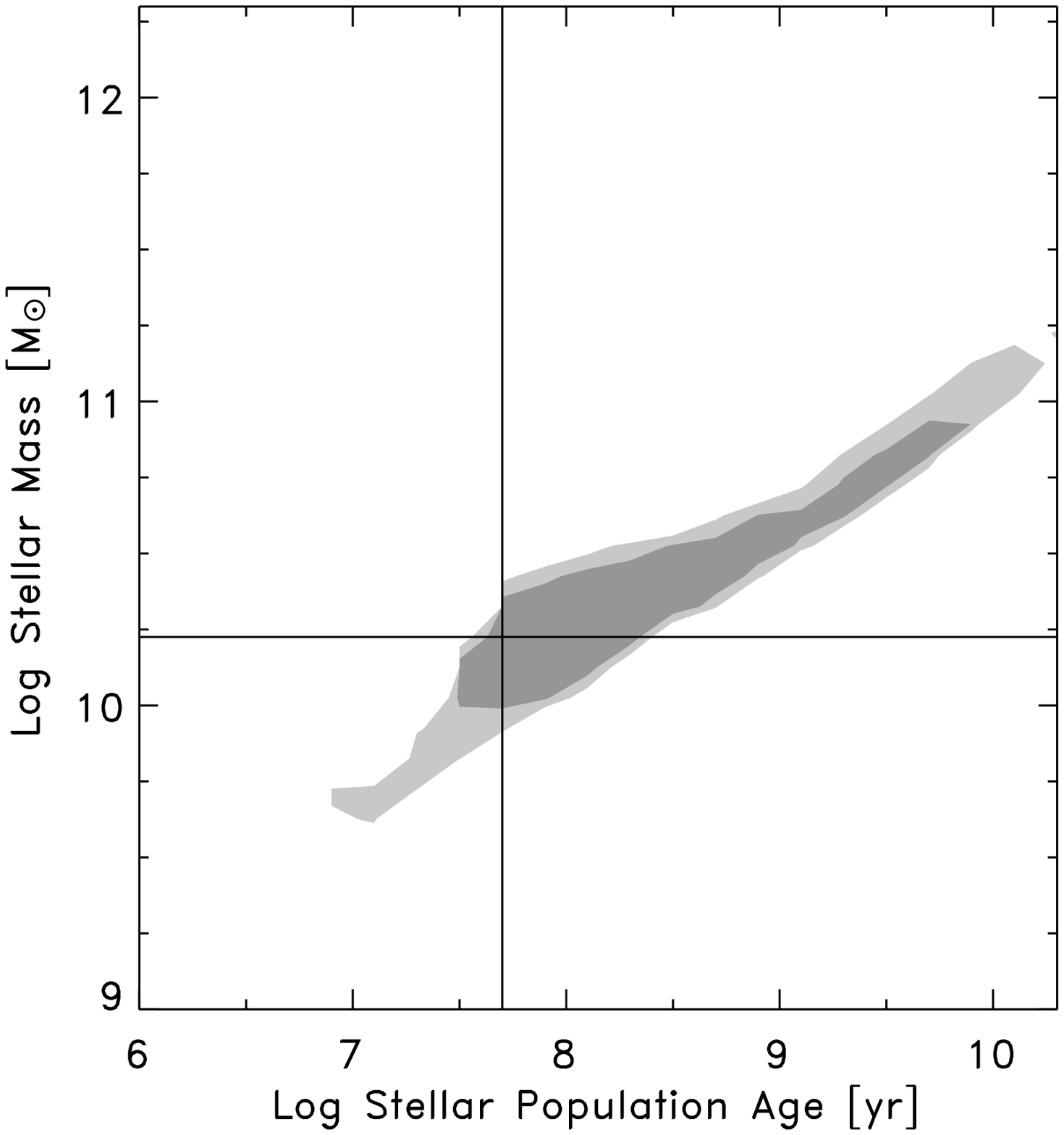}
\epsscale{0.75}
\plotone{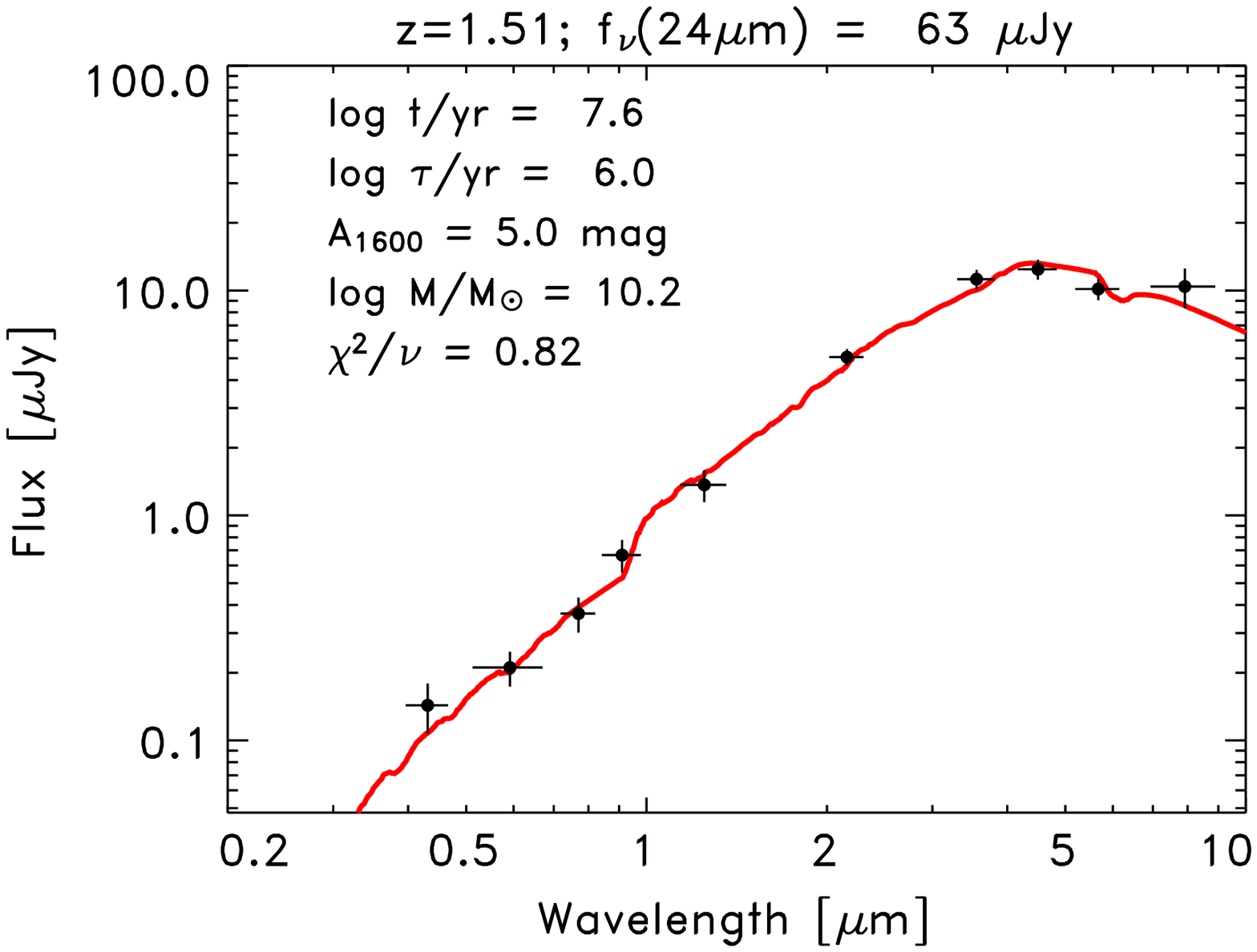}
\epsscale{1.0}
\caption\figcapsedfittwo
\end{figure}
\fi

Figures~\ref{fig:sedfitone} and \ref{fig:sedfittwo} present examples
of the results for two of the DRGs using the single--component
star--formation history models.   The examples are typical of the two
populations identified by the $(\jmk)_\mathrm{Vega} > 2.3$~mag
selection.  Each figure shows the probability distribution function in
two--dimensional projections of the age--extinction, and age--mass
planes.  The figures also show the best--fit model spectrum over all
parameters (i.e., the model with the minimum $\chi^2$ from all models)
overplotted on the data.

Figure~\ref{fig:sedfitone} shows a galaxy with a strong
4000~\AA/Balmer break between the $J$-- and \ks--bands presumably due
to a substantial mature stellar population.   The galaxy also has a
small amount of ongoing star--formation that produces the rest--frame
UV light.  The best--fit model corresponds to a star--formation
history that has undergone roughly $t/\tau\simeq 2$ $e$--folding
times.   For this model, the galaxy formed most of its stellar mass in
the past at a substantially higher SFR.  The early--type (OB type)
stars from the early onset of star formation have died off, and thus
most of the stellar mass resides in later--type stars, which formed
well in the past and now dominate the optical and near--IR rest--frame
light.  The best--fit model requires substantial dust extinction:
$>2$~mag at 1600~\AA\ (68\% confidence).   The estimated stellar mass
is quite robust for this set of star--formation histories.  The 68\%
confidence range  on the stellar mass ranges from $1.6-2.7\times
10^{11}$~\msol, with a most--likely value of $2.4\times 10^{11}$~\msol.

Figure~\ref{fig:sedfittwo} shows a galaxy with a heavily extincted
recent starburst.   The best--fit model to this galaxy is a young
stellar population formed roughly instantaneously in a burst 40~Myr in
the past.   However, the 68\% confidence range on the
stellar--population age ranges from $\sim 30$~Myr to several Gyr.  For
all possible ages, the modeled star--formation history has undergone
many $e$--folding times and has substantial dust extinction, which is
required to produce the red UV--to--near-IR rest--frame colors.   This
best--fit model requires 5~mag of extinction at 1600~\AA.   Owing
primarily to the larger model degeneracies in the age and dust
extinction for this galaxy, the stellar mass is less well constrained
compared to the example in Figure~\ref{fig:sedfitone}.   The 68\%
confidence region on the stellar mass is $9\times 10^9$ to $8.4\times
10^{10}$~\msol, with a most--likely value of $1.6\times
10^{10}$~\msol.

\subsection{Stellar population models with double component
star--formation histories}\label{section:twocomp}

We also fit the DRG photometry models with a two--component
star--formation history characterized by a passively evolving stellar
population formed in a previous ``burst'' with $z_\mathrm{form} =
\infty$, summed with the exponentially--decaying--SFR model above.
For these models, the stellar--population age is the time since the
onset of star formation in the monotonically evolving component.
These models check the effects of discrete bursts on the derived
parameters.   Our choice of a burst at $z_\mathrm{form} = \infty$ is a
proxy for bursts at all times before the observed redshift, and
reducing $z_\mathrm{form}$ would not strongly affect our conclusions.
Placing the burst at $z_\mathrm{form} = \infty$ provides a limit on
the maximum stellar mass because it has the maximal mass--to--light
ratio possible at the observed redshift.

\ifsubmode
\begin{figure}
\epsscale{1.1}
\plottwo{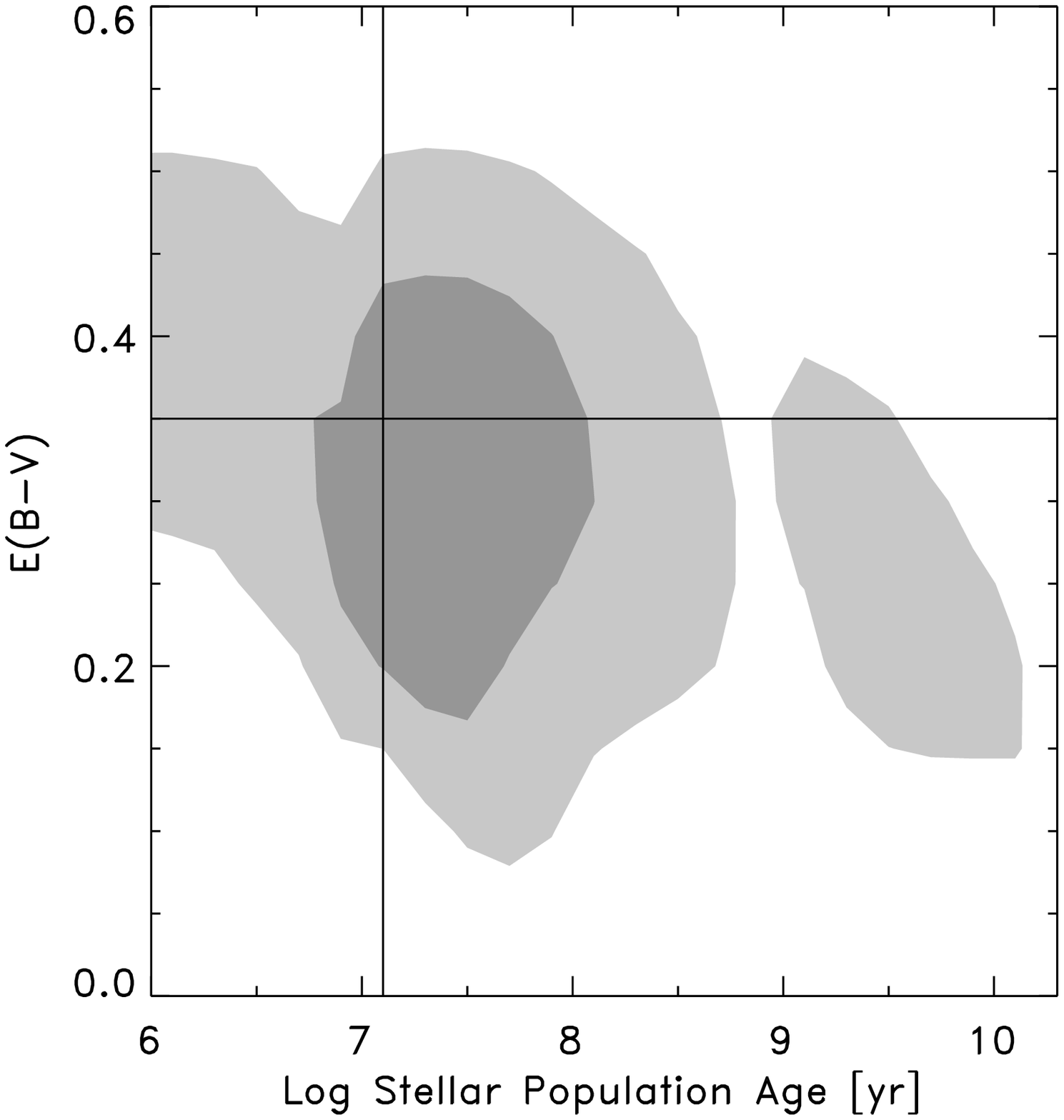}{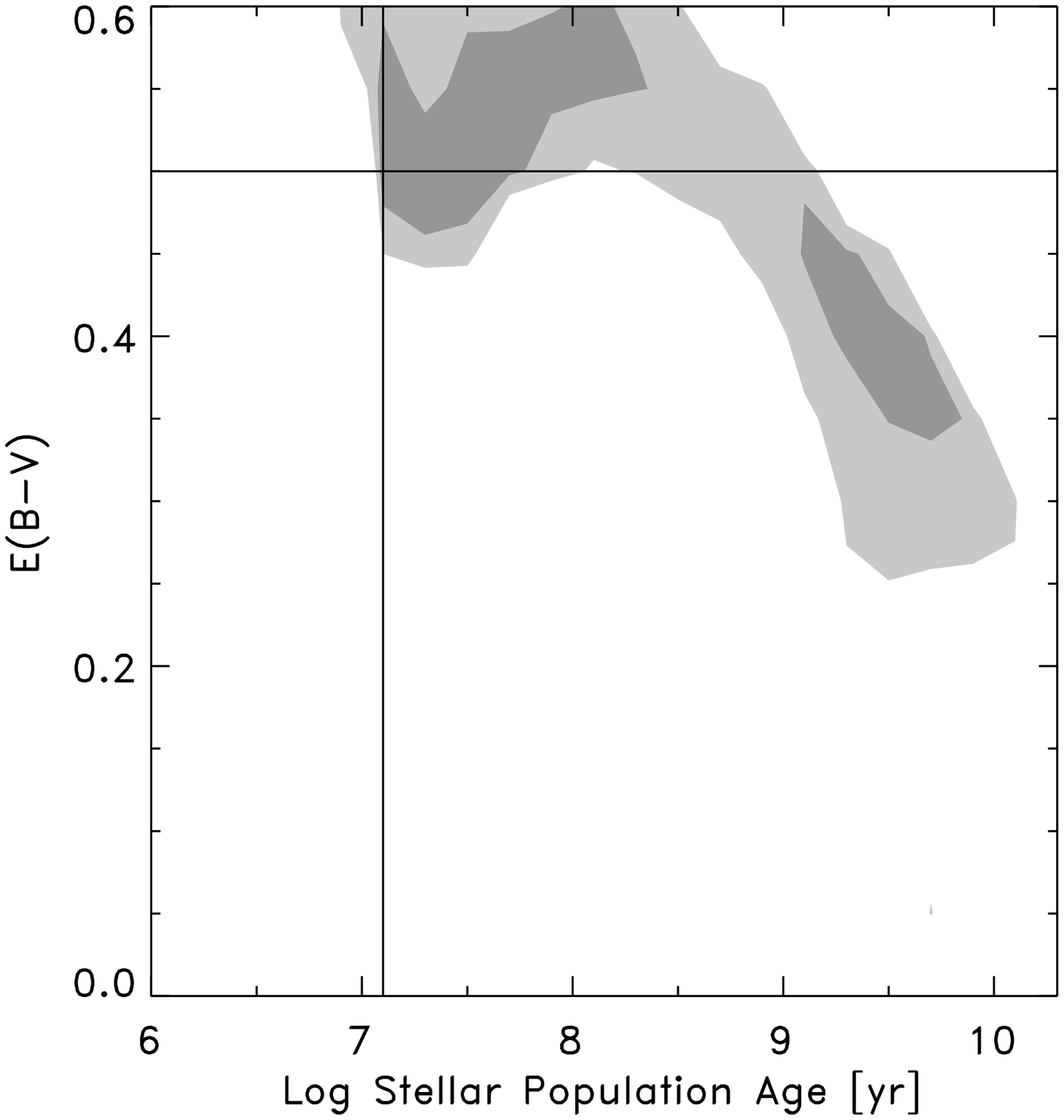}
\plottwo{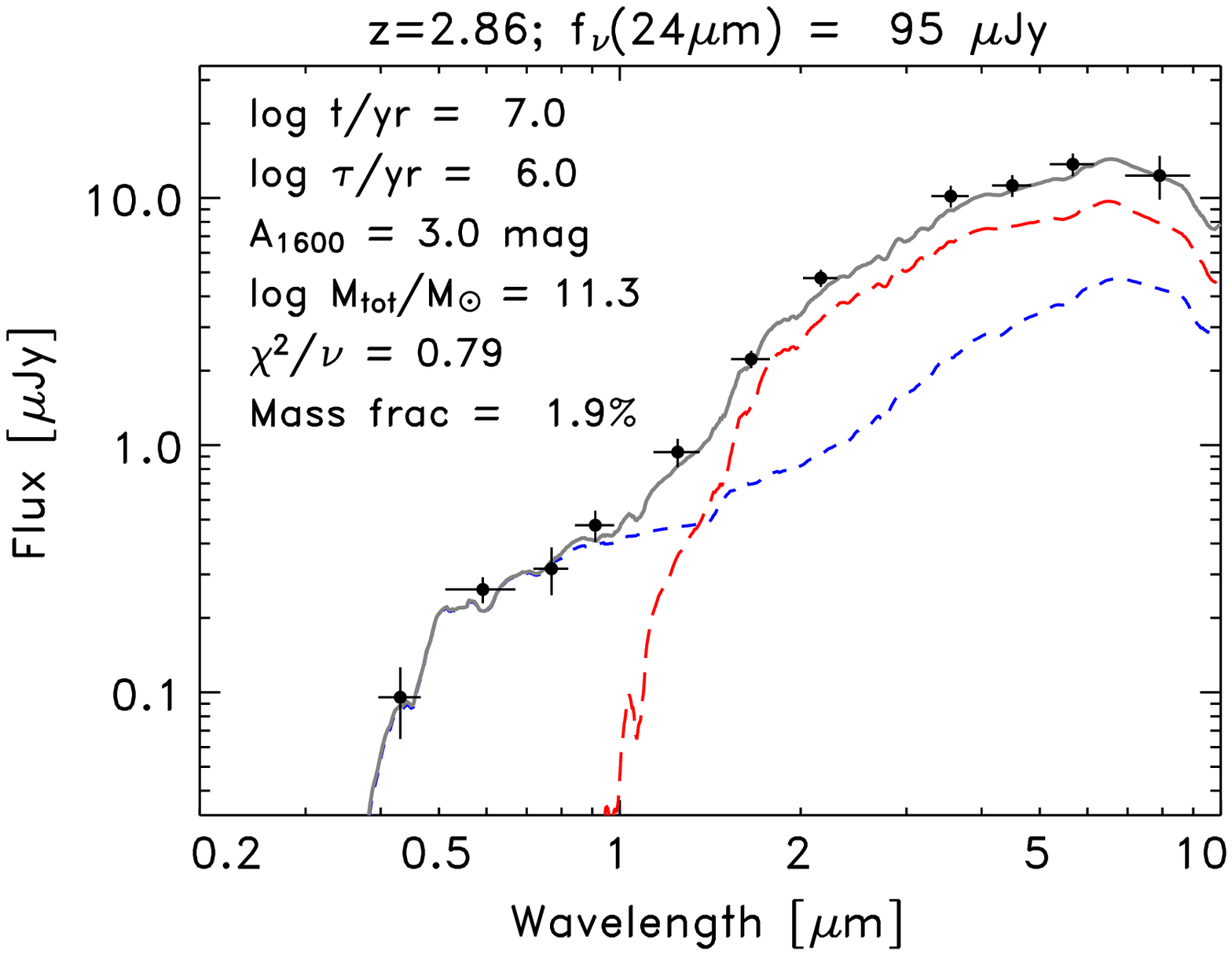}{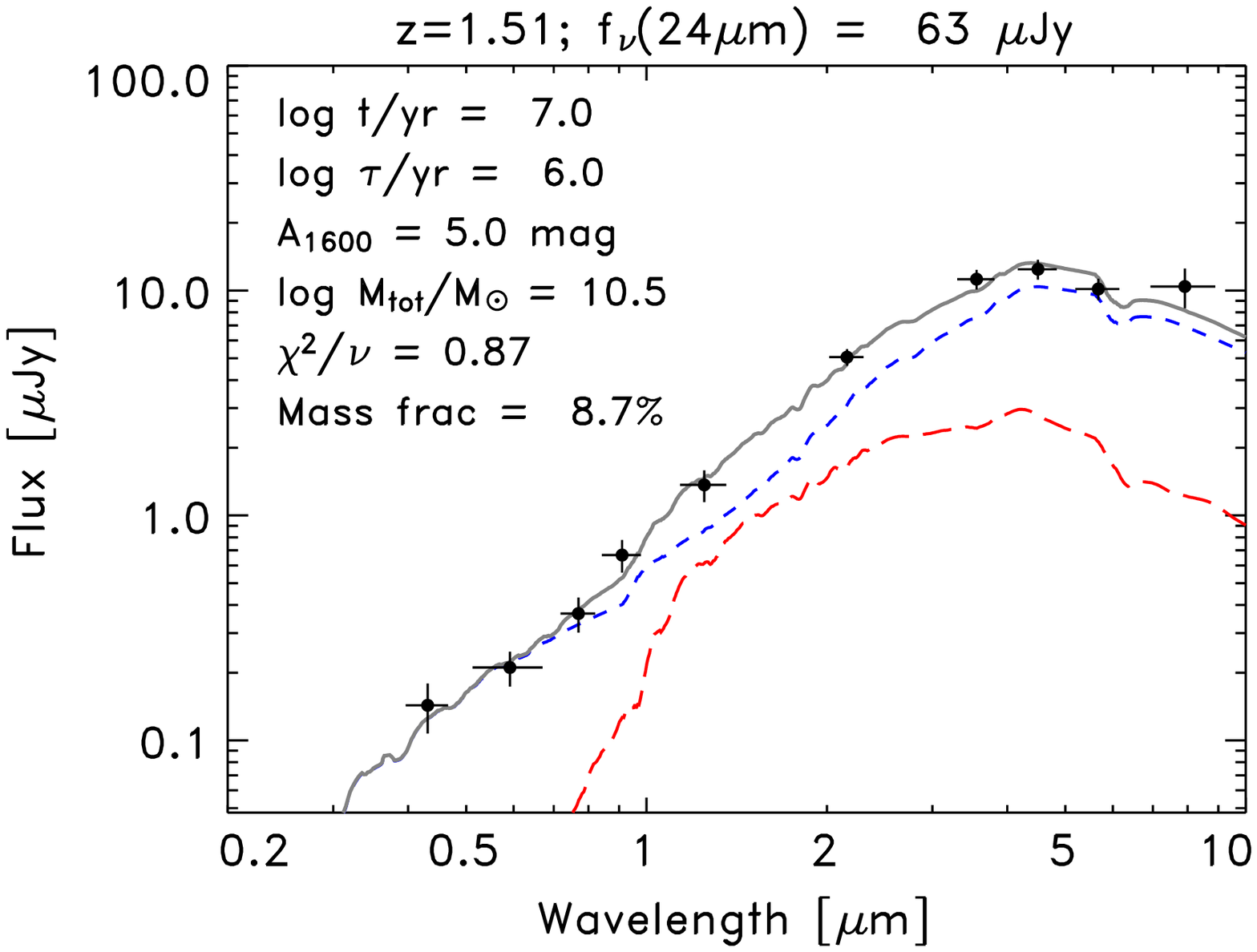}
\epsscale{1.0}
\caption\figcapsedfittwocomp
\end{figure}
\fi

Figure~\ref{fig:sedfittwocomp} shows the results for the
two--component model fits to the two DRGs described above.  For the
galaxy shown in Figure~\ref{fig:sedfitone}, a very young ($t\sim
10$~Myr), heavily extincted ($A_{1600} \sim 3$~mag) starburst
dominates the UV rest--frame emission.    The optical and near--IR
rest--frame emission is produced by the previously formed population,
which is now quite old.  However, the 68\% confidence range on the age
and dust extinction  shows that a large region of the parameter space
fits the data equally well.  This is in contrast to the region
permitted by the single--component model fits in
Figure~\ref{fig:sedfitone}.  Clearly the interpretation of the ages
and extinction depend strongly on the assumed star--formation history.
However, the derived stellar masses are fairly robust.   The total
stellar mass is nearly unchanged in this model relative to the
single--component model above; the 68\% confidence range is
$1.6-2.3\times 10^{11}$~\msol.   The upper range of the confidence
region drops simply because in the previous model we did not enforce
the constraint that the age of the model be less than the age of the
Universe at the given redshift.   Such a constraint is imposed (by
construction) on the two--component fits, and as a result there is
less time for stellar populations to evolve, produce large
mass--to--light ratios, and increase the stellar mass.

\ifsubmode
\else
\begin{figure}
\plotone{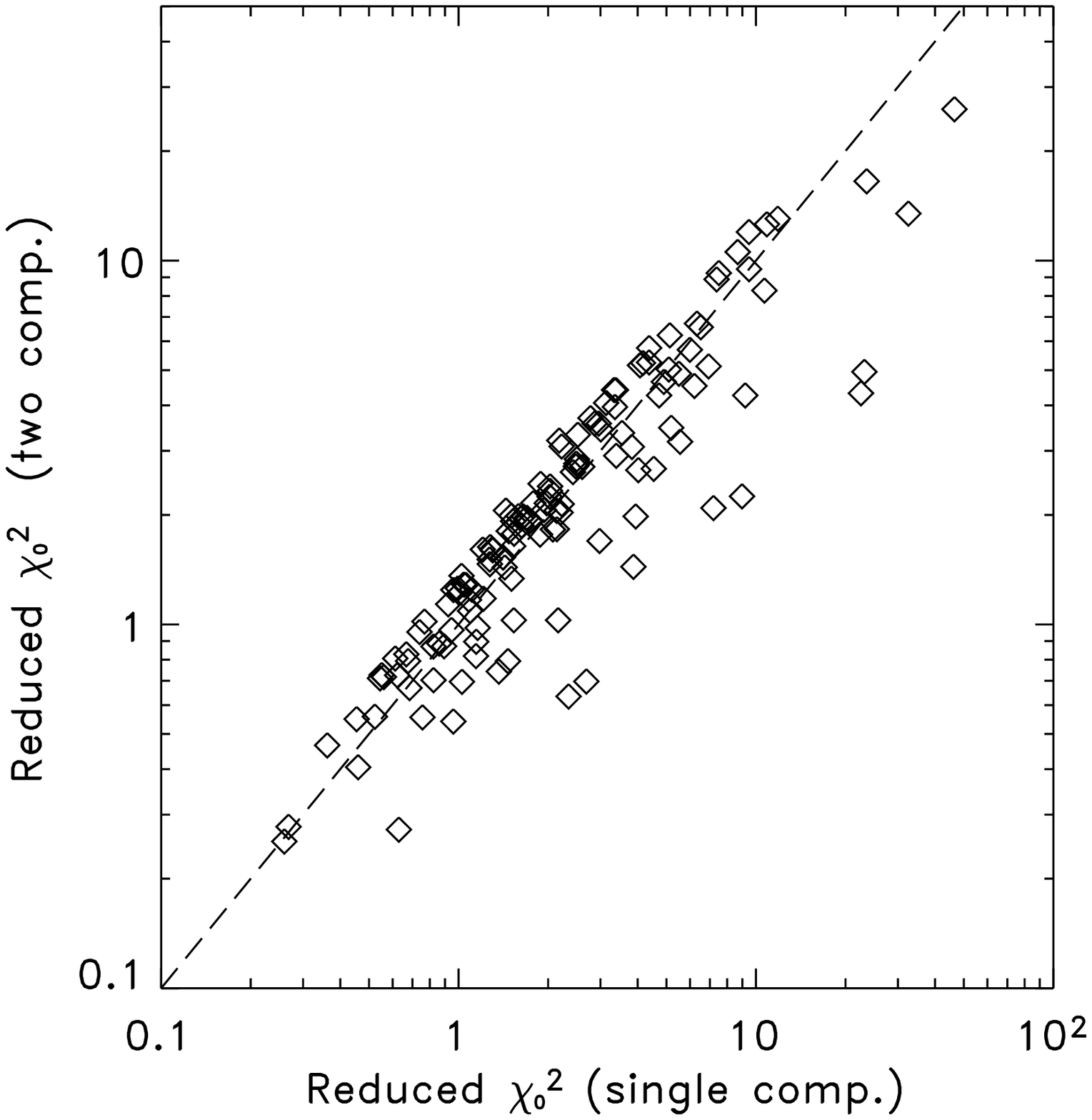}
\caption\figcapchisq
\end{figure}
\fi

The best--fit two--component star--formation history model for the
galaxy in Figure~\ref{fig:sedfittwo} shows little change compared to
the single--component fit.  In this case, young, reddened stars
dominate the light at most wavelengths for both types of models.
Thus, for galaxies of this sort, adding the second component only
provides an upper limit on the inferred stellar mass content.

\ifsubmode
\begin{figure}
\plotone{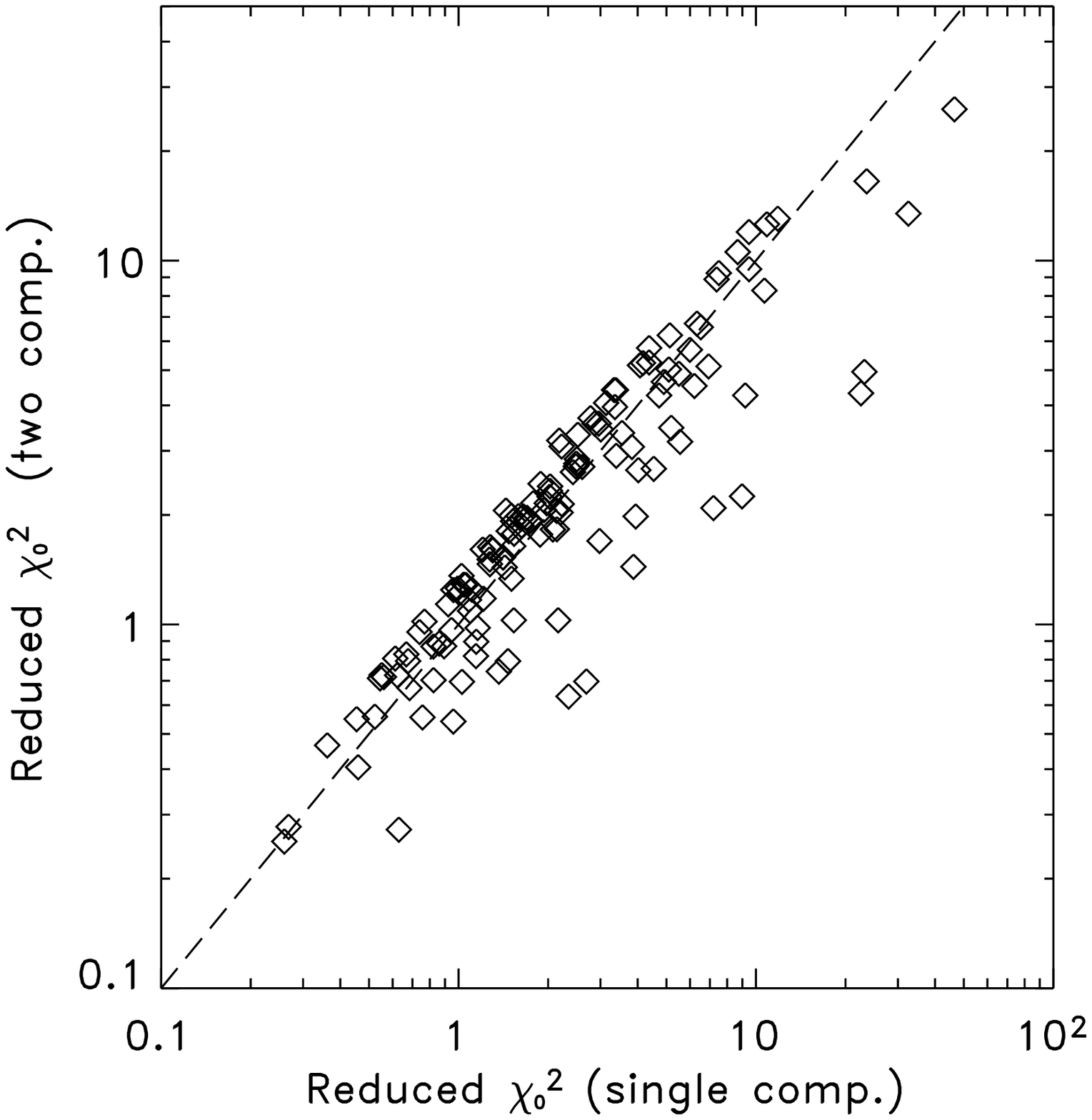}
\caption\figcapchisq
\end{figure}
\fi

In general, the two--component fits for the DRGs appear more
consistent with the data than the less complex, single--component
models.   Figure~\ref{fig:chisq} compares the reduced $\chi^2_0$ from
the fits to the single--component models to the reduced $\chi^2_0$
from the fits to the two--component models for all the DRGs.  For many
DRGs (91/152), adding the second model component to the fit has little
effect on the minimum reduced $\chi^2$, and these points lie near the
unity relation indicated in the plot.  For a large fraction of DRGs
adding the second burst lowers the reduced minimum $\chi^2$
significantly.  It is possible that some of this
effect results from the fact that we have not included the
contribution of emission lines in the various bandpasses.  However, as
noted earlier, this contribution should be small, contributing $\lsim
0.1$~mag based on the observed emission--line equivalent widths
\citep{vandok04,vandok05}, except perhaps for galaxies with AGN, for
which the models may not apply.  Thus, in general this effect does not
account for the smaller reduced $\chi^2$ values observed in our
galaxies.   For example, the best--fitting single--component model
for the galaxy in Figure~\ref{fig:sedfitone} deviates from the
$J$--band photometry at $\sim 1$~\micron\ by nearly $2\,\sigma$,
whereas the two--component model is better able to reproduce this data
point (see Figure~\ref{fig:sedfittwocomp}).

The photometry in many of the DRGs is better represented by
star--formation histories that are more complex than the simple,
monotonically evolving exponentially decaying SFR.   Formally, the
$\chi^2$ statistic rejects the single--component model for 13/152 DRGs
at the $3\,\sigma$ level, while not rejecting the double--component
model at this significance level.    In contrast, the $\chi^2$
statistic never rejects the double--component model in favor of the
single--component model at this significance level.   Furthermore,
many of the best--fit single--component models favor stellar
population ages that are older than the age of the Universe for the
measured redshift (see \S~\ref{section:ages}).   Restricting these
models to ages less than the age of the Universe increases the minimum
$\chi^2$ value, making the difference between single-- and
two--component fits more pronounced.  We interpret this behavior to
indicate that in general the DRG population has star--formation
histories that are more complex than simple monotonically evolving
stellar populations.  A similar scenario has been suggested based on
modeling the SEDs of LBGs at these redshifts
\citep[\eg,][]{saw98,pap01,pap04a,sha05}, and is likely consistent
with hierarchical models \citep{delucia05,nag05}.

\section{Discussion}\label{section:discussion}

The ensemble properties of the DRGs span a range of the
stellar--population model parameter space.  Broadly speaking, the
$(\jmk)_\mathrm{Vega} > 2.3$~mag color selection identifies galaxies
whose rest--frame optical and near--IR light is dominated by
later--type stars, and galaxies whose light is dominated by heavily
extincted starbursts \citep{fra03}.   In this way, it is not dissimilar to
traditional $R-K$ or $I-K$ selection criteria for EROs, but it tends
to pick out objects at higher redshifts.  In addition, roughly $\simeq
15$\% of the DRGs are luminous in X--rays, implying that some of the
UV to IR emission may stem from SMBH accretion processes rather than star
formation. Here we consider the implications that our analysis has for
star--formation in massive galaxies at high redshifts.  We briefly
consider AGN activity in these galaxies, with a more detailed analysis
to be presented in L.~A.~Moustakas et al. (in preparation).

\subsection{Stellar Populations and Star--Formation Histories of DRGs}

\subsubsection{Dust Extinction}\label{section:dust}

The best--fitting single--component DRG models span extinctions of
$\sim 0-6$~mag at 1600~\AA.    The mean dust extinction from the best
fits to the ACS to IRAC photometry is $\langle A_{1600} \rangle =
3.1$~mag, with a standard deviation of 1.5~mag.   The mean is 
substantially larger than that inferred for UV--selected LBGs at
comparable redshifts, for which \citet{pap01} and \citet{sha01} find
average extinction values of 1.2 and 1.6~mag at 1600~\AA,
respectively.  \citet{for04} find that DRGs in the FIRES fields have
median extinction of $\simeq 6.2$~mag at 1600~\AA\ for the Calzetti
extinction law (averaged by number), but restricted to models with constant
SFR, and these values change little after including \spitzer/IRAC
data \citep{lab05}.   Although on the surface this appears at odds
with our findings, F\"orster--Schreiber \etal\ found that using
exponentially declining models reduces the measured extinction by
roughly a factor of two, in better agreement with our results.

To study differences between the sub--populations of DRGs (i.e., those
with substantial mass in late--type stars and those dominated by
highly dust--extincted starbursts), we divided the sample into subsets
with $E(B-V) \leq 0.35$ and $E(B-V) > 0.35$.  We have used the results
from the fits to the single--component models for this selection, but
there is little change if we use the two--component models.  This
division point is the approximate upper bound on values inferred for
UV--selected LBGs (largely as a result of the UV--selection itself;
e.g., Adelberger \& Steidel 2000).

DRGs with $E(B-V) > 0.35$ comprise roughly 40\%
(61/152) of the full sample. These galaxies have a mean redshift of
$\langle z \rangle \simeq 1.7$ with standard deviation $\sigma(z) =
0.5$.  The mean dust extinction is 4.6~mag at 1600~\AA\ for the
Calzetti law.  We estimate the resulting IR luminosities from these
extinction values using the empirical relations between the UV
luminosity, extinction, and far--IR luminosity \citep{meu99,cal00}.
In these galaxies, the observed UV luminosities and model extinction
(using the 68\% confidence limits) correspond to IR luminosities of
$10^{11--12.5}$~\lsol.  These are generally less than the
24~\micron--derived IR luminosity or the 24~\micron\ upper limit for
undetected galaxies).   However, because the 68\% confidence interval
on galaxy extinction is large, $\delta A(1600\AA)\sim 1-2$~mag, in
roughly $\sim$50\% of the objects the extinction from the upper 68\%
confidence bound yields a \lir\ comparable or exceeding the
24~\micron\ value.   This is consistent with the fraction of DRGs in
figure~\ref{fig:uvir} with IR/UV ratios and UV spectral slopes near
the local relation from \citet{meu99}.

The DRGs with best--fit $E(B-V) \leq 0.35$ have a mean extinction of
2.0~mag at 1600~\AA.  Later--type stars dominate the rest--frame
optical and near--IR light in these galaxies, while the UV rest--frame
emission stems from small amounts (by mass) of ongoing star--formation
with low to moderate extinction.  The mean redshift for the DRGs with
$E(B-V) \leq 0.35$ is $\langle z \rangle \simeq 2.5$, noticeably
larger than that for the DRGs with higher extinction.  This higher
redshift range arises because galaxies with a strong Balmer/4000~\AA\
break satisfy the $(\jmk)_\mathrm{Vega} > 2.3$~mag selection only for
$z\gsim 2.0$ as this break is moving through the $J$ and $H$ bands
\citep{fra03}, whereas heavily reddened galaxies can enter the sample
at lower redshifts.  The lower $E(B-V)$ estimates from the model
fitting would smaller IR luminosities relative to the DRGs with higher
extinction described above.  Unlike the case for the
higher--extinction DRGs, the IR--luminosity for the the DRGs with
$E(B-V) \leq 0.35$ are less than the 24~\micron--derived
\lir\ by factors of $\approx 2-20$.   Therefore, the observed
24~\micron\ emission in DRGs with relatively low dust extinction from
the SED modeling does not originate only from the extincted stellar
populations that dominate the UV and optical rest--frame light.
These galaxies require either additional embedded star formation, or
an obscured AGN, or both, and these obscured components contribute
negligibly at bluer wavelengths.

\subsubsection{Stellar Population Ages and Star Formation
Histories}\label{section:ages} 

Our stellar--population modeling of the DRGs is more sensitive to the
ratio of model age, $t$, to the star--formation $e$--folding time,
$\tau$, than on $t$ or $\tau$ individually, and we discuss them
simultaneously.   From the models with a single, exponentially
decaying SFR the median age is 1.1~Gyr, and on average a DRG has
undergone $t/\tau \simeq 4$ $e$--folding times.   In an analysis of
the 34 DRGs in the FIRES fields, \citet{for04} tested constant star
formation histories ($\tau \gg t$) and found  median stellar
population ages of $t \sim 1.7-2$~Gyr.    The lower median ages for
our models result because they favor star--formation histories with a
smaller number of $t/\tau$ $e$--folding times than the constant star
formation histories.  Indeed, F\"orster--Schreiber et al.\ also
considered models with $\tau=300$~Myr, for which they derive a median age
of 1~Gyr, implying $t/\tau \sim 3$.  This is consistent with our
result taken over a wider range of star--formation histories.

There are 17 DRGs ($\simeq 11$\% of the sample) where the lower limits
from the 68\% confidence range from the single--component--model fits
are older than the Universe.   Of these, nine (53\%) show indications
of AGN activity either because they are detected in the
\textit{Chandra} data,  they have $\lir > 10^{13}$~\lsol, and/or they have
have IRAC colors satisfying the AGN selection of \citet[see also
\S~\ref{section:agn}]{stern05}.  We suspect an AGN may influence the
observed photometry and the model fits.  The remaining DRGs have SEDs
qualitatively similar to the example shown in
Figure~\ref{fig:sedfitone}, but the model fits favor older ages.   In
all these cases the two--component model fits have solutions with ages
less than the age of the Universe within the 68\% confidence interval,
and are thus more physical.   Moreover the reduced minimum $\chi^2$
values for the two--component fits are significantly improved for more
than half of these, implying such models are better realizations of
the data.

Thus, including the constraint that the age of the model stellar
population be less than that of the Universe favors models with more
complicated star--formation histories than the single--component,
monotonically evolving models.  This reinforces our conclusion that
the two--component models better describe the star--formation
histories of the DRGs based on their lower reduced $\chi^2_0$ values
(see \S~\ref{section:twocomp}). The real star--formation histories are
probably even more complex.   It is unlikely that the star--formation
histories of  DRGs are consistent with scenarios where all their mass
formed in the distant past with subsequent passive evolution
\citep[see, \eg][]{cim02b}.   Massive galaxies at these redshifts
continue forming stars up to the epoch we are observing them.  Recent
semi--analytic prescriptions of the star--formation histories of
massive galaxies typically involve both stochastic ``burst'' modes, and
quasi--continuous ``quiescent'' modes of star formation
\citep[\eg,][]{som01,delucia05,nag05}, qualitatively consistent with
our analysis. It seems that the earliest star--formation episodes in
DRGs did occur in the distant past, and that much of their stellar
mass was assembled much earlier than the epoch at which we are
observing them.  Taking the median age of the stellar population as a
fiducial value for the onset of star formation, DRGs began forming
stars at $z_\mathrm{form} \gsim 3.5$.   Their progenitors may be the
star--forming galaxies observed at those earlier epochs as
UV--dropouts at $z\gsim 3.5$ \citep[\eg,][]{pap04a}.

\subsubsection{Stellar Masses}\label{section:mass}

\ifsubmode
\else
\begin{figure*}
\plottwo{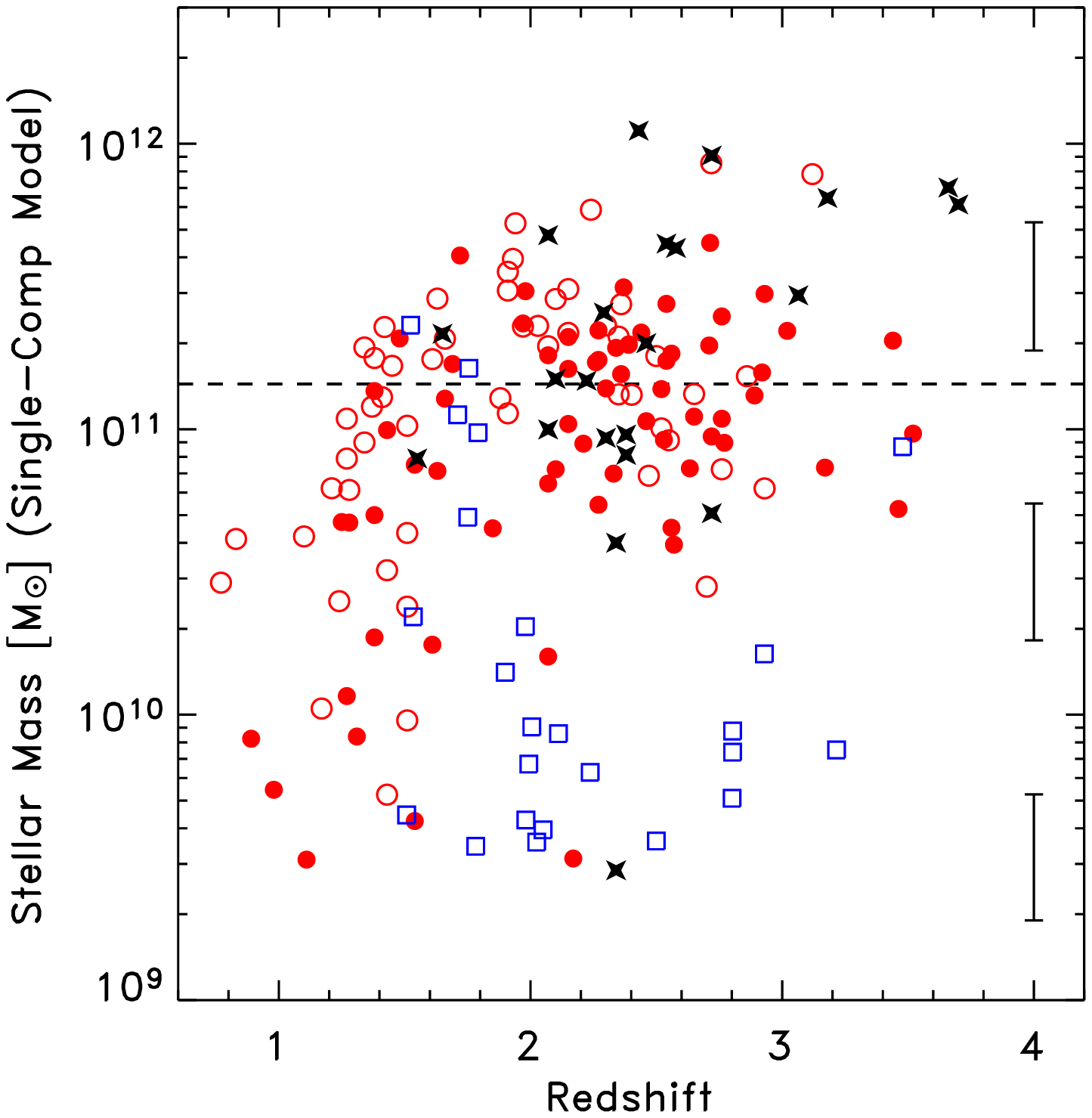}{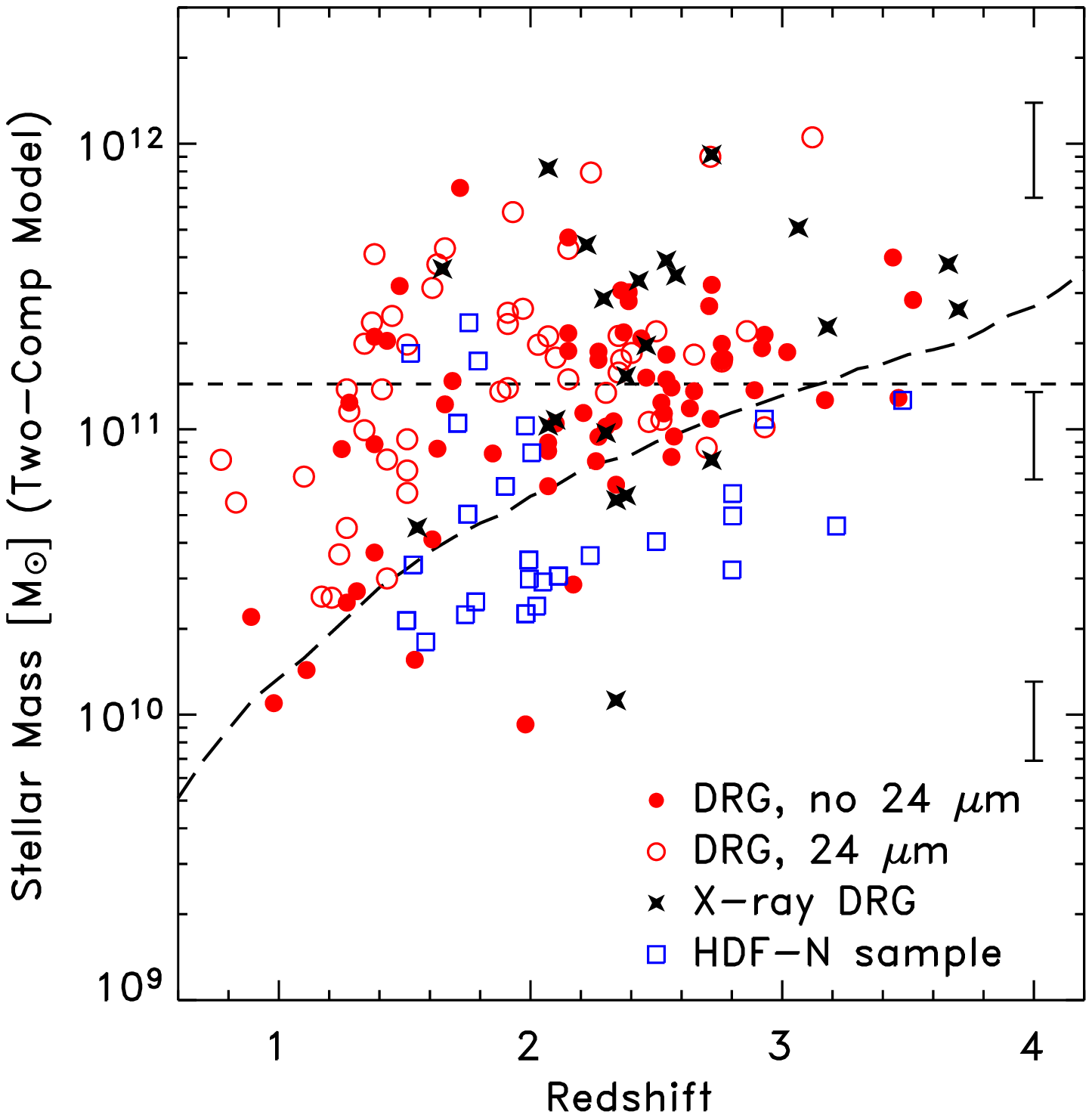}
\caption\figcapmassz
\vspace{8pt}
\end{figure*}
\fi
\ifsubmode
\begin{figure}
\epsscale{1.12}
\plottwo{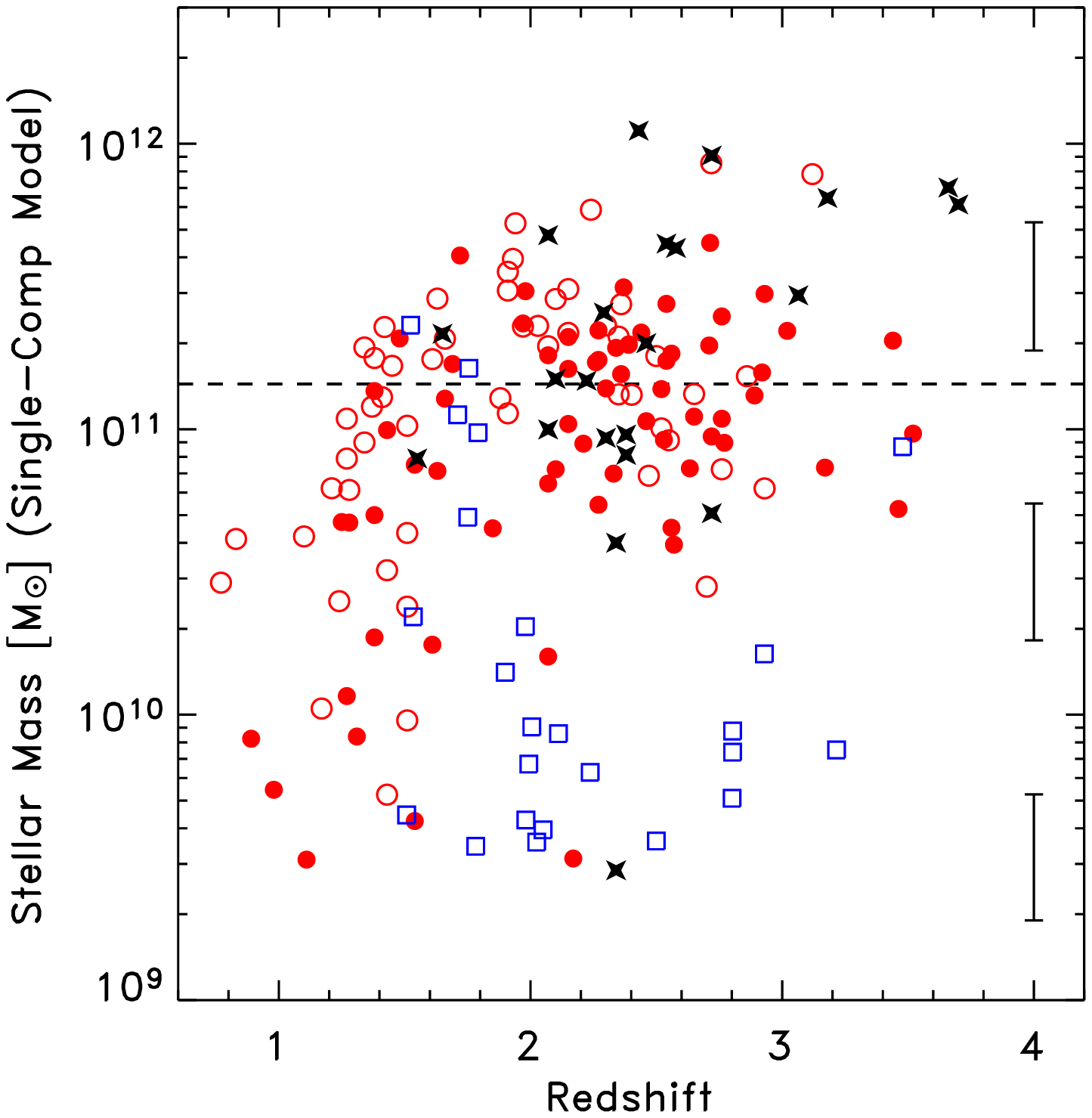}{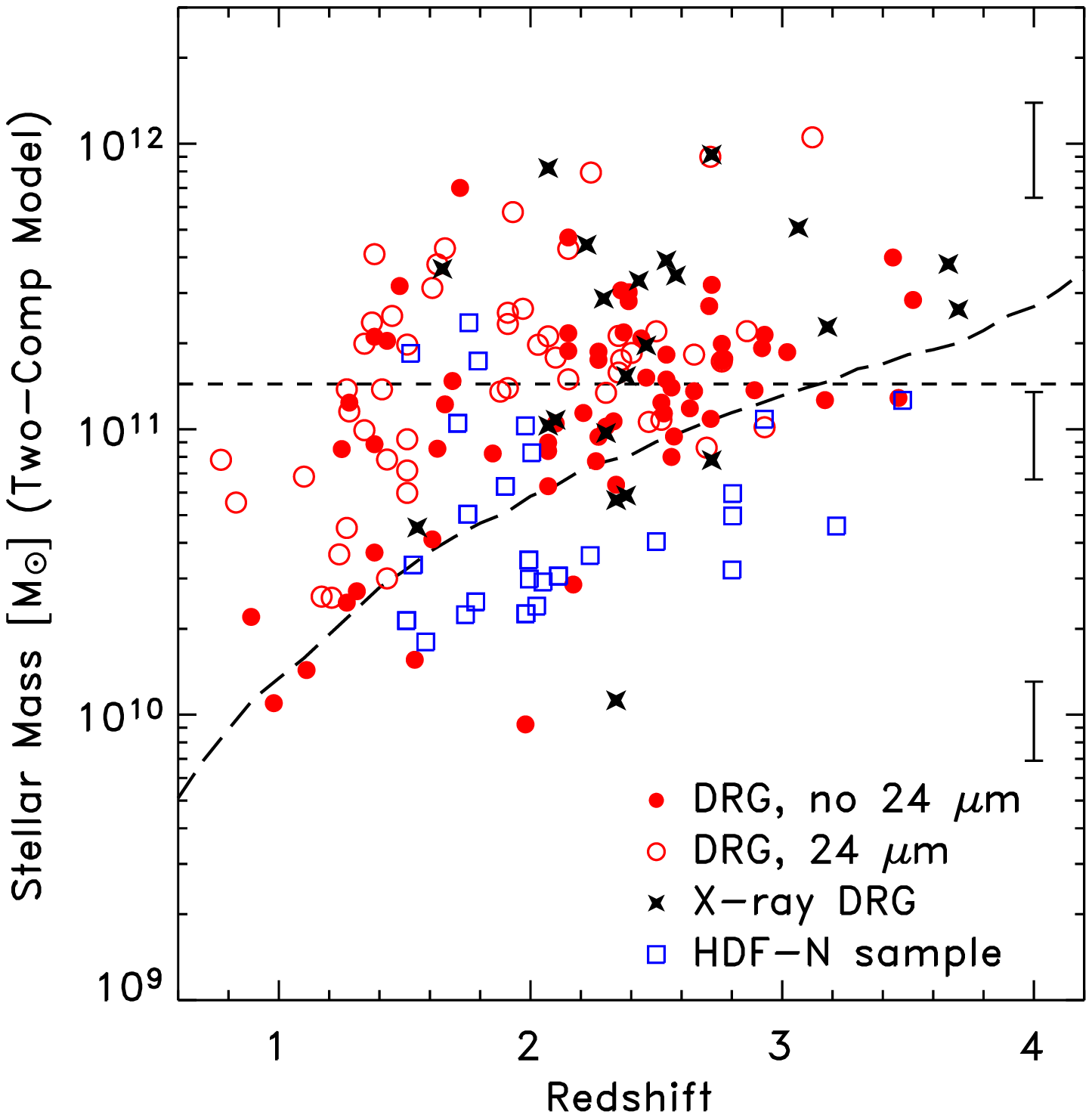}
\epsscale{1.0}
\caption\figcapmassz
\end{figure}
\fi

The SED modeling provides relatively robust estimates of the
stellar-mass--to--light ratios of these galaxies, providing good
measures of their stellar masses.  Figure~\ref{fig:massz} shows the
distribution of the stellar masses for the single-- and two--component
star--formation histories for both the DRGs and the galaxies in the
\hdfn.  Typical errors are 0.1--0.3~dex (for a given IMF and
metallicity).

The median stellar masses for the GOODS--S DRGs are 1.1 and 1.7
$\times 10^{11}$~\msol\ for the single-- and two--component
star--formation history models, respectively.  There is little
difference between the stellar masses of DRGs detected and undetected
at 24~\micron.   The interquartile range (containing the inner 50\% of
the sample) spans from 0.13 to $3.3\times 10^{11}$~\msol\ for the
single--component models, and from 0.29 to $4.6\times 10^{11}$~\msol\
for the two--component models.  The main effect of the two--component
models is to increase the masses of objects with low masses from the
one--component models.  Higher mass objects typically have larger
stellar-mass--to--light ratios, $\mcal/L$, and are less effected
because the uncertainties on $\mcal/L$ are smaller
\citep[\eg,][]{dic03}. For the 34 DRGs in the FIRES fields,
\citet{for04} found median stellar masses of $\mcal = 0.8-1.6\times
10^{11}$~\msol, with an interquartile range of $\sim 0.6-3\times
10^{11}$~\msol\ (using different assumptions for the star--formation
history). Because our models are taken over a wider range of
star--formation histories, we expect our stellar masses to be slightly
lower with respect to those derived for constant star--formation
models (F\"orster--Schreiber et al.\ 2004).   Thus the range and
average of the DRG stellar--mass distribution seem generally
consistent with those in the FIRES fields \citep[see also][]{lab05}.
The tail of lower--mass DRGs in our sample probably also arises from
the greater number of objects with lower redshifts.  Restricting our
sample to a higher minimum redshift would further increase the
lower--bound of our interquartile range.

The long--dashed line in Figure~\ref{fig:massz} shows the stellar mass
limit for a passively evolving stellar population formed in a single
burst at $z=\infty$ and with $\ks = 23.2$~mag.   Our DRG selection is
approximately complete in stellar mass for passively evolving galaxies
with $\mcal \ge 10^{11}$~\msol\ to $2 \leq z \leq 3$, because no
galaxy  can have a mass--to--light ratio higher than a maximally old
stellar population.  Bluer galaxies with $\mcal \ge 10^{11}$~\msol\
and lower mass--to--light ratios such as UV--luminous LBGs could be
excluded by the $(\jmk)_\mathrm{Vega} > 2.3$~mag selection.   However,
\citet{sha05} infer that the majority of LBGs at $z\sim 2$ with $\mcal
\ge 10^{11}$~\msol\ would satisfy $(\jmk)_\mathrm{Vega} > 2.3$~mag
\citep[see also,][]{red05}.   At $z\leq 2$, the \jmk\ color selection
may miss some passively evolving galaxies as the Balmer/4000~\AA\
break shifts to wavelengths below the blue--edge of the $J$--band
filter \citep{dad04}. We consider the DRG selection roughly complete
in stellar mass to $\mcal \geq 10^{11}$~\msol\ over the majority of
the redshift range $2 \leq z \leq 3$.

Based on the fits to the single--component models, the integrated DRG
stellar mass density for objects at $2 \leq z \leq 3$ with $\mcal \geq
10^{11}$~\msol\ is $2.4_{-0.4}^{+0.5} \times 10^{7}$~\msol\
Mpc$^{-3}$.   This increases to $2.8_{-0.5}^{+0.6} \times
10^{7}$~\msol\ Mpc$^{-3}$ using the stellar masses derived from the
two--component model fits.   In both cases the uncertainties are
estimated using a bootstrap resampling of the dataset, which
constructs random samples of DRGs with the sample size taken from the
Poisson, counting uncertainties, and stellar masses drawn from the
measured distribution (with replacement) modulated by the inferred
stellar--mass errors.   However, many of the DRGs show evidence for an
AGN, either based on X--ray detections, IR luminosity, or rest--frame
near-- and mid--IR colors (see \S~\ref{section:agn}). These objects
have relatively higher derived stellar masses
(Figure~\ref{fig:massz}), which could imply that AGN tend to reside in
the most massive galaxies at these redshifts.   Alternatively it could
be that AGN may contribute to the rest--frame UV--to--near-IR
emission, leading to unduly larger stellar mass estimates.  To bound
this effect, we recalculate the stellar mass densities excluding all
DRGs with X--ray detections, $\lir \ge 10^{13}$~\lsol, or with
AGN--like rest--frame near--IR colors.  In this case, the stellar mass
density decreases to $2.0_{-0.4}^{+0.4}$ and $2.4_{-0.4}^{+0.4} \times
10^{7}$~\msol\ Mpc$^{-3}$ for the single-- and two--component models,
respectively.   In this case, the DRGs contribute 25--70\% to the
total stellar mass density integrated over all galaxies at $2 \leq z
\leq 3$, in reasonable agreement with the conclusions of \citet{rud03}
and \citet{fon03}.

\subsubsection{Star--Formation Rates and ``Dead''
Objects}\label{section:sfrs} 

The 24~\micron--detected DRGs span SFRs from $\sim 100
- 1000$~\msol\ yr$^{-1}$, excluding those objects directly detected in
the \chandra\ data (see Figure~\ref{fig:sfrsfr}).  The mean SFR for
these sources is $\simeq 500$~\msol\ yr$^{-1}$, with a 
systematic uncertainty in the 24~\micron\ to SFR conversion of
$\approx 0.5$~dex.  This value includes objects with $\lir
\geq 10^{13}$~\msol\ or rest--frame near--IR colors indicative of AGN
(see \S~\ref{section:agn}).  Excluding these objects, the average SFR
of the 24~\micron--detected DRGs drops to $\simeq 400$~\msol\
yr$^{-1}$.   The mean SFR for the complete DRG sample
(including those not detected with MIPS at 24~\micron) is
lower still.   Taking the conservative limit that the
24~\micron--undetected DRGs have no ongoing star formation, the mean
SFR is $\simeq 220$~\msol\ yr$^{-1}$ excluding the X--ray sources.
This does not change if we also exclude those sources with infrared
luminosities or colors indicative of AGN.

The mean SFR for the DRGs reported here is somewhat larger, but
comparable to recent measurements of the ``stacked'' X--ray and
sub--mm emission from DRG samples.  \citet{rub04} and
\citet{red05} find the average SFR is $\sim 100-300$~\msol\ yr$^{-1}$
based on statistical X--ray detections for DRGs with $\ks \lsim 23$~AB
mag.   The inferred mean SFR for the various DRG populations is
strongly dependent on the limiting $\ks$--band of the survey, and in
the case of the X--ray derived measurements, on indications of AGN
activity within the sample (see the discussion in Reddy \etal\ 2005).
However, the X--ray SFR calibration has a systematic uncertainty on the
order of a factor of five \citep{ran03,per04}, owing to assumptions on
the formation timescales of X--ray binaries in starbursts.  A higher
SFR--to--X-ray luminosity calibration may be appropriate for galaxies
in the more intense starbursts such as those for the DRGs
\citep{per04,ten05}, so some scatter is expected in the conversion is
expected. \citet{knu05} report an average SFR of 130~\msol\ yr$^{-1}$,
based on a stacked sub--mm 850~\micron\ flux density of DRGs with $\ks
\leq 24.4$~AB mag, but is sensitive to the assumed average dust
temperature \citep[\eg,][]{cha05}.  Their sample extends to DRGs
roughly a magnitude deeper in the \ks--band, and \citet{red05}
demonstrate that the mean SFR of all types of star--forming galaxies
at $z\sim 2$ (including DRGs) decreases with decreasing $\ks$--band
flux density (see also Daddi et al.\ 2005b).   Given the systematic
uncertainties in the SFR calibrations, and the varying limiting
magnitudes of the different DRG samples, we conclude that the mean SFR
we derive using the MIPS 24~\micron\ data for the GOODS--S DRGs is in
broad agreement with these other values.  The ``typical'' DRG is
forming stars at rates in excess of $\gsim 200$~\msol\ yr$^{-1}$.

We can set an upper limit on the lifetime of a starburst in the DRGs
at the gas--consumption timescale, defined to be the ratio of the gas
mass to the SFR.  For a massive galaxy with a molecular gas mass of
$\mcal(H_2) \lsim 10^{11}$~\msol\
\citep[comparable to the gas reservoirs of the massive $z\gsim 3$
radio galaxies,][]{deb03,deb05,gre04}, the gas--consumption timescale
is $\lsim 10^8-10^9$~yr for SFRs of  $\sim 10^2 - 10^3$~\msol\
yr$^{-1}$ (although these timescales would be shorter for lower gas
masses, such as those in local ULIRGs, see Downes \& Solomon 1998).
This timescale is generally shorter than or comparable to the median
ages we derive for the DRGs (consistent also with findings in
F\"orster--Schreiber et al.\ 2004).   Recent theoretical work suggests that
high--redshift, massive galaxies quasi-continuously form stars at high
rates for periods of 1--2 Gyr, with shorter periods ($\lsim 100$~Myr)
of boosted star formation (Nagamine et al. 2005; Finlator et al.\
2005; De Lucia et al.\ 2005).  
If the DRGs sustain these the high SFRs
for more than $\sim 10^8-10^9$~yr, then they will exhaust their gas
supply unless fresh cold gas is repeatedly or continuously accreted.
\citet{ker05} predict that cold--gas accretion may dominate in massive
galaxies at high redshifts, which may advocate such a scenario exists
for the massive DRGs. 

Assuming a lifetime of $\sim 10^8-10^9$ yr, a galaxy
with a SFR of $10^2-10^3$~\msol\ yr$^{-1}$ would assemble
$10^{10-12}$~\msol\ in stars.  If LBGs are forming stars rapidly
and/or experiencing recurrent bursts at redshifts higher than that of
the DRGs, $z\gsim 3.5$, then eventually they will assemble a
sufficient population of late--type stars to dominate the optical to
near--IR rest--frame light.   It thus takes of order 1~Gyr to produce
a DRG (unless high--redshift starbursts have an IMF weighted towards
higher--mass stars, in which case even more time is required; Ferguson
\etal\ 2002).   Under these star--formation histories, the first
progenitors of DRGs at z $\sim$ 2 to 3 should appear as starburst
galaxies at z $\gsim$ 3 to 5.   Massive star--forming galaxies appear
to exist by $z\simeq 5$ from UV--dropout surveys
\citep{eyl05,yan05}, and such objects could be forming the stellar
populations that evolve to become the old components at the redshifts
of the DRGs.

Based on the model fits, we find several candidates among the DRGs for
passively evolving galaxies at high redshifts (see also
\S~\ref{section:jmk}).  There are 15 DRGs with best--fit models that
favor older ages ($t \geq 1$~Gyr) that have undergone several
$e$--folding times ($t/\tau \geq 3$) with low dust extinction ($E[B-V]
\leq 0.1$) and no X--ray or 24~\micron\ emission.  Massive, ``dead''
DRGs are either galaxies that are between starbursts but already with
a massive old stellar population, or galaxies that have completed the
intense phase of their assembly.   These galaxies span $1.9 \leq z
\leq 3.0$, and have $\acsi - \ks > 2.6$ and $\ks - \mtwo < 1.2$ (see
also Figure~\ref{fig:jmkcc} and discussion in \S~\ref{section:jmk}).
These objects contribute $\simeq 4.9\pm 1.4~\times 10^{6}$~\msol\
Mpc$^{-3}$ to the global stellar mass density at these redshifts.
Such massive, ``dead'' objects make up only a fairly low fraction of
the stellar mass density  ($\lsim 10$\%) integrated over all galaxies
at $z\sim 2-3$, and therefore are relatively rare at these redshifts.
In contrast, early--type, massive ($\geq 10^{11}$~\msol) galaxies
account for $\sim 30$\% of the total local stellar--mass density for
the  stellar--mass function of Bell et al.\ (2003, see also Baldry
\etal\ 2004).   Thus, the fraction of the global stellar--mass density
in such galaxies increases threefold from $z\gsim 2$ to 0. 

\subsection{The Relation between Star Formation and Stellar Mass in
Galaxies at $z\sim 1.5-3.0$}

\ifsubmode
\else
\begin{figure}
\plotone{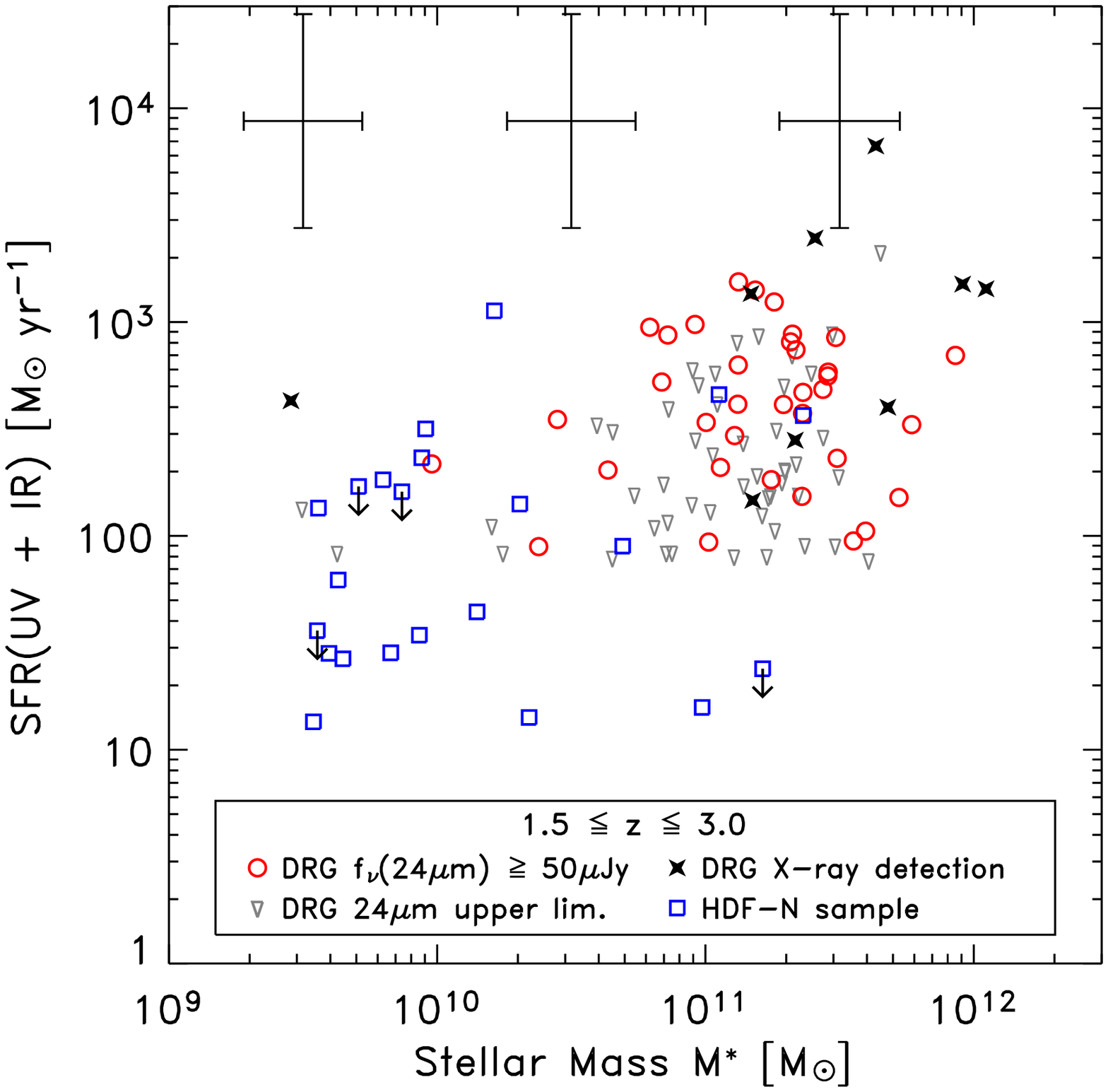}
\caption\figcapsfrmass
\end{figure}
\fi

Figure~\ref{fig:sfrmass} shows the SFRs  as a function of the stellar
mass for the DRGs and the \hdfn\ sample.   The figure
shows the stellar masses derived from the models with
single--component star--formation histories, although using the
two--component fits does not strongly change any of the conclusions.
DRGs not detected at 24~\micron\ by MIPS are shown as upper limits
assuming that their 24~\micron\ flux density is less than the 50\%
completeness limit, $f_\nu(24\micron) \leq 60$~$\mu$Jy.  We also
restrict the DRG and \hdf\ samples to those objects with $1.5 \leq z
\leq 3.0$, where we are approximately complete for massive galaxies.
Furthermore, restricting the sample to $z\leq 3$ removes few galaxies
from the samples, and facilitates the contrast with the
lower--redshift samples (see \S~\ref{section:spec_sfr}).

\ifsubmode
\begin{figure}
\plotone{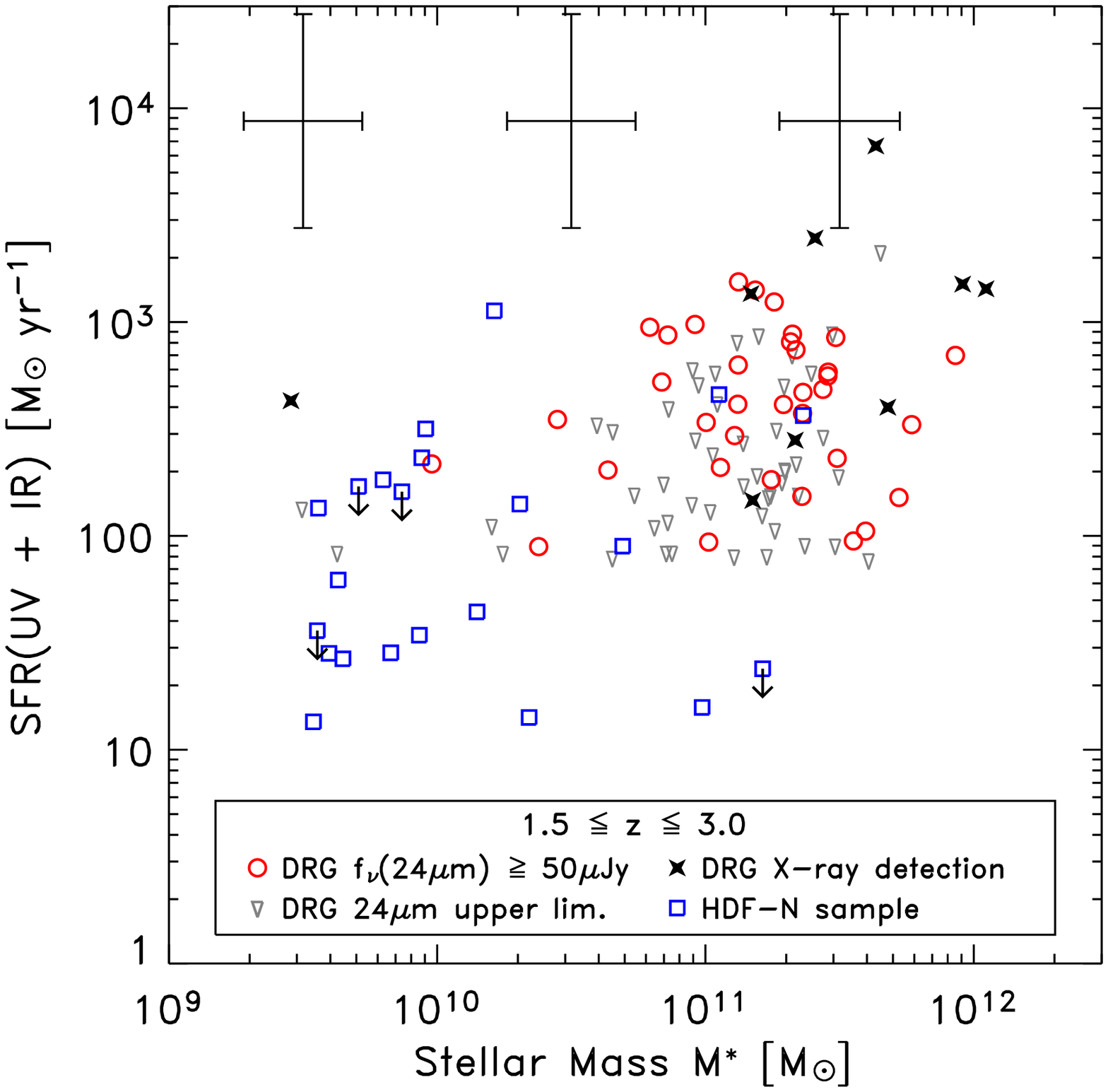}
\caption\figcapsfrmass
\end{figure}
\fi

The DRGs have higher stellar masses and SFRs relative to the \hdfn\
24~\micron--selected sample, owing primarily to the fainter flux limit
of the 24~\micron\ data in the \hdfn.  There is a weak correlation
between SFR and stellar mass, although this trend may be softened if
the SFRs of DRGs with upper limits are in fact lower.   Objects with
stellar masses greater than $10^{11}$~\msol\ span 1.5~dex in SFR,
excluding the X-ray--detected DRGs with inferred $\lir \gsim
10^{13}$~\lsol.   The upper envelope of SFRs decreases for lower
mass galaxies, and such objects should be detected in our data if
present.   Therefore, at these redshifts galaxies with the highest
stellar masses also appear to be experiencing the highest SFRs.

\ifsubmode
\else
\vspace{8pt}
\fi

\subsection{Growth of Supermassive Black Holes in Massive,
High--Redshift Galaxies}\label{section:agn}

Many of the DRGs with the highest stellar masses and SFRs are detected
in the deep X--ray data.  Twelve of the 13 X--ray detected DRGs with
$\mcal^\ast \ge 10^{11}$~\msol\ have X--ray to optical flux ratios,
$\log f_\mathrm{X}/f_\mathrm{R} \geq -1.0$ (the remaining one has
$\log f_\mathrm{X}/f_\mathrm{R} \sim -1.4$), and these imply the
presence of an AGN with $L_\mathrm{X} \gsim 10^{42}$~erg s$^{-1}$ for
DRGs at $z\gsim 1.5$ \citep[\eg,][]{hor01,ale03}.  In addition,
\citet{wor05} conclude that as much as 40\% of the hard X--ray
background stems from heavily obscured AGN at $z \gsim 0.5$.   Recent
\spitzer\ IR observations are finding many AGN candidates whose
X--ray, UV,  and optical emission is heavily obscured by gas and dust,
and missed in deep X--ray surveys \citep[\eg,][]{alo05,don05}.
\citet{alo05} find that as many as 50\% of MIPS 24~\micron\ sources
with red IRAC colors are not detected in the deep X--ray data.
This suggests that AGN may be hidden behind sufficiently
Compton--thick material that none of the direct X--ray emission
escapes \citep{brandt05}.  Models of the X--ray background
support this possibility \citep{gilli04}.

At $1.5 \lsim z \lsim 3.5$, galaxies dominated by the light from
stellar photospheres should have relatively blue $\mtwo-\mfour$ colors
(see \S~\ref{section:jmk}).  Alternatively galaxies dominated by AGN
emission in the rest--frame near--IR should have redder IRAC colors
\citep[see, \eg,][]{rie78,neu79,lac04,alo05,stern05}.   \citet{lac04}
found that Seyfert 1 galaxies and obscured AGN have red $\mtwo -
\mfour$ colors, similar to the red rest--frame $J-K_s$ colors of
Parkes Quasars in 2MASS at $z\sim 0-0.5$ \citep{fra04}.   Similarly,
of the non--X-ray detected DRGs with IRAC counterparts, roughly 25\%
(26/106) satisfy the IR color selection for AGN of
\citet{stern05}. Upon inspection of their individual SEDs, most are 
plausible AGN candidates with red IRAC colors.  Four of these DRGs are
star--forming objects at $z\gsim 2.9$ whose IRAC colors mimic those of
AGN candidates at lower redshift.  These galaxies are optically
fainter and have higher redshifts than typical galaxies in the Stern
\etal\ sample, and are an additional source of contamination in IR
selection of AGN.  An additional 10 of these DRGs have $z\simeq
1.0-1.4$ and their SEDs are consistent with heavily extincted
starbursts whose \mfour\ magnitudes may be augmented by the 3~\micron\
PAH emission feature.\footnote{Stern \etal\ (2005) find that 17\% of
objects satisfying their IRAC color--color selection are
spectroscopically classified as galaxies, roughly consistent with the
galaxy contamination here, modulo differences in survey limiting
magnitudes.}   Excluding these 14 objects, approximately 10\% (10/125)
of the X-ray--undetected DRGs  may have dust--enshrouded AGN that
contribute some fraction of the rest--frame near-- and mid--IR light.
Combined with the 15\% of X--ray--detected DRGs, perhaps a quarter of
the DRG population host AGN, which implies they are actively growing
their SMBHs.

The X-ray--detected fraction of DRGs apparently rises with increasing
IR luminosity ($>$50\% at $\lir \gsim 10^{13}$~\lsol), a trend
also seen in the analysis of X--ray emission from high--redshift
sub--mm galaxies \citep{alex05}. The GOODS--S field lacks deep
sub--mm observations, which would allow us to explore any direct
overlap between these coeval populations. Locally, the most luminous
ULIRGs and HyLIRGs frequently host AGN sufficiently powerful to affect
their global properties \citep{vei95}.   If the IR--luminous DRGs are
similar, then the high incidence of AGN is expected, especially for
those with inferred $\lir > 10^{13}$~\lsol.   This implies that the
massive galaxies at these epochs may be both assembling their stellar
populations and growing SMBHs (see also L.~Moustakas et al.\ in
preparation). Locally, the space density of ULIRGs is $\approx
10^{-7}\,h_{70}$~Mpc$^{-3}$ \citep{san03}, while the DRGs with
ULIRG--like luminosities have a space density of $5.8 \pm 0.9\times
10^{-5}\,h_{70}^{-3}$~Mpc$^{-3}$ --- an increase by a factor of 600
\citep[and similar to the conclusion for $BzK$--selected
objects,][]{dad05b}.   Thus, such objects are a much more common
phenomenon at high redshift.

\ifsubmode
\else
\vspace{8pt}
\fi

\subsection{Evolution of the Star Formation Rate as Function of
Stellar Mass}\label{section:spec_sfr}

\ifsubmode
\else
\begin{figure*}
\epsscale{0.45}
\plotone{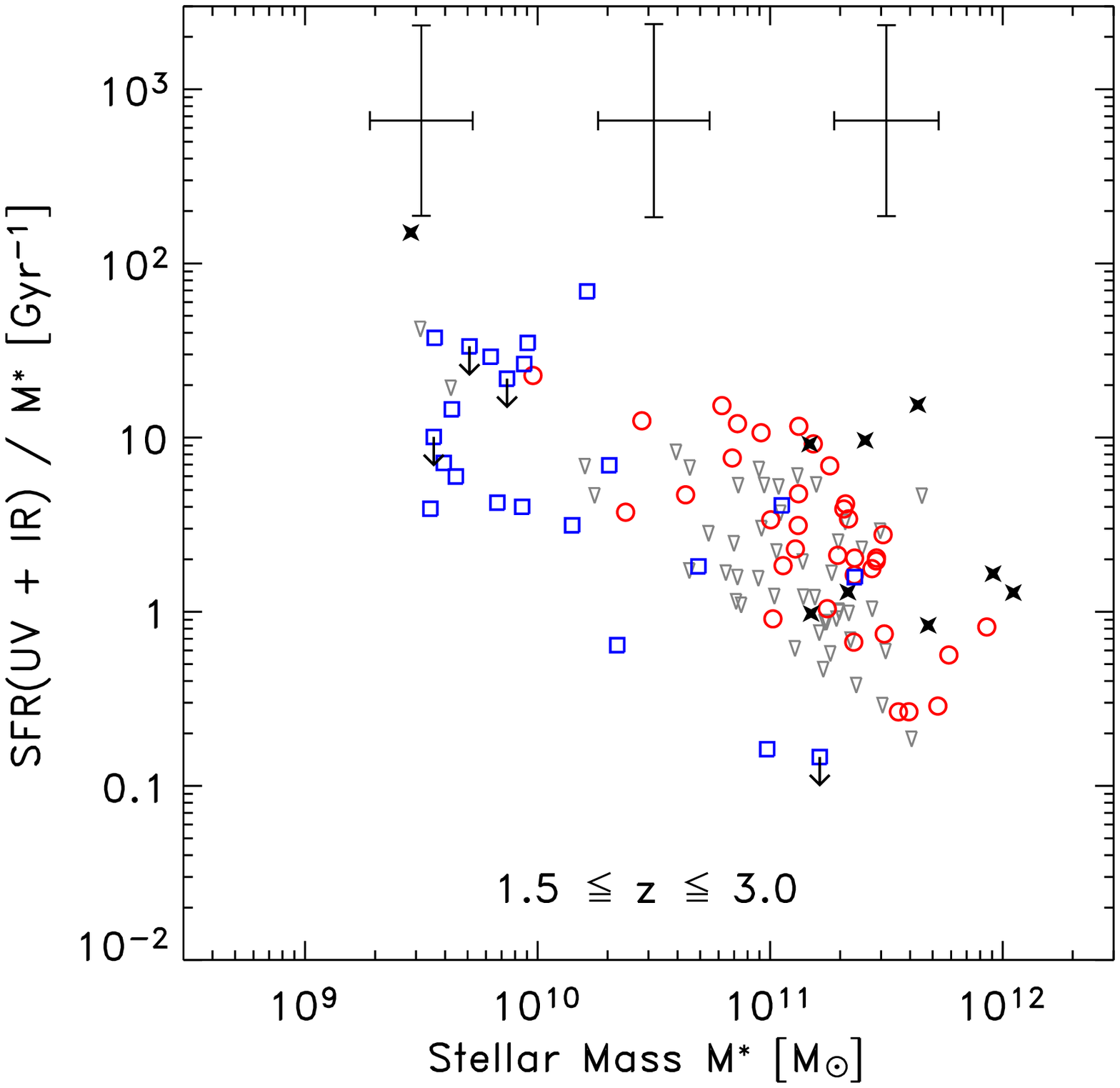}

\epsscale{1.0}
\plottwo{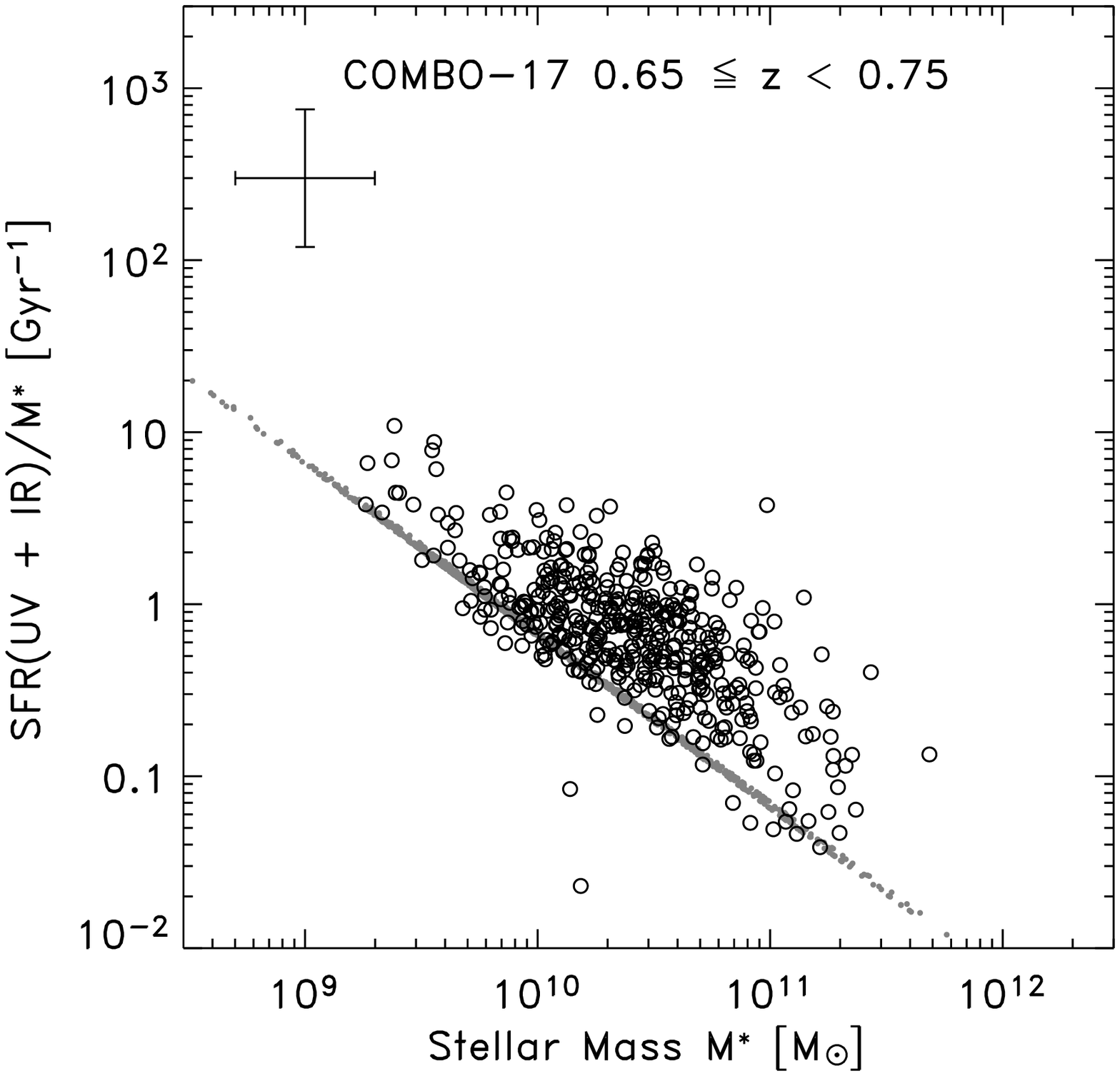}{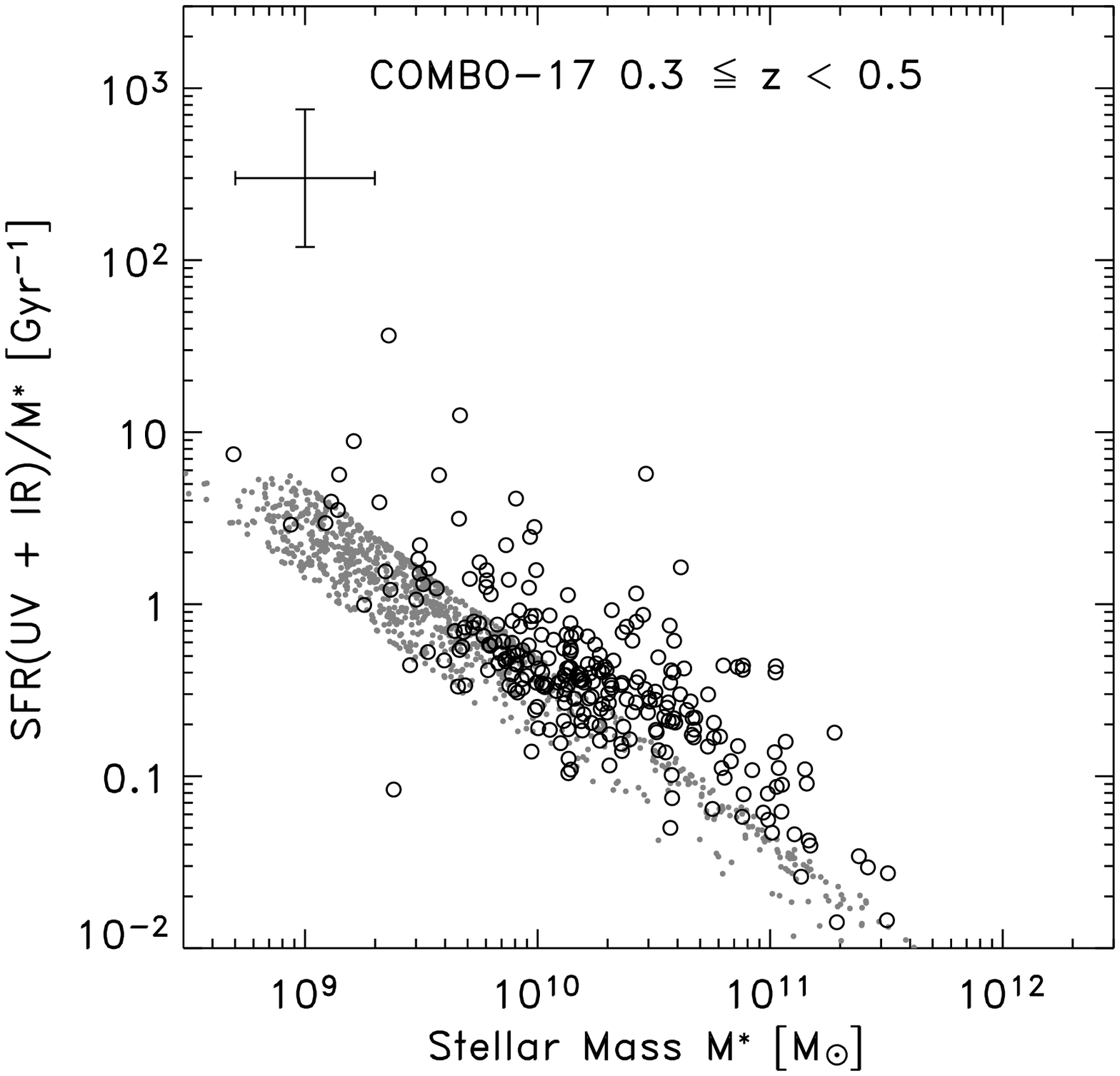}
\epsscale{1.0}
\caption\figcapspecsfrmass
\end{figure*}
\fi

The DRGs have higher specific SFRs (SFR per unit stellar mass,
$\Psi/\mcal$) than galaxies with comparable mass at lower redshifts
($z\lsim 1$). Figure~\ref{fig:specsfrmass} shows the specific SFRs for
DRGs and \hdf\ galaxies at $1.5 \leq z \leq 3.0$, comparing them to
galaxies at lower redshift in the COMBO--17 survey (see
\S~\ref{section:combo17}).   The SFRs for the COMBO--17 galaxies are
calculated using the MIPS 24~\micron\ imaging and rest--frame UV
emission in an analogous manner as for the DRGs.   Masses for
COMBO--17 galaxies were estimated from their rest--frame $M(V)$ and
$U-V$ colors, and have a typical uncertainty of 0.3~dex.   This method
will tend to \textit{overestimate} the masses of galaxies involved in
starbursts, but it is fairly robust for quiescent galaxies such as
those that dominate our conclusions here (see Bell et al.\ 2004, 2005a).

\ifsubmode
\begin{figure}
\epsscale{0.495}
\plotone{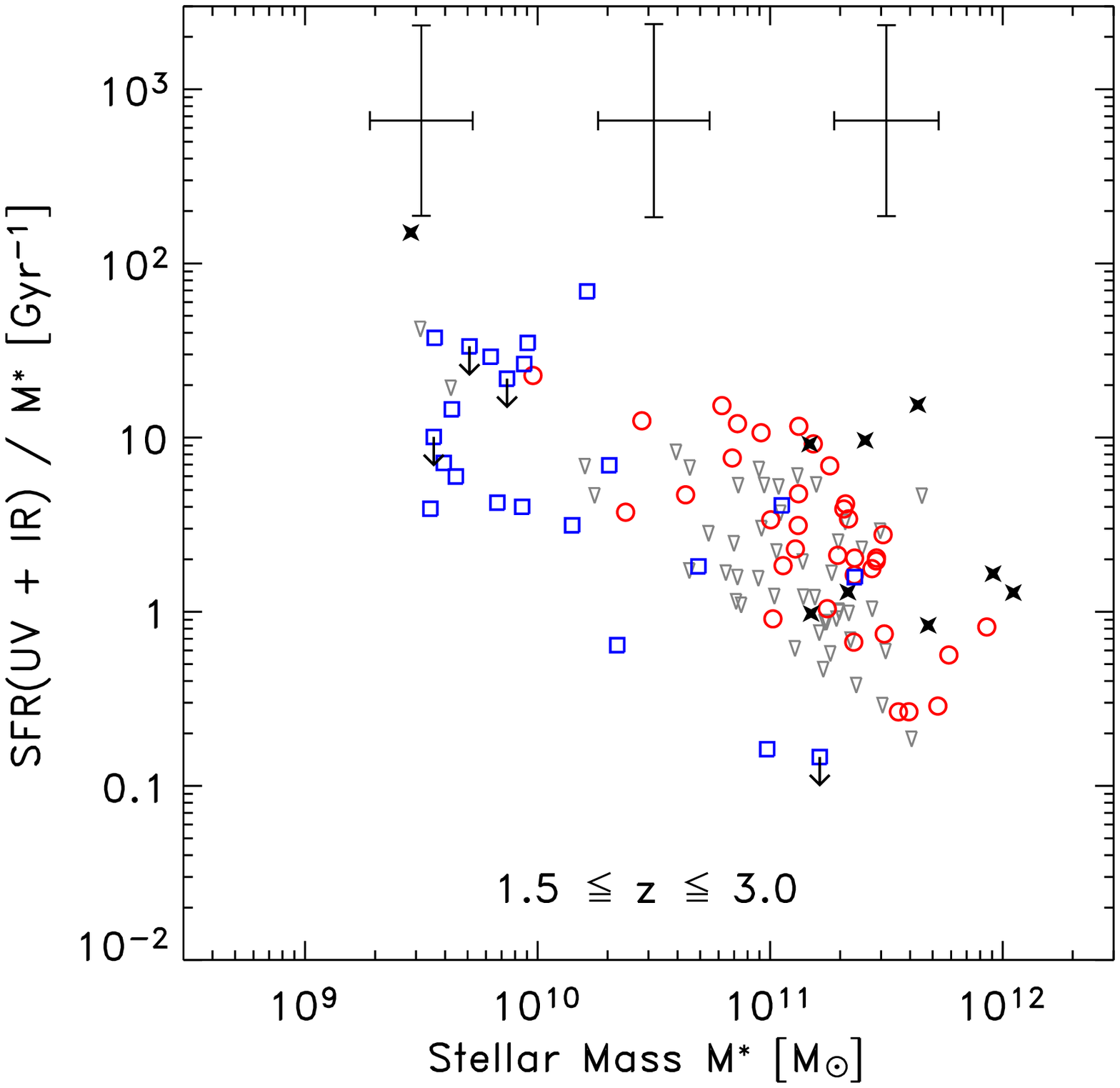}
\epsscale{1.1}
\plottwo{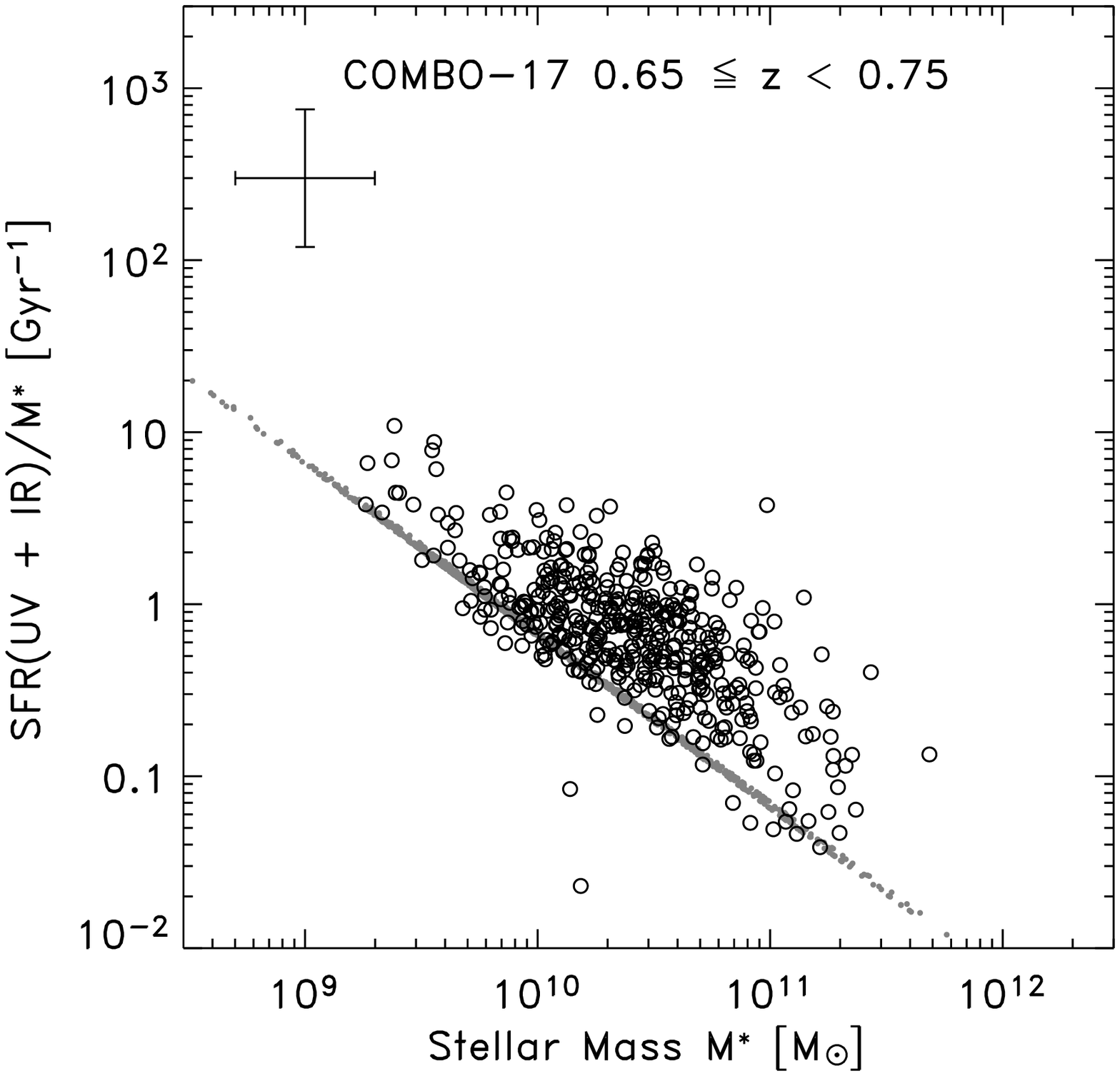}{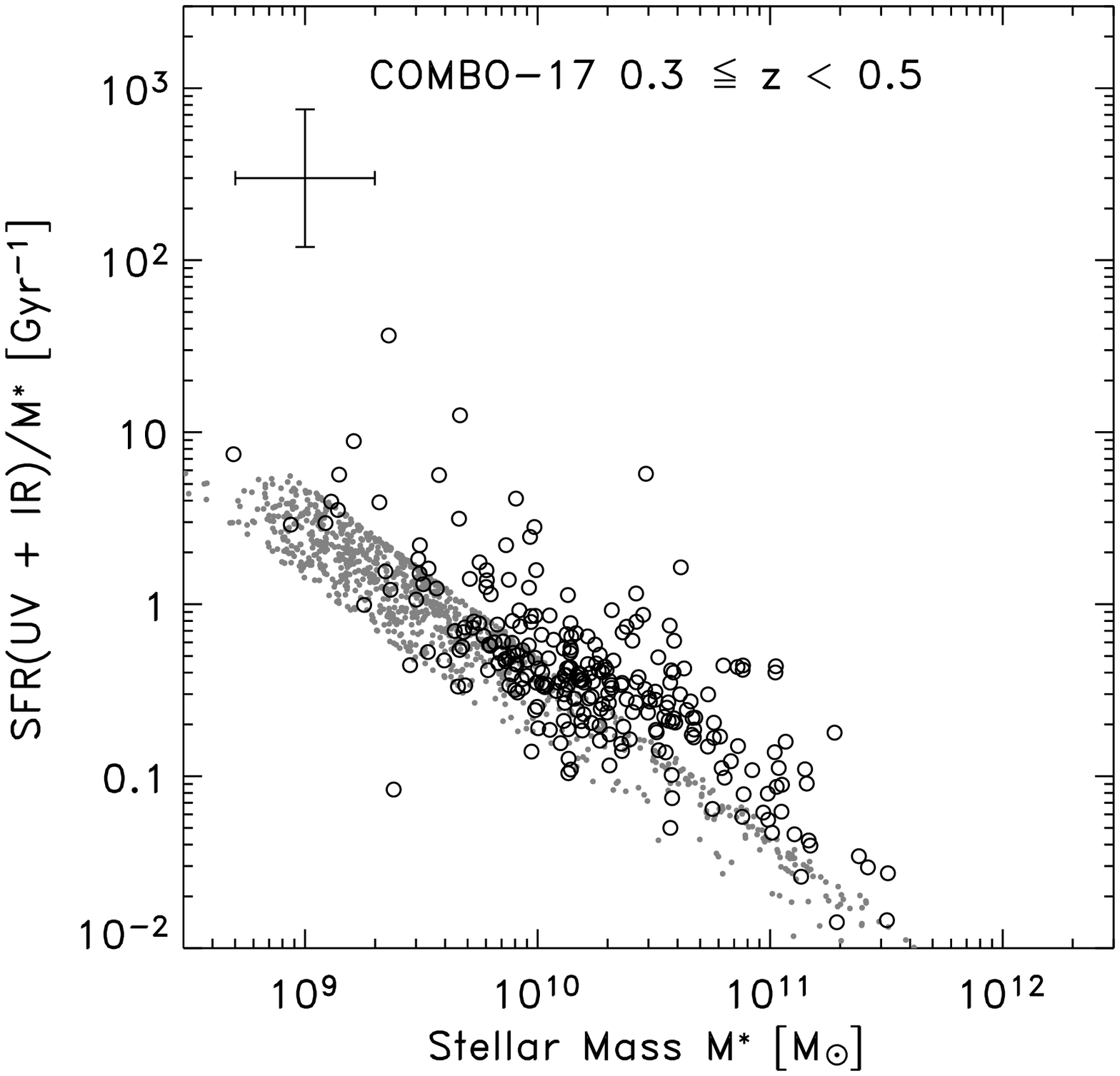}
\epsscale{1.0}
\caption\figcapspecsfrmass
\end{figure}
\fi

We are biased against galaxies with low stellar masses and low
specific SFRs, which causes the lower ``envelope'' in all the panels
of Figure~\ref{fig:specsfrmass}.  This selection effect can produce
the apparent anticorrelation between specific SFR and stellar mass
(most apparent in the COMBO--17 galaxy plots).  The galaxies at
$z\lsim 1$ show an upper envelope in the sense that there is a lack of
galaxies with high specific SFRs and high stellar masses.   In
contrast, the DRGs at $1.5 \leq z \leq 3$ show a nearly unchanging
range of specific SFRs at all stellar masses, although there is a hint
that the DRGs at the high--mass end ($\mcal \geq 10^{11}$~\msol) show
an upper envelope on their specific SFRs.  Even then the massive galaxies
have systematically higher specific SFRs than the $z\lsim 1$ galaxies
in COMBO--17.   Quantitatively, the DRGs with $\mcal > 10^{11}$~\msol\
and $1.5 \leq z \leq 3$ have specific SFRs of $\Psi/\mcal \sim 0.2-10$
Gyr$^{-1}$, with a mean value of $\sim 2.4$~Gyr$^{-1}$   (excluding
those with X--ray detections).  By $z\sim 0.7$ galaxies with $\mcal
\geq 10^{11}$~\msol\ have $\Psi/\mcal \sim 0.1-1$ Gyr$^{-1}$ and at
$z\sim 0.4$ galaxies with $\mcal \geq 10^{11}$~\msol\ have $\Psi/\mcal
\lsim 0.5$ Gyr$^{-1}$ --- an order of magnitude lower than for the
massive DRGs. 

This downward evolution in the specific SFRs of massive galaxies has
been referred to as ``downsizing'' (see \S~\ref{section:intro}).  The
hypothesis is that massive galaxies host most of the SFR density at
high redshifts, and that galaxy formation shifts to less--massive
systems at lower redshifts.  It is unclear if downsizing is a proper
description as nearly \textit{all} galaxies at $1.5 \lsim z\lsim 3.0$
have higher specific SFRs at higher redshift, not just the most
massive.   ``Downsizing'' is not equivalent to a global change in
$\Psi/\mcal$ with redshift. However, our results indicate that
star--formation in massive galaxies is reduced for $z\lsim 1$ as
galaxies with lower stellar masses have higher specific SFRs.  It
appears that massive galaxies are largely done assembling their
stellar mass in high specific--SFR events by $z\sim 1.5$ (although
mergers accompanied by low specific SFRs, i.e., ``dry'' mergers, could
still take place at lower redshift, see, e.g., Bell \etal\ 2005b;
Faber \etal\ 2005; van Dokkum 2005).

\ifsubmode
\else
\begin{figure*}
\epsscale{0.9}
\plotone{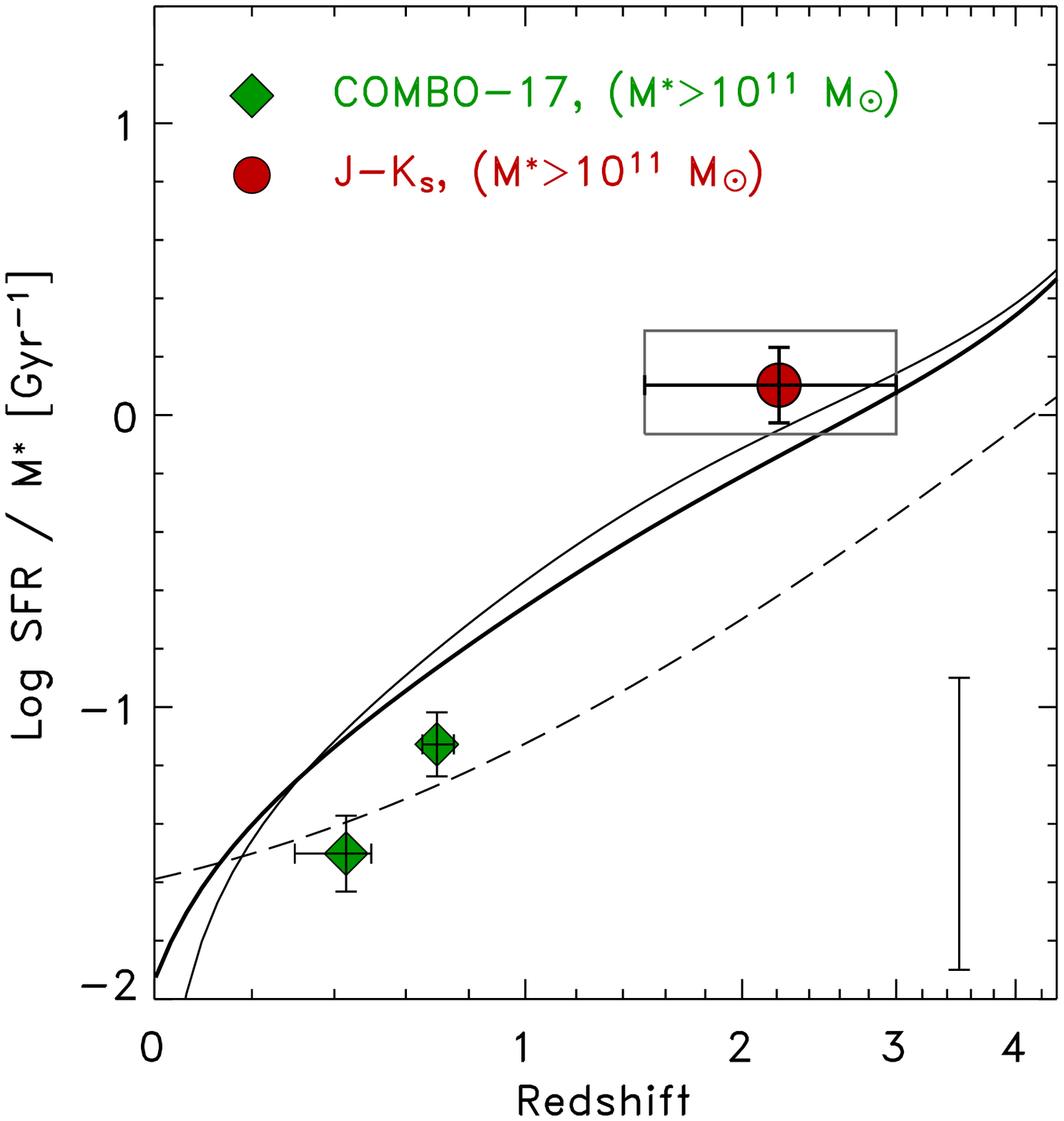}
\caption\figcapspecsfrevol
\epsscale{1.0}
\end{figure*}
\fi

We define the integrated specific SFR, $\Upsilon$, as the ratio of the
sum of the SFRs, $\Psi_i$, to the sum of their stellar masses,
$\mcal_i$, over all $i$ galaxies,
\begin{equation}\label{eqn:int_sfr}
\Upsilon  = \frac{\sum_i \Psi_i}{\sum_i
\mcal_i}.
\end{equation}
This is essentially just the ratio of the SFR density to the stellar
mass density for a volume--limited sample of galaxies.
Figure~\ref{fig:specsfrevol} shows the integrated specific SFRs for
DRGs at $1.5 \leq z \leq 3.0$ and COMBO--17 galaxies at $z\sim 0.4$
and 0.7 with $\mcal \geq 10^{11}$~\msol. The data point for the DRGs
includes all DRGs with $\mcal \geq 10^{11}$~\msol, and assumes that
the DRGs without 24~\micron\ detections have a no star--formation
($\Psi = 0$~\msol\ yr$^{-1}$).  The error box illustrates how changes
in our assumptions affect the result.  The bottom bound of the box
shows the value if we exclude those objects with X--ray detections, or
with IR luminosities or colors indicative of AGN (see
\S~\ref{section:agn}). The upper bound of the box shows the value for
DRGs in this redshift and mass range, and if we calculate SFRs for the
DRGs without 24~\micron\ detections assuming they have
$f_\nu(24\micron) = 60$~\ujy.   Note that for this calculation we have
used the stellar masses from the single--component fits.  Using those
from the two--component fits increases the DRG point and error box by
a small amount, $<$0.05~dex.

The integrated specific SFR in massive galaxies declines by more than
an order of magnitude from $z\sim 1.5-3$ to redshifts $z\sim 0.7$ and
0.4.   Using the rest--frame UV luminosities and [\ion{O}{2}]
emission--line fluxes of galaxies in the Gemini Deep--Deep Survey,
\citet{jun05} found that the specific SFRs of galaxies with $\mcal \ge
10^{10.8}$~\msol\ decline by roughly a factor of six from $z\sim 2$ to
1.  Our results suggest that this value may be underestimated when the
IR emission from these galaxies is included in the SFR determination.

\ifsubmode
\begin{figure}
\epsscale{0.75}
\plotone{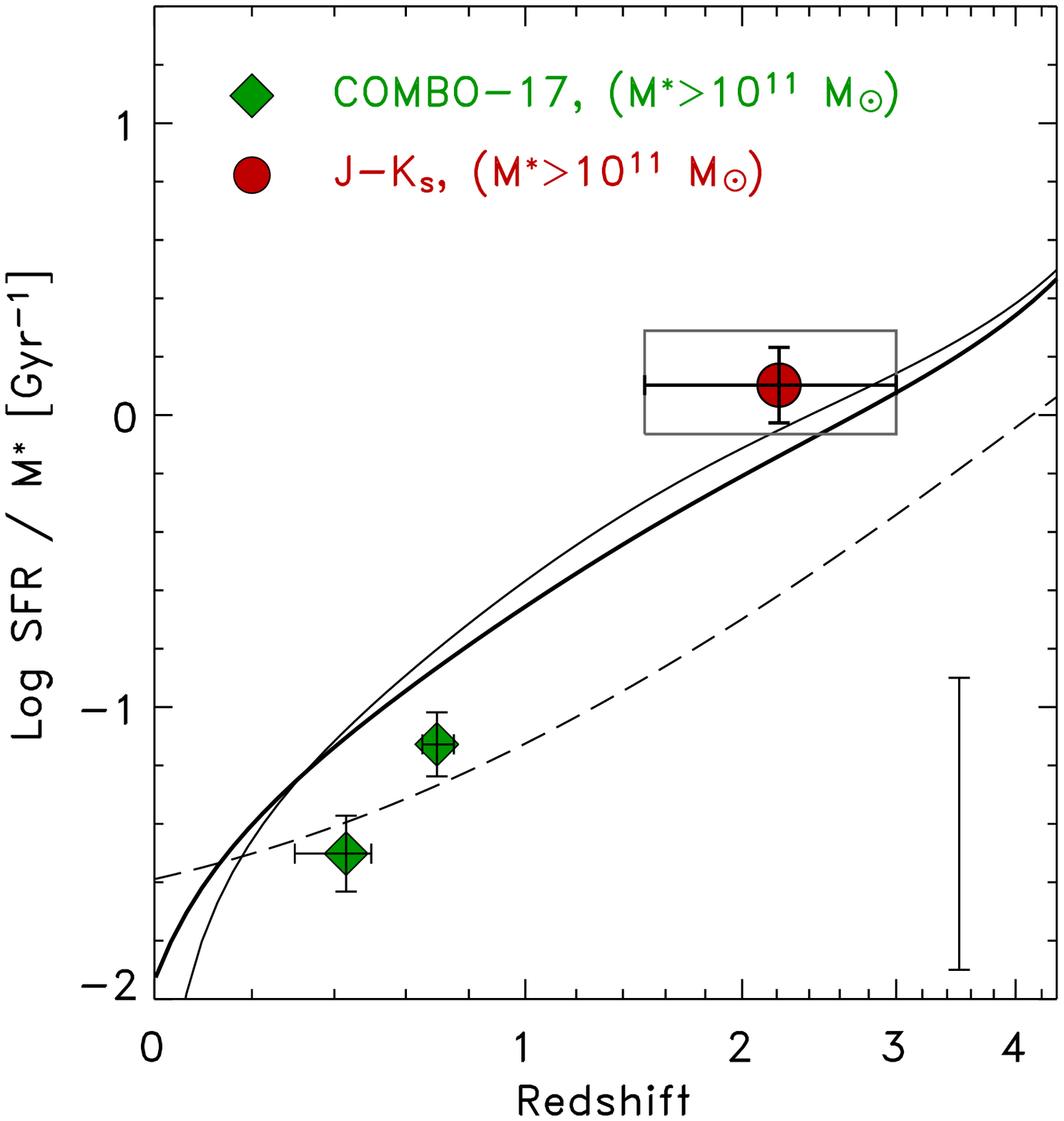}
\epsscale{1.0}
\caption\figcapspecsfrevol
\end{figure}
\fi

Figure~\ref{fig:specsfrevol} shows the specific SFR integrated over
all galaxies by taking the ratio of the cosmic SFR density to the
integrated SFR density from \citet{col01}, i.e., $\Upsilon =
\dot{\rho}_\ast / \int \dot{\rho}_\ast\, dt$ and also from the
predictions from \citet{her03}.   The integrated specific SFR over all
galaxies declines steadily with decreasing redshift, because the SFR
density peaks between $z\sim 1-3$ \citep{hop04}, corresponding to
when  the stellar mass density appears to have grown the most rapidly
\citep{dic03}.  That is, there is a decrease in the global specific
SFR.

The evolution in the integrated specific SFR in massive galaxies is
accelerated relative to the integrated global value for all galaxies.
Galaxies with $\mcal \ge 10^{11}$~\msol\ at $z\sim 1.5-3$ were forming
stars at or slightly above the rate integrated over all galaxies.
These massive galaxies contribute between 10--30\% to the
global SFR density at $z\sim 1.5-3$ \citep[comparing to][]{hop04}.
In contrast, by $z\lsim 1$, galaxies with $\mcal \ge 10^{11}$~\msol\
have an integrated specific SFR much lower than value integrated over
all galaxies.   Thus, our results indicate that by  $z\lsim 1$ massive
galaxies have formed most of their stellar mass, and lower--mass
galaxies dominate the cosmic SFR density.  If we tried to track
individual galaxies we would need to account for their growth in mass
from $z\sim 3$ to $z\lsim 1$, and as a result this trend would be even
stronger.
 
Our conclusion would be undermined if there were a substantial
population of passively evolving, massive galaxies at high redshifts
that we have missed in our selection.  The GOODS--S DRG sample is
approximately mass--limited for passively evolving galaxies to
$10^{11}$~\msol\ at $2 \leq z \leq 3$, although we are presumably
missing some at $z\leq 2$ (see Daddi \etal\ 2004, and
\S~\ref{section:mass}).  However, if we restrict the calculation to
those DRGs with $2 \leq z \leq 3$, the integrated specific SFR
\textit{increases} by roughly 0.1~dex.  Including those sources at
$z<2$ provides a more conservative measure of the evolution in the
specific SFRs of massive galaxies, so we keep these objects here.
The DRG selection is potentially biased against other star--forming
galaxies, most notably UV--luminous LBGs.  However, LBGs with these
stellar masses have  high inferred SFRs, comparable even to those of
the DRGs (see, e.g., Shapley et al.\ 2003; and Figure~\ref{fig:uvir}).
Moreover, Shapley et al.\ find that the inferred $\jmk$ colors of most
UV--selected LBGs at $z\sim 2$ with $\mcal \geq 10^{11}$~\msol\
satisfy the DRG selection criterion.   Therefore, including massive
LBGs into the integrated specific SFR would have little effect on our
conclusions, and we are confident that the results presented here
apply to the general galaxy population at $z\sim 1.5-3$ with $\mcal
\geq 10^{11}$~\msol.

This decline in the specific SFR of massive galaxies seems to signify
the point where massive galaxies have formed the bulk of their stellar
populations.  This may happen because they have converted most of
their gas reservoirs into stellar mass, and thus can no longer sustain
high SFRs.   This may be supported by the timescales for the galaxies
to sustain their current SFRs discussed in \S~\ref{section:sfrs}.
Alternatively, because some of the massive DRGs show indications that
they are fueling powerful AGN, it may be that AGN feedback provides a
means to suppress further star formation in these objects
\citep{hop05}. To understand the rapid evolution in the specific SFRs
of massive galaxies at these redshifts, we will need to better
understand the coeval assembly of stellar mass and AGN, and the
energetic feedback from the processes.

\section{Conclusions}\label{section:conclusions}

We have investigated the properties of massive, star--forming galaxies
at $z\sim 1-3.5$ using observations from the \textit{Hubble Space
Telescope}, ground--based near--infrared (IR) imaging, and IR
observations from the \textit{Spitzer Space Telescope} at
3--24~\micron.   From a \ks--selected galaxy sample over a $\simeq
130$~arcmin$^2$ field in GOODS--S, we identified 153 DRGs with
$(\jmk)_\mathrm{Vega} \ge 2.3$ to $\ks \leq 23.2$~mag, which is
approximately complete in stellar mass to $10^{11}$~\msol\ for $z\leq
3$.    The majority of DRGs (90\%) are  detected in deep \spitzer/IRAC
imaging at $3.6-8.0$~\micron, and the remainder are confused with
nearby sources within the IRAC beam. Roughly half of the DRGs are
detected by \spitzer/MIPS at 24~\micron\ with $f_\nu(24\micron) \geq
50$~\ujy.  At $z\sim 1-3.5$, these 24~\micron\ flux densities
correspond to IR luminosities of $\lir \sim 10^{11-13}$~\lsol.

Based on the full suite of photometry, we find that the DRG selection
identifies galaxies whose optical rest--frame light is dominated by a
population of evolved stars combined with trace amounts of ongoing
star formation ($z_\mathrm{med}\sim 2.5$), and galaxies whose light is
dominated by heavily extincted ($A_{1600} \gsim 4-6$~mag) starbursts
($z_\mathrm{med} \sim 1.7$), in roughly equal proportions.   Only a
very small fraction of the DRGs ($\lsim 10$\%) in the sample have SEDs
consistent with pure old ($\gsim 1$~Gyr) stellar populations with no
indication of current star formation.  We estimate that the number
density of massive, passively evolving DRGs at $2 \leq z \leq 3$ is
$3.2\times 10^{-5}$~$h_{70}^{-3}$ Mpc$^{-3}$, which is nearly an order
of magnitude lower than that for DRGs in the HDF--S (Labb\'e et al.\
2005), although some of this discrepancy likely results from the fact
that the HDF--S sample extends to fainter \ks--band sources.  The
candidates for passively--evolving DRGs at $2 \leq z \leq 3$ with
$\mcal \ge 10^{11}$~\msol\ contribute only a small fraction ($\lsim
10$\%) to the global stellar--mass density integrated over all
galaxies at these redshifts.  Thus, massive, passively evolving
objects are rare at $z\gsim 2$.  However, the fraction of the
stellar--mass density in massive, passively evolving galaxies
increases threefold from $z\gsim 2$ to 0.

The UV--derived SFRs for the DRGs are lower than those that
include the reradiated IR emission  by up to two orders of magnitude.
The DRGs have IR/UV luminosity ratios typically in excess of what
is expected from their rest--frame UV spectral slope, consistent with
observations of local LIRGs and ULIRGs.

We compare the DRG photometry to stellar population synthesis models
to estimate the stellar masses and to study the properties of the
stars that dominate the rest--frame UV through near--IR light.  Models
allowing for a previous stellar population formed in an instantaneous
burst in the distant past accompanied by additional on--going
star--formation generally provide better fits to the data.  We
conclude that the DRGs have complex and stochastic star--formation
histories consistent with other star--forming galaxies at these
redshifts and locally, and with predictions from hierarchical models.

DRGs at $z\sim 1.5-3$ with stellar masses greater than
$10^{11}$~\msol\ have specific SFRs ranging from $0.2-10$ Gyr$^{-1}$.
This is  more than an order of magnitude larger than that derived for
galaxies with stellar masses above $10^{11}$~\msol\ at $z\sim 0.7$ and
0.4.  Simultaneously, the most luminous and massive DRGs show indications
for the presence of AGN either based on X--ray luminosity, IR
luminosities, or IR colors. We find that as many as one--quarter
of the DRG population contain AGN, and therefore the  growth
of SMBHs coincides with the formation of massive galaxies at $z\gsim
1.5$.  Further implications from the  prevalence of AGN in DRGs will
be discussed in L.~Moustakas et al.\ (in preparation).

At $z\gsim 1.5$, galaxies with stellar masses $> 10^{11}$~\msol\ are
forming stars at rates at or slightly higher than the global value
integrated over all galaxies at this epoch.   In contrast, galaxies at
$z\lsim 0.7$ with $\mathcal{M} \geq 10^{11}$~$\msol$ are forming stars
at rates less than the global value for all galaxies. The evolution in
the specific SFRs of massive galaxies occurs at an accelerated rate
compared to that of all galaxies. The bulk of star formation in
massive galaxies occurs at early cosmic epochs and that further mass
assembly in these galaxies is accompanied by low specific SFRs. At
$z\lsim 1$ massive galaxies have formed most of their stellar mass,
and lower mass galaxies dominate the cosmic SFR density.

\acknowledgements

We wish to thank our colleagues for stimulating conversations, the
other members of the \spitzer\ MIPS/GTO and GOODS teams who
contributed to many aspects of this program, and the SSC and STScI
staffs for their optimal planning of the observations and efficient
processing of the \spitzer\ and \hst\ data, respectively.  We are
grateful to A.~Alonso--Herrero, T.~Budav\'ari, R.--R.~Chary,
R.~Dav\'e, R.~Kennicutt, K.~Misselt, J.~Monkiewicz, and D.~Stern for
assistance, and for helpful comments and suggestions. We also wish to
thank the anonymous referee, whose comments improved the clarity and
presentation of this work.  This research has made use of the
NASA/IPAC Extragalactic Database (NED) which is operated by the Jet
Propulsion Laboratory, California Institute of Technology, under
contract with the National Aeronautics and Space Administration.
Partial support for this work was provided by NASA through the Spitzer
Space Telescope Fellowship Program, through a contract issued by the
Jet Propulsion Laboratory (JPL), California Institute of Technology
(Caltech) under a contract with NASA.  The work of LAM and PRME was
carried out at JPL/Caltech, under NASA.   DMA thanks the Royal Society
for support.   Partial support for this work was provided by NASA
through an award issued by JPL/Caltech, and by NASA as part of the
\spitzer\ Legacy Science Program.   Partial support was provided by
NASA through grant GO09583.01-96A from STScI, which is operated by
AURA, Inc., under NASA contract NAS5-26555.




\begin{thebibliography}

\bibitem[Adelberger \& Steidel(2000)]{ade00} Adelberger, 
K.~L., \& Steidel, C.~C.\ 2000, \apj, 544, 218

\bibitem[Alexander et al.(2003)]{ale03} Alexander, D.~M., et 
al.\ 2003, \aj, 126, 539 

\bibitem[Alexander et al.(2005)]{alex05} Alexander, D.~M., 
Smail, I., Bauer, F.~E., Chapman, S.~C., Blain, A.~W., Brandt, W.~N., \& 
Ivison, R.~J.\ 2005, \nat, 434, 738


\bibitem[Alonso--Herrero \etal(2005)]{alo05} Alonso--Herrero, A.,
\etal, 2005, \apj, submitted

\bibitem[Appleton et al.(2004)]{app04} Appleton, P.~N., et 
al.\ 2004, \apjs, 154, 147

\bibitem[Baldry \& Glazebrook(2003)]{bal03} Baldry, I.~K., \& 
Glazebrook, K.\ 2003, \apj, 593, 258 

\bibitem[Baldry et al.(2004)]{bal04} Baldry, I.~K., Glazebrook, K.,
Brinkmann, J., Ivezi{\' c}, {\v Z}., Lupton, R.~H., Nichol, R.~C., \&
Szalay, A.~S.\ 2004, \apj, 600, 681  

\bibitem[Barger et al.(2000)]{bar00} Barger, A.~J., Cowie, 
L.~L., \& Richards, E.~A.\ 2000, \aj, 119, 2092 

\bibitem[Bauer \etal(2005)]{bauer05} Bauer, A.~E., Drory, N., Hill,
G.~J., \& Feulner, G. 2005, \apj, 621, L89

\bibitem[Baugh \etal(1998)]{bau98} 
Baugh, C.~M., Cole, S., Frenk, C.~S., \& Lacey, C.~G. 1998, \apj, 498, 504

\bibitem[Baugh et al.(2005)]{bau05} Baugh, C.~M., Lacey, 
C.~G., Frenk, C.~S., Granato, G.~L., Silva, L., Bressan, A., Benson, A.~J., 
\& Cole, S.\ 2005, \mnras, 356, 1191 

\bibitem[Bell(2003)]{bel03} Bell, E.~F.\ 2003, \apj,
586, 794 

\bibitem[Bell et al.(2003)]{bel03b} Bell, E.~F., McIntosh, 
D.~H., Katz, N., \& Weinberg, M.~D.\ 2003, \apjs, 149, 289

\bibitem[Bell \etal(2004)]{bel04}
Bell, E.~F., \etal\ 2004, \apj, 608, 752 

\bibitem[Bell \etal(2005a)]{bel05}
Bell, E.~F., \etal\ 2005a, \apj, 625, 23

\bibitem[Bell \etal(2005b)]{bel05b}
Bell, E.~F., \etal\ 2005b, \apj, submitted (astro--ph/0506425)

\bibitem[Bertin \& Arnouts(1996)]{ber96}
Bertin, E., \& Arnouts, S. 1996, \aaps, 117, 393

\bibitem[Blain et al.(2002)]{bla02} Blain, A.~W., Smail, I., 
Ivison, R.~J., Kneib, J.-P., \& Frayer, D.~T.\ 2002, \physrep, 369, 111 

\bibitem[Blain et al.(2004)]{bla04} Blain, A.~W., Chapman, 
S.~C., Smail, I., \& Ivison, R.\ 2004, \apj, 611, 725 

\bibitem[Bouwens et al.(2004)]{bou04} Bouwens, R.~J., et al.\ 
2004, \apjl, 606, L25 

\bibitem[Brandt \& Hasinger(2005)]{brandt05} Brandt, W.~N., \& Hasinger,
G. 2005, \araa, in press (astro--ph/0501058)

\bibitem[Brinchmann \& Ellis(2000)]{bri00} Brinchmann, J., \& Ellis,
R.~S. 2000, \apj, 536, L77

\bibitem[Bruzual \& Charlot(2003)]{bru03}
Bruzual, G.~A., \& Charlot, S. 2003, \mnras, 344, 1000

\bibitem[Budav\'ari \etal(2000)]{bud00} Budav\'ari, T., Szalay, A.~S.,
Connolly, A.~J., Csabai, I., \& Dickinson, M.\ 2000, \aj, 120, 1588


\bibitem[Buat et al.(2005)]{bua05} Buat, V., et al.\ 2005, 
\apjl, 619, L51 

\bibitem[Calzetti et al.(2000)]{cal00} Calzetti, D., Armus, 
L., Bohlin, R.~C., Kinney, A.~L., Koornneef, J., \& Storchi-Bergmann, T.\ 
2000, \apj, 533, 682 

\bibitem[Calzetti et al.(1994)]{cal94} Calzetti, D., Kinney, 
A.~L., \& Storchi-Bergmann, T.\ 1994, \apj, 429, 582 


\bibitem[Caputi et al.(2005)]{cap05} Caputi, K., et al.\ 2005, ApJ,
submitted 

\bibitem[Chary \& Elbaz(2001)]{cha01}
Chary, R.~R., \& Elbaz, D. 2001, \apj, 556, 562

\bibitem[Chapman et al.(2003)]{cha03}
Chapman, S. C., Helou, G., Lewis, G. F., \& Dale, D. A. 2003, \apj,
588, 186

\bibitem[Chapman et al.(2005)]{cha05} Chapman, S.~C., Blain, 
A.~W., Smail, I., \& Ivison, R.~J.\ 2005, \apj, 622, 772 

\bibitem[Charmandaris et al.(2004)]{cha04a} Charmandaris, V., 
Le Floc'h, E., \& Mirabel, I.~F.\ 2004, \apjl, 600, L15 
\bibitem[Cimatti et al.(2002a)]{cim02} Cimatti, A., et al.\ 
2002a, \aap, 381, L68

\bibitem[Cimatti et al.(2002b)]{cim02b} Cimatti, A., et al.\ 
2002b, \aap, 391, L1

\bibitem[Cole \etal(2001)]{col01}  Cole, S., \etal\ 2001, MNRAS, 326, 255

\bibitem[Coleman et al.(1980)]{col80} Coleman, G.~D., Wu, 
C.-C., \& Weedman, D.~W.\ 1980, \apjs, 43, 393 

\bibitem[Cowie et al.(1999)]{cow99} Cowie, L.~L., Songaila, 
A., \& Barger, A.~J.\ 1999, \aj, 118, 603

\bibitem[Daddi et al.(2003)]{dad03} Daddi, E., \etal\ 2003, \apj, 588, 50 

\bibitem[Daddi et al.(2004)]{dad04} Daddi, E., Cimatti, A., 
Renzini, A., Fontana, A., Mignoli, M., Pozzetti, L., Tozzi, P., \& 
Zamorani, G.\ 2004, \apj, 617, 746

\bibitem[Daddi et al.(2005a)]{dad05} Daddi, E., \etal\ 2005a, \apj,
626, 680

\bibitem[Daddi et al.(2005b)]{dad05b} Daddi, E., \etal\ 2005b, \apjl,
631, L13

\bibitem[Dale et al.(2001)]{dal01}
Dale, D. A., Helou, G., Contursi, A., Silbermann, N. A., \& Kolhatkar,
S., 2001, \apj, 549, 215

\bibitem[Dale \& Helou(2002)]{dal02} Dale, D.~A., \& Helou, 
G.\ 2002, \apj, 576, 159

\bibitem[De Breuck et al.(2003)]{deb03} De Breuck, C., \etal\ 2003,
\aap, 401, 911

\bibitem[De Breuck et al.(2005)]{deb05} De Breuck, C., 
Downes, D., Neri, R., van Breugel, W., Reuland, M., Omont, A., \& Ivison, 
R.\ 2005, \aap, 430, L1

\bibitem[De Lucia et al.(2005)]{delucia05} De Lucia,~G., Springel,~V.,
White,~S.~D.~M., Croton,~D., \& Kauffmann,~G.~K.\ 2005, \mnras,
submitted (astro--ph/0509725)

\bibitem[Di Matteo et al.(2005)]{dimat05} Di Matteo, T., 
Springel, V., \& Hernquist, L.\ 2005, \nat, 433, 604 

\bibitem[Dickinson \etal(2000)]{dic00} Dickinson, M.~\etal\ 2000,
\apj, 531, 624 

\bibitem[Dickinson \etal(2003)]{dic03} Dickinson, M., Papovich, C.,
  Ferguson, H.~C., \& Budav\'ari, T. 2003, \apj, 587, 25

\bibitem[Donley et al.(2005)]{don05} Donley, J. L., Rieke, G. H.,
Rigby, J., \& P\'erez--Gonz\'alez, P.~G.\ 2005, \apj, in press
(astro-ph/0507676) 

\bibitem[Downes \& Solomon(1998)]{dow98} Downes, D., \& 
Solomon, P.~M.\ 1998, \apj, 507, 615 
 
\bibitem[Dunlop \etal(1996)]{dun96} Dunlop,~J., Peacock,~J.,
Spinrad,~H., Dey,~A., Jimenez,~R., Stern,~D., \& Windhorst, R.~A.\
1996, \nat, 381, 581

\bibitem[Eggen, Lynden--Bell, \& Sandage(1962)]{egg62} Eggen, O. J.,
Lynden--Bel, D., \& Sandage, A.~R., 1962, \apj, 136, 748

\bibitem[Elbaz et al.(2002)]{elb02} Elbaz, D., Cesarsky, 
C.~J., Chanial, P., Aussel, H., Franceschini, A., Fadda, D., \& Chary, 
R.~R.\ 2002, \aap, 384, 848 


\bibitem[Eyles et al.(2005)]{eyl05}
Eyles, L., Bunker A., Stanway, E., Lacy, M., \& Ellis, R.\ 2005,
\mnras, in press (astro--ph/0502385)

\bibitem[Faber et al.(2005)]{fab05}
Faber, S.~M., \etal\ 2005, \apj, submitted (astro--ph/0506044)

\bibitem[Ferguson et al.(2002)]{fer02} Ferguson, H.~C., 
Dickinson, M., \& Papovich, C.\ 2002, \apjl, 569, L65 

\bibitem[Finlator et al.(2005)]{fin05} Finlator, K., Dav\'e,~R.,
Papovich,~C., \& Hernquist,~L.\ 2005, \apj, submitted (astro--ph/0507719)

\bibitem[Fontana \etal(2003)]{fon03} Fontana, A., et al.\ 2003, \apj,
594, L9

\bibitem[Fontana \etal(2004)]{fon04} Fontana, A., et al.\ 2004, \aap, 424, 23 

\bibitem[F\"{o}rster--Schreiber et al.(2003)]{for03}
F\"orster--Schreiber, N.~M., Genzel, R., Lutz, D., \& Sternberg, A.\
2003, \apj, 599, 193

\bibitem[F\"{o}rster--Schreiber et al.(2004)]{for04}
F\"{o}rster--Schreiber, N.~M. \etal, 2004, \apj, 616, 40

\bibitem[Francis et al.(2004)]{fra04} Francis, P.~J., Nelson, 
B.~O., \& Cutri, R.~M.\ 2004, \aj, 127, 646

\bibitem[Franx et al.(2003)]{fra03} Franx, M., et al.\ 2003, 
\apjl, 587, L79

\bibitem[Gebhardt et al.(2000)]{geb00}  Gebhardt, K., et al.\ 
2000, \apjl, 539, L13 

\bibitem[Giacconi \etal(2002)]{giac02}
Giacconi, R., et al.\ 2002, \apjs, 139, 369

\bibitem[Giavalisco(2002)]{giav02} Giavalisco, M.\ 2002, 
\araa, 40, 579 

\bibitem[Giavalisco et al.(2004a)]{gia04a} Giavalisco, M., et al.\ 2004a,
\apjl, 600, L93

\bibitem[Giavalisco et al.(2004b)]{gia04b} Giavalisco, M., et 
al.\ 2004b, \apjl, 600, L103 

\bibitem[Gilli(2004)]{gilli04} Gilli, R.\ 2004, Adv.\ Space Res., 34, 2470

\bibitem[Glazebrook et al.(2004)]{gla04} Glazebrook, K., et 
al.\ 2004, \nat, 430, 181 

\bibitem[Goldader et al.(2002)]{gol02} Goldader, J.~D., 
Meurer, G., Heckman, T.~M., Seibert, M., Sanders, D.~B., Calzetti, D., \& 
Steidel, C.~C.\ 2002, \apj, 568, 651

\bibitem[Gordon \etal(2005)]{gor05} Gordon, K., et al.\ 2005, \pasp,
117, 503

\bibitem[Granato et al.(2001)]{gran01} Granato, G.~L., Silva, 
L., Monaco, P., Panuzzo, P., Salucci, P., De Zotti, G., \& Danese, L.\ 
2001, \mnras, 324, 757 

\bibitem[Greve, Ivison, \& Papadopoulos(2004)]{gre04} Greve, T.~R.,
Ivison, R.~J., Papadopoulos, P.~P.\ 2004, \aap, 419, 99

\bibitem[Haas et al.(2003)]{haa03} Haas, M., et al.\ 2003, 
\aap, 402, 87 

\bibitem[Heavens et al.(2004)]{hea04} Heavens, A., Panter, 
B., Jimenez, R., \& Dunlop, J.\ 2004, \nat, 428, 625

\bibitem[Hernquist \& Springel(2003)]{her03}
Hernquist, L., \& Springel, V. 2003, \mnras, 341, 1253

\bibitem[Hopkins(2004)]{hop04} Hopkins, A. 2004, ApJ, 615, 209

\bibitem[Hopkins \etal(2005)]{hop05}
Hopkins, P. F., Hernquist, L., Cox, T.~J., Di Matteo, T., Martini, P.,
Robertson, B., \& Springel, V. 2005, \apj, 625, 71

\bibitem[Hornschemeier \etal(2001)]{hor01} Hornschemeier, A., et al.\
2001, \apj, 554, 742

\bibitem[Im et al.(2002)]{im02} Im, M., Yamada, T., Tanaka, 
I., \& Kajisawa, M.\ 2002, \apjl, 578, L19 

\bibitem[Juneau et al.(2005)]{jun05} Juneau, S., et al.\ 
2005, \apjl, 619, L135 

\bibitem[Kauffmann \& Charlot(1998)]{kau98} Kauffmann, G., \& 
Charlot, S.\ 1998, \mnras, 294, 705

\bibitem[Kauffmann et al.(2004)]{kau04} Kauffmann, G., White, 
S.~D.~M., Heckman, T.~M., M{\' e}nard, B., Brinchmann, J., Charlot, S., 
Tremonti, C., \& Brinkmann, J.\ 2004, \mnras, 353, 713 

\bibitem[Kennicutt(1998)]{ken98} Kennicutt, R.~C.\ 1998, 
\araa, 36, 189

\bibitem[Kere{\v s} et al.(2005)]{ker05} Kere{\v s}, D., 
Katz, N., Weinberg, D.~H., \& Dav{\'e}, R.\ 2005, \mnras, 823 

\bibitem[Kinney et al.(1996)]{kin96} Kinney, A.~L., Calzetti, 
D., Bohlin, R.~C., McQuade, K., Storchi-Bergmann, T., \& Schmitt, H.~R.\ 
1996, \apj, 467, 38 

\bibitem[Knudsen et al.(2005)]{knu05} Knudsen, K.~K., \etal\ 2005,
ApJ, in press (astro--ph/0509104)

\bibitem[Kong et al.(2004)]{kon04} Kong, X., Charlot, S., 
Brinchmann, J., \& Fall, S.~M.\ 2004, \mnras, 349, 769 



\bibitem[Labb\'e \etal(2005)]{lab05}
Labb\'e, I.~et al.\ 2005, \apjl, 624, L81

\bibitem[Lacy \etal(2004)]{lac04}
Lacy, M., \etal\ 2004, \apjs, 154, 166

\bibitem[Larson(2005)]{lar05} Larson, R.~B. 2005, \mnras, 359, 211

\bibitem[Le F{\` e}vre et al.(2004)]{leferve04} Le F{\` e}vre, 
O., et al.\ 2004, \aap, 428, 1043 

\bibitem[Le Floc'h et al.(2001)]{lef01} Le Floc'h, E., 
Mirabel, I.~F., Laurent, O., Charmandaris, V., Gallais, P., Sauvage, M., 
Vigroux, L., \& Cesarsky, C.\ 2001, \aap, 367, 487 

\bibitem[Le Floc'h et al.(2005)]{lef05} Le Floc'h, E., et 
al.\ 2005, \apj, in press (astro--ph/0506462)



\bibitem[Madau et al.(1998)]{mad98} Madau, P., Pozzetti, L., \&
Dickinson, M.\ 1998, \apj, 498, 106

\bibitem[Magorrian et al.(1998)]{mag98} Magorrian, J., et 
al.\ 1998, \aj, 115, 2285 

\bibitem[Marcillac et al.(2005)]{mar05} Marcillac, D., Elbaz,~D.,
Chary,~R.~R., Dickinson,~M., Galliano,~F., \& Morrison,~G.\ 2005,
\aap, submitted

\bibitem[McCarthy(2004)]{mccar04} McCarthy, P.~J.\ 2004, \araa, 
42, 477 

\bibitem[McCarthy \etal(2004)]{mccar04b} McCarthy, P.~J., et al.\
2004, \apj, 614, L9

\bibitem[Meurer \etal(1999)]{meu99}
Meurer, G., Heckman, T.~M., \& Calzetti, D. 1999, \apj, 521, 64

\bibitem[Mignoli et al.(2005)]{mig05} Mignoli, M., \etal\ 2005, \aap,
437, 883

\bibitem[Mobasher et al.(2004)]{mob04} Mobasher, B., et al.\ 
2004, \apjl, 600, L167

\bibitem[Moustakas \& Somerville(2002)]{mou02} Moustakas, 
L.~A., \& Somerville, R.~S.\ 2002, \apj, 577, 1

\bibitem[Moustakas et al.(2004)]{mou04} Moustakas, L.~A., et 
al.\ 2004, \apjl, 600, L131

\bibitem[Nagamine et al.(2005)]{nag05} Nagamine, K., Cen, R.,
Hernquist, L, Ostriker, J. P., \& Springel, V. 2005, \apj, 627, 608

\bibitem[Neugebauer \etal(1979)]{neu79} Neugebauer, G., Oke, J. B.,
Becklin, E. G., \& Matthews, K. 1979, \apj 230, 79

\bibitem[Papovich, Dickinson, \& Ferguson(2001)]{pap01}
Papovich, C., Dickinson, M., \& Ferguson, H.~C. 2001, \apj, 559, 620

\bibitem[Papovich \& Bell(2002)]{pap02} Papovich, C., \& 
Bell, E.~F.\ 2002, \apjl, 579, L1 

\bibitem[Papovich \etal(2004a)]{pap04a}
Papovich, C., et al.\ 2004a, \apjl, 600, L111 

\bibitem[Papovich \etal(2004b)]{pap04b}
Papovich, C., et al.\ 2004b, \apjs, 154, 70

\bibitem[Papovich \etal(2005)]{pap05}
Papovich, C., Dickinson, M., Giavalisco, M., Conselice, C.~J., \&
Ferguson, H.~C. 2005, \apj, 631, 101

\bibitem[P\'erez--Gonz\'alez \etal(2005)]{per05} 
P\'erez--Gonz\'alez, P.~G. \etal\ 2005, ApJ, 630, 82

\bibitem[Persic et al.(2004)]{per04} Persic, M., Rephaeli, 
Y., Braito, V., Cappi, M., Della Ceca, R., Franceschini, A., \& Gruber, 
D.~E.\ 2004, \aap, 419, 849 

\bibitem[Ranalli et al.(2003)]{ran03} Ranalli, P., Comastri, 
A., \& Setti, G.\ 2003, \aap, 399, 39

\bibitem[Reddy et al.(2005)]{red05} Reddy, N.~A., Erb,~D.~K.,
Steidel,~C.~C., Shapley,~A.~E., Adelberger,~K.~L., \& Pettini,~M.\
2005, \apj, in press (astro-ph/0507264)

\bibitem[Rieke(1978)]{rie78}
Rieke, G. H.\ 1978, \apj, 226, 550


\bibitem[Roussel et al.(2001)]{rou01} Roussel, H., Sauvage, 
M., Vigroux, L., \& Bosma, A.\ 2001, \aap, 372, 427 

\bibitem[Rubin et al.(2004)]{rub04} Rubin, K.~H.~R., van 
Dokkum, P.~G., Coppi, P., Johnson, O., F{\" o}rster Schreiber, N.~M., 
Franx, M., \& van der Werf, P.\ 2004, \apjl, 613, L5

\bibitem[Rudnick \etal(2003)]{rud03}
Rudnick, G., et al.\ 2003, \apj, 599, 847

\bibitem[Sanders et al.(1988)]{san88} Sanders, D.~B., Soifer, 
B.~T., Elias, J.~H., Neugebauer, G., \& Matthews, K.\ 1988, \apjl, 328, L35 
 
\bibitem[Sanders \& Mirabel(1996)]{san96} Sanders, D.~B., \& 
Mirabel, I.~F.\ 1996, \araa, 34, 749 

\bibitem[Sanders et al.(2003)]{san03} Sanders, D.~B., 
Mazzarella, J.~M., Kim, D.-C., Surace, J.~A., \& Soifer, B.~T.\ 2003, \aj, 
126, 1607 

\bibitem[Sawicki \& Yee(1998)]{saw98} Sawicki, M.~\& Yee, 
H.~K.~C.\ 1998, \aj, 115, 1329 

\bibitem[Sawicki(2002)]{saw02} Sawicki, M.\ 2002, \aj, 124, 3050  


\bibitem[Shapley \etal(2001)]{sha01} Shapley, A.~E., Steidel, C.~C.,
Adelberger, K.~L., Dickinson, M., Giavalisco, M., \& Pettini, M. 2001,
\apj, 562, 95

\bibitem[Shapley et al.(2004)]{sha04} Shapley, A.~E., Erb, 
D.~K., Pettini, M., Steidel, C.~C., \& Adelberger, K.~L.\ 2004, \apj, 612, 
108 

\bibitem[Shapley et al.(2005)]{sha05} Shapley, A.~E., 
Steidel, C.~C., Erb, D.~K., Reddy, N.~A., Adelberger, K.~L., Pettini, M., 
Barmby, P., \& Huang, J.\ 2005, \apj, 626, 698

\bibitem[Simpson \& Eisenhardt(1999)]{sim99} Simpson, C., \& 
Eisenhardt, P.\ 1999, \pasp, 111, 691 
 

\bibitem[Silva \etal(1998)]{sil98} Silva, L., Granato, 
G.~L., Bressan, A., \& Danese, L.\ 1998, \apj, 509, 103 

\bibitem[Smail et al.(2002)]{sma02} Smail, I., Owen, F.~N., 
Morrison, G.~E., Keel, W.~C., Ivison, R.~J., \& Ledlow, M.~J.\ 2002, \apj, 
581, 844 

\bibitem[Soifer et al.(1995)]{soi95} Soifer, B.~T., Cohen, 
J.~G., Armus, L., Matthews, K., Neugebauer, G., \& Oke, J.~B.\ 1995, \apjl, 
443, L65 

\bibitem[Somerville, Primack, \& Faber(2001)]{som01} 
Somerville, R.~S., Primack, J.~R., \& Faber, S.~M. 2001, \mnras, 320, 504

\bibitem[Spinoglio et al.(1995)]{spi95} Spinoglio, L., 
Malkan, M.~A., Rush, B., Carrasco, L., \& Recillas-Cruz, E.\ 1995, \apj, 
453, 616 

\bibitem[Spinrad et al.(1997)]{spi97} Spinrad, H., Dey, A., 
Stern, D., Dunlop, J., Peacock, J., Jimenez, R., \& Windhorst, R.\ 1997, 
\apj, 484, 581

\bibitem[Spoon \etal(2004)]{spo04} Spoon, H.~W.~W., 
Moorwood, A.~F.~M., Lutz, D., Tielens, A.~G.~G.~M., Siebenmorgen, R., \& 
Keane, J.~V.\ 2004, \aap, 414, 873 

\bibitem[Springel et al.(2005a)]{spr05a} Springel, V., Di 
Matteo, T., \& Hernquist, L.\ 2005a, \apjl, 620, L79 

\bibitem[Springel \etal(2005b)]{spr05b} Springel, V., et al.\ 2005b,
\nat, 435, 629

\bibitem[Steidel \etal(1996)]{ste96} Steidel, C.~C., Giavalisco, M.,
Pettini, M., Dickinson, M., \& Adelberger, K.~L. 1996, \apj, 462, L17

\bibitem[Steidel et al.(1999)]{ste99} Steidel, C.~C., 
Adelberger, K.~L., Giavalisco, M., Dickinson, M., \& Pettini, M.\ 1999, 
\apj, 519, 1

\bibitem[Stern et al.(2005)]{stern05} Stern, D., et al.\ 2005, \apj,
631, 163

\bibitem[Szokoly et al.(2004)]{szo04} Szokoly, G.~P., et al.\ 
2004, \apjs, 155, 271 

\bibitem[Teng \etal(2005)]{ten05} Teng, S.~H., Wilson, A.~S.,
Veilleux,~S., Young,~A.~J., Sanders,~D.~B., \& Nagar,~N.~M.\ 2005,
\apj, in press (astro--ph/0508112)

\bibitem[Totani et al.(2001)]{tot01} Totani, T., Yoshii, Y., 
Iwamuro, F., Maihara, T., \& Motohara, K.\ 2001, \apjl, 558, L87 
 
\bibitem[Treu et al.(2005)]{treu05} Treu, T., Ellis, R.~S., 
Liao, T.~X., \& van Dokkum, P.~G.\ 2005, \apjl, 622, L5 

\bibitem[van Dokkum et al.(2003)]{vandok03} van Dokkum, P.~G., 
et al.\ 2003, \apjl, 587, L83 

\bibitem[van Dokkum et al.(2004)]{vandok04} van Dokkum, P.~G., 
et al.\ 2004, \apj, 611, 703 

\bibitem[van Dokkum(2005)]{vandok05b} van Dokkum, P.~G.\ 2005, ApJ,
submitted (astro-ph/0506661)

\bibitem[van Dokkum et al.(2005)]{vandok05} van Dokkum, P.~G., Kriek,
M., Rodgers, B., Franx, M., \& Puxley, P.\ 2005, \apjl, 622, L33

\bibitem[Vanzella et al.(2005)]{vanz04} Vanzella, E., et al. 2005,
\aap, 434, 53

\bibitem[Veilleux et al.(1995)]{vei95} Veilleux, S., Kim, 
D.-C., Sanders, D.~B., Mazzarella, J.~M., \& Soifer, B.~T.\ 1995, \apjs, 
98, 171 

\bibitem[Veilleux et al.(1999)]{vei99} Veilleux, S., Kim, 
D.-C., \& Sanders, D.~B.\ 1999, \apj, 522, 113 

\bibitem[Williams \etal(1996)]{wil96}
Williams, R.~E., \etal\ 1996, \aj, 112, 1335

\bibitem[Wilson et al.(2004)]{wil04} Wilson, G., et al.\ 
2004, \apjs, 154, 107 

\bibitem[Wolf et al.(2003)]{wol03}
Wolf, C., Meisenheimer, K., Rix, H.--W. Borch, A., Dye, S., \&
Kleinheinrich, M. 2003, A\&A, 401, 73

\bibitem[Wolf et al.(2004)]{wol04}
Wolf, C., et al.\ 2004, \aap, 421, 913

\bibitem[Worsley et al.(2005)]{wor05} Worsley, M.~A., et al.\ 
2005, \mnras, 357, 1281 

\bibitem[Yan et al.(2004a)]{yan04a} Yan, H., et al.\ 2004a, 
\apj, 616, 63 

\bibitem[Yan et al.(2005)]{yan05} Yan, H., et al.\ 2005, \apj, in
press (astro-ph/0507673)

\bibitem[Yan et al.(2004b)]{yan04b} Yan, L., et al.\ 2004b, 
\apjs, 154, 75 

\end{thebibliography}
\end{document}